\documentclass[preprint>]{elsarticle}
\usepackage{amssymb,amsmath,bm,mathtools,xcolor,float}
\numberwithin{equation}{section}
\usepackage[ruled,linesnumbered,vlined]{algorithm2e}
\usepackage{graphicx}
\usepackage[colorlinks=true,linkcolor=blue]{hyperref}
\usepackage{geometry}
\usepackage{subcaption}	
\begin{document}

\captionsetup[subfigure]{skip=1pt, singlelinecheck=false}

\title{Emerging contact force heterogeneity in ordered soft granular media}
\author[1]{Liuchi Li\corref{cor1}%
 \fnref{fn1}}
\ead{liuchili@alumni.caltech.edu}
\cortext[cor1]{Corresponding author}
\fntext[fn1]{Present address: Hopkins Extreme Materials Institute, Johns Hopkins University, Baltimore, MD 21218, USA.}
\address[1]{Earth and Environmental Science Area, Lawrence Berkeley National Laboratory, Berkeley, CA 94720, USA}
\author[2]{Konstantinos Karapiperis}
\address[2]{Mechanics \& Materials Lab, Department of Mechanical and Process Engineering, ETH Z\"{u}rich, 8092 Z\"{u}rich, Switzerland}
\author[3]{Jos\'{e} E. Andrade}
\address[3]{Division of Engineering and Applied Science, California Institute of Technology, Pasadena, CA 91125, USA}
\begin{abstract}
Under external perturbations, inter-particle forces in disordered granular media are well known to form a heterogeneous distribution with filamentary patterns. Better understanding these forces and the distribution is important for predicting the collective behavior of granular media, the media second only to water as the most manipulated material in global industry. However, studies in this regard so far have been largely confined to granular media exhibiting only geometric heterogeneity, leaving the dimension of mechanical heterogeneity a rather uncharted area. Here, through a FEM contact mechanics model, we show that a heterogeneous inter-particle force distribution can also emerge from the dimension of mechanical heterogeneity alone. Specifically, we numerically study inter-particle forces in packing of mechanically heterogeneous disks arranged over either a square or a hexagonal lattice and under quasi-static isotropic compression. Our results show that, at the system scale, a hexagonal packing exhibit a more heterogeneous inter-particle force distribution than a square packing does; At the particle scale, for both packing lattices, preliminary analysis shows the consistent coexistence of outliers (i.e., softer disks sustaining larger forces while stiffer disks sustaining smaller forces) in comparison to their homogeneous counterparts, which implies the existence of nonlocal effect. Further analysis on the portion of outliers and on spatial contact force correlations suggest that the hexagonal packing shows more pronounced nonlocal effect over the square packing under small mechanical heterogeneity. However, such trend is reversed when assemblies becomes more mechanically heterogeneous. Lastly, we confirm that, in the absence of particle reorganization events, contact friction merely plays the role of packing stabilization while its variation has little effect on inter-particle forces and their distribution.
 \end{abstract}
 \begin{keyword}
Soft granular media; Contact mechanics; Finite element method; Mechanical heterogeneity; Inter-particle force distribution; Spatial force correlation
 \end{keyword}
 \maketitle
\section{Introduction}
Upon external perturbations, inter-particle contact forces in disordered granular media are well known to form, both experimentally \cite {majmudar2005contact} and numerically \cite {radjai1996force}, a spatially heterogeneous distribution with filamentary patterns (i.e., force chains). These forces and together with the distribution are well known to play a pivotal role in determining how granular media collectively behave (e.g., shear banding \cite {kawamoto2018all} and solid-liquid phase transitioning \cite {andreotti2013granular, li2020identifying}) and interact with external stimuli (e.g., intruder impact \cite {clark2015nonlinear} and wave propagation \cite {zhai2020influence}). Understanding them is therefore relevant to many applications in engineering (e.g., designing adaptive devices \cite {daraio2006energy, wangli2021}) and (geo-) physics (e.g., mitigating geophysical hazards \cite {johnson2005nonlinear}).

Numerous studies have shown that features of inter-particle forces and the distribution depend non-trivially and sensitively on the specific packing structure of a granular media \cite {dantu1957contribution, gendelman2016determines, PhysRevLett.117.159801, hurley2017linking, kollmer2019betweenness}, with the packing structure itself being also heterogeneous (``geometric heterogeneity") and depends on various properties of the constituent particles such as particle shape \cite {kawamoto2018all, azema2013packings, li2019capturing, karapiperis2020investigating, wang2020shear}, size polydispersity \cite {nguyen2014effect}, friction \cite {blair2001force, binaree2020combined}. These studies leveraged either advanced or numerical techniques to quantify contact forces and study their distribution formed within a granular packing under external mechanical perturbations: On the experiment side, photo-elastic experiments \cite {majmudar2005contact, drescher1972photoelastic} using rubber-like birefringent materials have been playing a pivotal role in quantify contact forces in deformable particle packings; On the simulation side, the explicit Discrete Element Method (DEM) \cite {cundall1979discrete} and the implicit Non-smooth Contact Dynamics (NSCD) \cite {jean1999non} method have been the two major means to quantify contact forces in rigid particle packings. Very recently, these two numerical methods are also being extended to study the collective compaction behavior of highly deformable \cite{vu2019numerical, cantor2020compaction} or compressible \cite{vu2021effects} particle packings, or that of bi-mixtures of rigid and deformable particle packings \cite{cardenas2020compaction}. These experimental and numerical studies found that, for disordered disk or sphere packings, both the normal and tangential (frictional) contact forces show exponential distributions for strong forces (i.e., those above the mean) and show power-law distributions for weak forces (i.e., those below the mean) \cite{majmudar2005contact,radjai1996force}. Further, the observation of exponentially distributed strong normal forces seems to be insensitive to particle shape variation based on studies investigating rigid polyhedron \cite{azema2009quasistatic} and deformable ellipse packings \cite{wang2021contact}, although the associated scaling exponent depends on particle shape. In addition, strong normal forces were found to gradually switch from showing an exponential distribution to showing a Gaussian distribution as a packing’s size polydispersity decreases \cite{voivret2009multiscale}, with the Gaussian distribution being discovered in ordered packings of frictionless and rigid disks \cite{van2007tail}. Lastly, it was found that, as far as normal contact forces at the boundaries of sphere packings are concerned, their show an exponential distribution whose shape is insensitive to contact friction \cite{blair2001force}. However, nearly all studies to date have been focusing on granular media composed of mechanically homogeneous (being either rigid or deformable) particles, leaving the aspect of mechanical heterogeneity a rather uncharted area. The first attempt to aim at exploring the aspect of mechanical heterogeneity, to the best of our knowledge, dates back to 1986 where a set of experiments were performed to investigate force transmissions in bi-mixtures of plexiglass and rubber particles arranged over a hexagonal lattice \cite {travers1986uniaxial}. This experimental study suggested that - though only qualitatively - heterogeneous contact forces can also be induced by only mechanical heterogeneity. Unfortunately, further investigations along this aspect have since remained largely undeveloped. As a result, it still remains unclear how contact forces are distributed in mechanically heterogeneous granular media.

In this paper, we attempt to study quantitatively via simulations inter-particle forces and the distribution within granular media composed of mechanically heterogeneous particles. We consider it an interesting and important problem, not only because of its relevance to many geophysical applications (where geo-materials can be highly heterogeneous mechanically), but due to a more fundamental aspect of opening a potential avenue of engineering novel granular media through a bottom-up perspective. Such bottom-up engineering may be achieved, in the future, through a tactical combination of mechanical heterogeneity and geometric heterogeneity, which in turn allows us to actively control the contact force distribution of granular media (e.g., achieving a homogeneous inter-particle force distribution). As a point of departure, in this paper we numerically approach this problem in its minimal dimension possible: we isolate the dimension of mechanical heterogeneity by considering packing of mono-sized disks (plane-strain cylinders) arranged over two canonical lattices: a square lattice and a hexagonal lattice. We also assume small deformation within every disk such that point-like contacts and contact forces can still be rationally defined. Lastly, we assume that no particle reorganization occurs upon quasi-static loading. Here the phrase ``particle reorganization" refers to particle movements that involve dynamical events and potential finite deformations that can also lead to the loss of inter-particle contacts. Examples are abrupt dynamical frictional slips between two contacting disks that may also cause large disk rotations and the loss of contacts between the two disks. Essentially, we restrain our scope to study disk assemblies where every disk is able to achieve static equilibrium given the surrounding contact tractions (normal and frictional tractions) in the limit of small deformation. Within the scope being defined, we wish to explore the following three questions:
\begin {itemize}
\item At the system scale, what does the inter-particle force distribution look like and how does it differ between the two packing lattices?
\item At the particle scale, how does the variation of the mechanical property of a disk alters the amount of force the disk sustains and how does it differ between the two packing lattices?
\item At both the system and the particle scale, does contact friction play a role for either lattice?
\end {itemize}

The rest of the paper is organized as follows. In section. \ref{section:modeling}, we briefly introduce an implementation of a 2D FEM multi-body contact mechanics algorithm whose details together with benchmark tests are presented in Appendix. In section. \ref{section:disks}, we apply the implemented algorithm to model ordered packing under quasi-static isotropic compression and analyze inter-particle forces on both the system scale and the particle scale. We also discuss the effect of contact friction. In section. \ref{section:summary}, we conclude with a brief summary of our findings and the inspired outlook for future work.

\section{Modeling methodology}
\label{section:modeling}
Within our scope of small deformation and no particle reorganization in the quasi-static limit, the implementation is greatly simplified. The central idea is to find iteratively the displacement field such that the resulting contact traction together with all other boundary conditions, equilibrate each solid body in a granular system under consideration. In turn, when equilibration is not possible, we take it as a sign of particles undergoing reorganization (e.g., induced by frictional instabilities) being inevitable\footnote{We acknowledge that, in actual implementation, discretization errors (e.g., poor mesh qualities) can also prevent equilibration from happening.}. We adopt the classical penalty formulation to model contacts between solid bodies. Using the penalty formulation allows us to readily detect contact between two adjacent disks. We pay extra attention to pick appropriate values of penalty parameters (the normal contact stiffness $k_n$ and the tangential contact stiffness $k_t$) that can capture reasonably well the contact physics but at the same time prevent numerical instabilities from happening. During the early stage of the implementation, we consulted the book \cite{laursen2013computational} and the paper \cite{simo1992augmented} for general theoretical perspectives. The source code is publicly accessible through \href{https://github.com/liuchili/2D-FEM-multibody-contact-mechanics.git}{\color{blue}{https://github.com/liuchili/2D-FEM-multibody-contact-mechanics.git}}. Details of the implementation are discussed from a top-down perspective in Appendices A and B with the corresponding pseudocode (Algorithm 1, 2, 3, 4, and 5) presented in Appendices C, D, E, and F. Interested readers can consult these contents for a quicker understanding of the implementation.

 \section{Modeling ordered packings of disks under quasi-static isotropic compression}
 \label{section:disks}
In this section, we use the developed implementation to model ordered packing of mono-size disks. We first introduce the model setup, the calibration of contact parameters, and the preparation of initial configurations. After that, we discuss the simulation results concerning inter-particle forces and the distribution on both the system scale and the particle scale.
 \subsection{Virtual experiment setup}
\begin{figure}[H]
\centering
\includegraphics[width=0.7\linewidth]{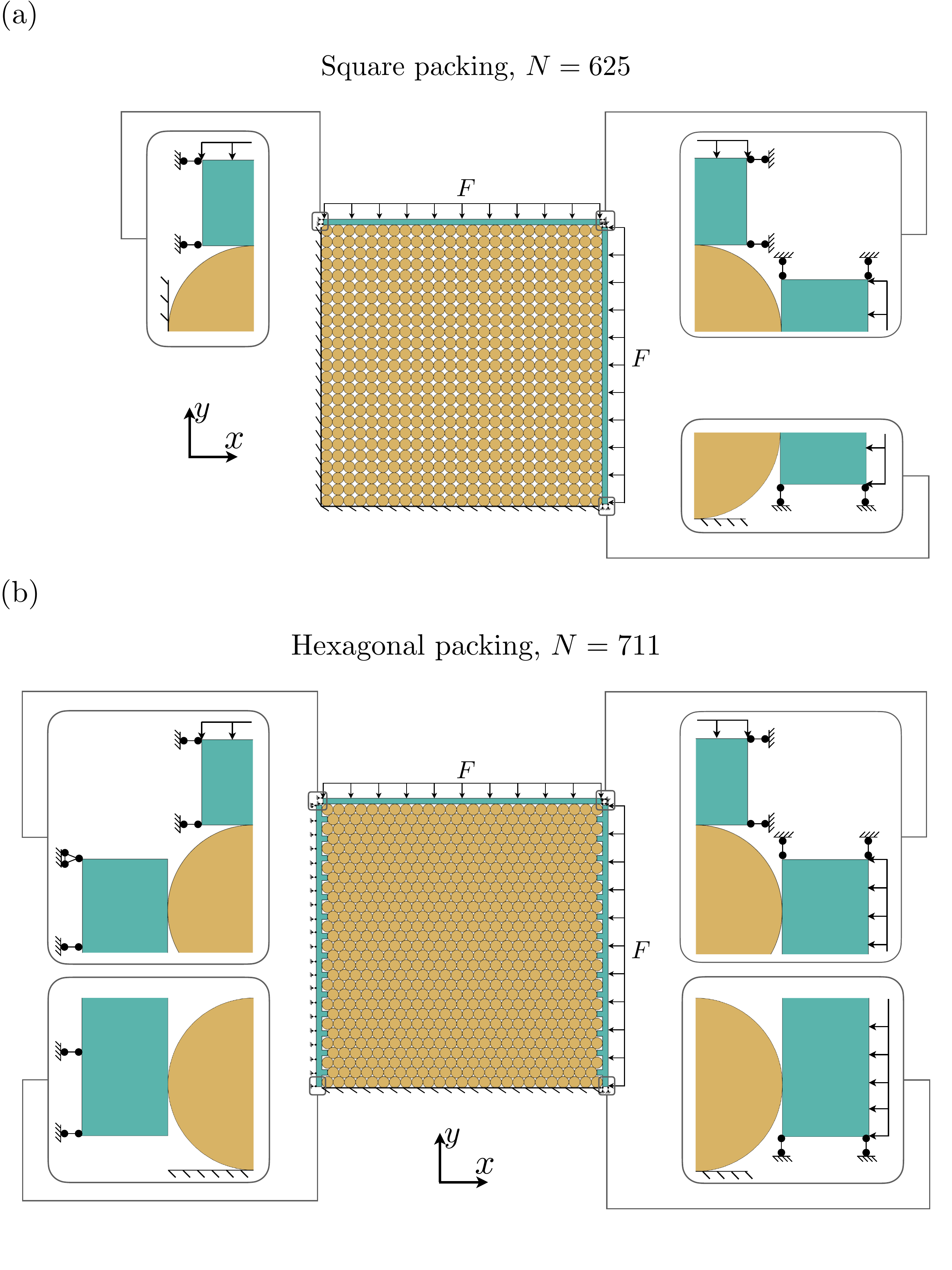}
\caption{Virtual experiment setup. (a) The considered square packing composed of 625 mono-size disks under isotropic compression. The top and right boundaries are modeled as very rigid rectangles that are both subjected to a compressive force $F$ while with their degrees of freedom being fixed along the direction perpendicular to the applied force $F$. The left and bottom boundaries are modeled as fixed rigid walls. (b) The considered hexagonal packing composed of 711 mono-size disks under isotropic compression. The top and bottom boundaries are treated as the same as those in the square packing. The left and right boundaries are designed to have a zig-zag shape for their surfaces touching the disks, which aims at providing a more homogeneous contact force distribution along the boundary. The right boundary is under the same constraints as the right boundary in the square packing, while the left boundary is constraint entirely from moving, to mimic the fixed rigid wall used as the left boundary in the square packing.}
 \label{latticeconfiguration}
\end{figure}

We consider mono-sized disks (with radius $R = 5\,\,\text{mm}$) arranged spatially over two canonical packing lattices: a square one with 625 disks (see Fig. \ref{latticeconfiguration}(a)) and a hexagonal one with 711 disks (see Fig. \ref{latticeconfiguration}(b)). These two assemblies are confined in two similar-sized rectangular domains respectively and subjected to quasi-static isotropic compression under plane-strain condition. We choose such a system size ($500 \sim1000$ particles) to be consistent with the commonly adopted system size in real experiments (e.g., photoelasticity) that investigate the particle-scale and meso-scale physics of granular materials \cite{bassett2015extraction,papadopoulos2016evolution}. For the square packing, the bottom and left boundaries are treated as stationary rigid walls, while the right and top boundaries are treated as rectangular-shaped solids and are both subjected to a constant compressive force $F$ and are constrained along the direction perpendicular to the direction of $F$, as shown in Fig. \ref{latticeconfiguration}(a). For the hexagonal packing, all boundaries are treated essentially the same as for the square packing case, except that the inner surfaces (those touching the disks) of the left and the right solids are changed to have zig-zag shapes, as shown in Fig. \ref{latticeconfiguration}(b). We make such a change to increase the homogeneity of contact forces for the hexagonal packing in the reference configuration where every disk has the same mechanical properties.
\subsection{Input of mechanical properties}
\begin{figure}[H]
\centering
\includegraphics[width=0.6\linewidth]{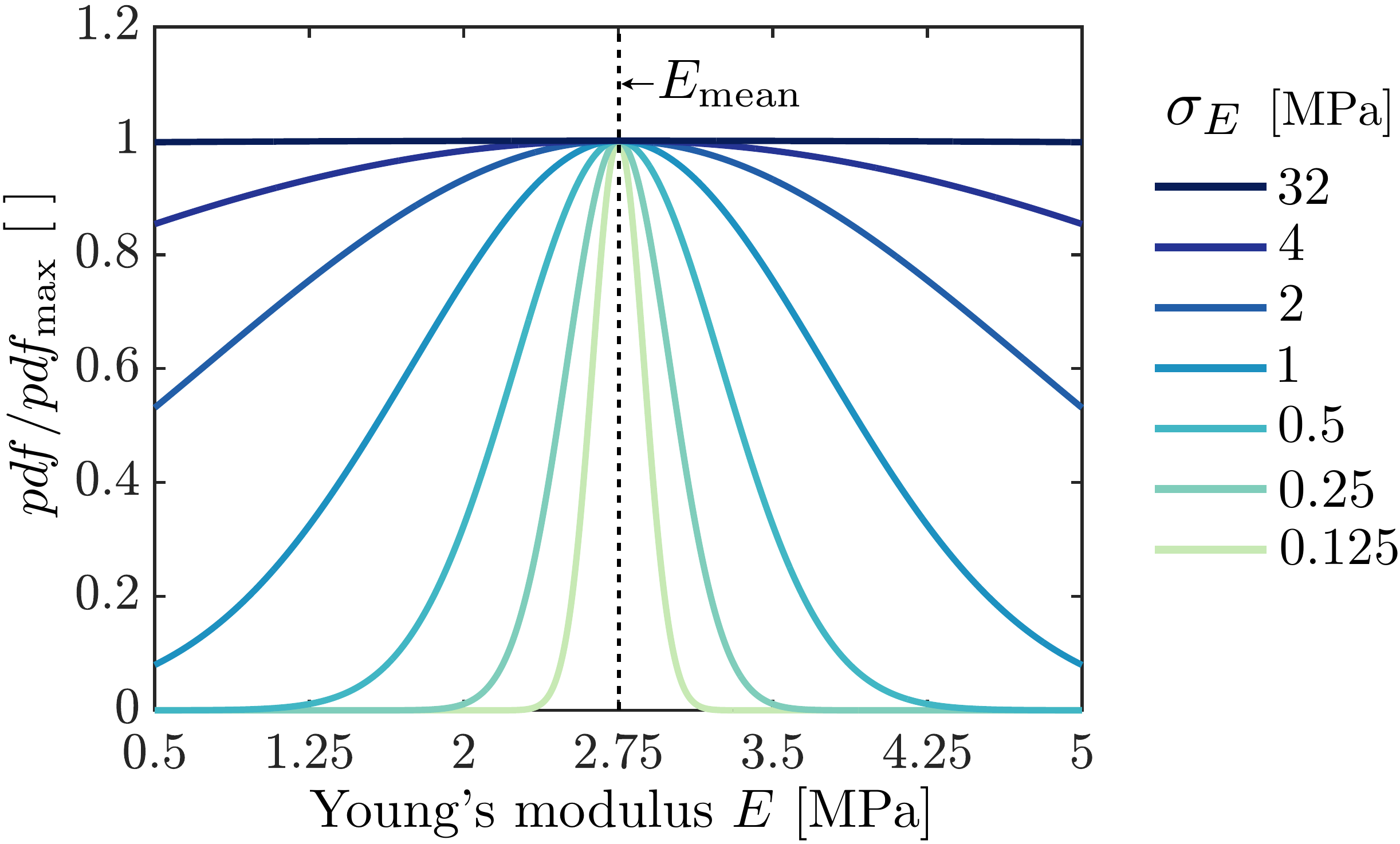}
\caption{The normalized truncated Gaussian distribution used to sample the Young's modulus $E$ of each disk.}
 \label{turncatednormal}
\end{figure}

We consider disks whose mechanical properties are close to those of rubber-like materials. Since rubber is mostly incompressible ($\nu_\text{rubber}= 0.5$), we consider a ``plane-strain-equivalent" material whose $E$ and $\nu$ satisfy $E = 8/9E_\text{rubber}$ and $\nu = 1/3$, where $E_\text{rubber}$ is the Young's modulus of the rubber. This relation can be deduced (see \cite{vu2019numerical} for detail) by matching the strain energy density between a rubber-like material and its ``plane-strain-equivalent" counterpart and demanding the in-plane principle stretches to be the same. In our virtual experiments, we remain $\nu = 1/3$ unchanged, and we sample $E$ for each disk from a truncated Gaussian distribution with a mean $E_\text{mean} = 2.75\,\,\text{MPa}$, a lower bound $E_\text{min} = 0.5\,\, \text{MPa}$ (which is close to the material used in \cite{vu2019numerical}), and an upper bound $E_\text{max} = 5\,\,\text{MPa}$ (which is close to the material used in \cite{hurley2014extracting}). We use the truncated Gaussian distribution for its convenient approximation of both a uniform distribution (by picking a relatively large standard deviation) and a Dirac delta distribution (by picking a relatively small standard deviation). Specifically, we vary the standard deviation $\sigma_E = 0.125, 0.25, 0.5, 1,  2, 4$, and $32$, and for each $\sigma_E$ we sample 10 different configurations to get meaningful statistics. Fig. \ref{turncatednormal} shows the normalized probability density function for each standard deviation $\sigma_E$. When $\sigma_E = 32$ we are essentially sampling from a uniform distribution from $E = 0.5\,\,\text{MPa}$ to $E = 5\,\,\text{MPa}$, whilst when $\sigma_E = 0.125$ we are essentially sampling from a peaked distribution that is very close to the Dirac-delta distribution $\delta(E_\text{mean})$.
\subsection{Determination of penalty parameters}
\begin{figure}[h]
\centering
\includegraphics[width=0.9\linewidth]{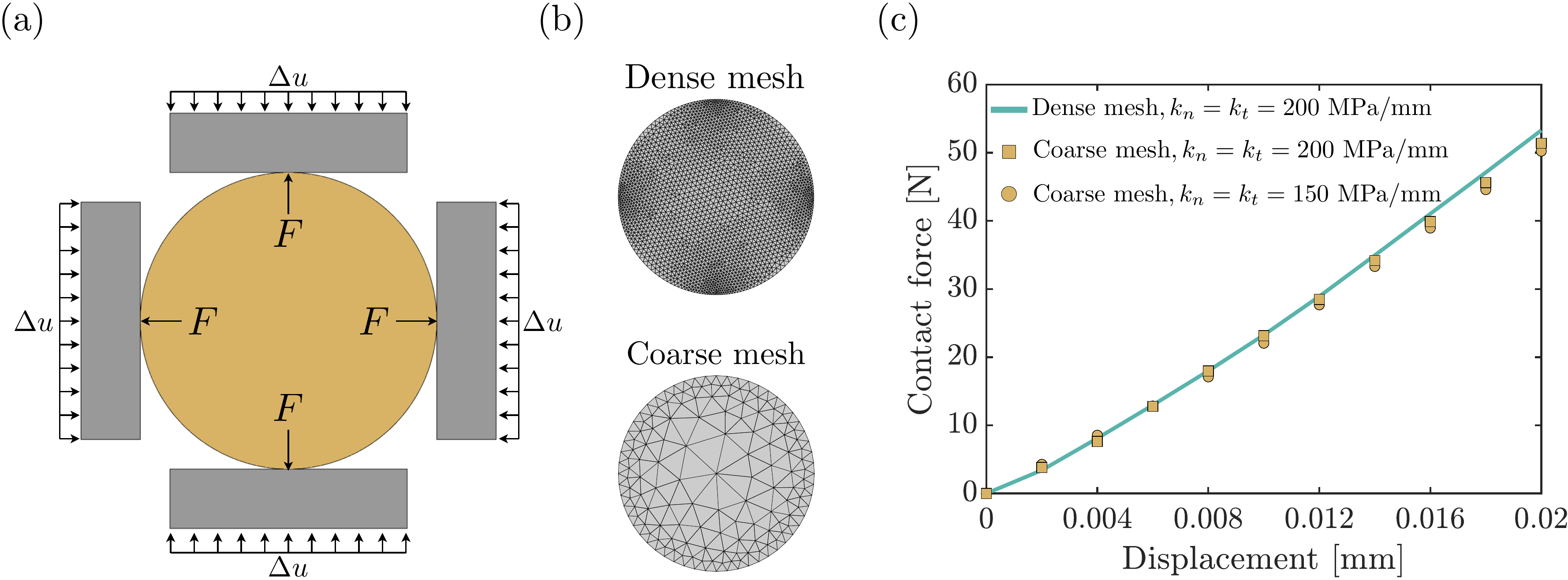}
\caption{(a) The isotropic compression setup used to calibrate the penalty parameters. (b) Visualizations of the dense mesh (top panel) and the coarse mesh (bottom panel). (c) Simulation results of boundary contact forces as a function of imposed displacement, for a disk with $E = E_\text {max} = 5\,\,\text {MPa}$ using both the dense and coarse mesh, and using different penalty parameters.}
 \label{meshcalibration}
\end{figure}

It is important to pick the appropriate penalization parameters ($k_n$ and $k_t$) for a contact problem simulation, especially in our cases where contacts happen among disks with different mechanical properties. Ideally, we will need to pick values for $k_n$ and $k_t$ that are as large as possible to approximate as close as possible the physical contact laws which require no normal inter-disk penetration and no tangential inter-disk slip when frictional tractions are below the thresholds set by normal tractions and the contact friction. Also, values of $k_n$ and $k_t$ needs to be larger when the considered contacting solids are stiffer (e.g., having a larger Young’s modulus). In our cases, if the values of $k_n$ and $k_t$ are large enough to physically capture the contact mechanics between stiffest disks ($E = 5$\,\, MPa), and at the same time if such values are not overly large so that the contact interaction between softest disks ($E = 0.5$\,\,MPa) is free from numerical instabilities, we will be able to accurately model contact mechanics between any disks that are between the softest and the stiffest. One more factor to consider is to pick the appropriate number of elements/nodes per disk that is computationally feasible for us. In light of these considerations, we perform displacement-controlled (with $\Delta u = 0.002\,\,\text{mm}$) isotropic compression tests on a single disk with 10 loading steps (Fig. \ref{meshcalibration}(a)), considering both $E = E_\text{min}$ and $E = E_\text{max}$, and considering both a dense mesh with 2321 nodes (top figure in Fig. \ref{meshcalibration}(b)) and a coarse mesh with 167 nodes (bottom figure in Fig. \ref{meshcalibration}(b)). The goal is to find appropriate values of $k_n$ and $k_t$ that can quantitatively capture contact forces $F$ using the coarse mesh by comparing to results obtained from the dense mesh. We find that when $k_n$ is larger than $200\,\,\text{MPa/mm}$ (taking $k_t = k_n$), for $E = E_\text{max}$ the resulting contact force no longer changes appreciably, at least for the range of considered loading steps.  We then apply the same $k_n$ and $k_t$ to a case using the coarse mesh and find good agreement (Fig. \ref{meshcalibration}(c)). However, this value of $200\,\,\text{MPa/mm}$ is too large for cases with $E=E_\text{min}$ to converge, and after calibration we find a value of $150 \,\,\text{MPa/mm}$ is a suitable choice, as (1) it can converge simulations with $E=E_\text{min}$ giving accurate results, and (2) it reduces negligibly the accuracy for simulations with $E=E_\text{max}$, at least for contact forces smaller than $30\,\,\text{N}$. Lastly, we note that due to the symmetry of the isotropic compression configuration, we find the above results insensitive to the specific value of $\mu_s$ (we tried with $\mu_s = 0$ and $\mu_s = 0.5$). Based on the above discussions, for our virtual isotropic compression experiments, we use $k_n=k_t=150\,\,\text{MPa/mm}$, and we apply an incremental load of $F = 250 \,\,\text{N}$ with three loading steps. At the last loading step with $F = 750\,\,\text{N}$ the resulting contact force will be around $30\,\,\text{N}$ which is in the accuracy range of using $k_n=k_t=150\,\,\text{MPa/mm}$. Lastly, we pick a contact friction $\mu_s = 0.5$ (a good choice for rubber-like material similar to \cite{li2019capturing}) between disks, and we set the contact friction between disks and the four boundaries to be zero.
\subsection{Preparation of initial configurations}
\begin{figure}[H]
\centering
\includegraphics[width=0.7\linewidth]{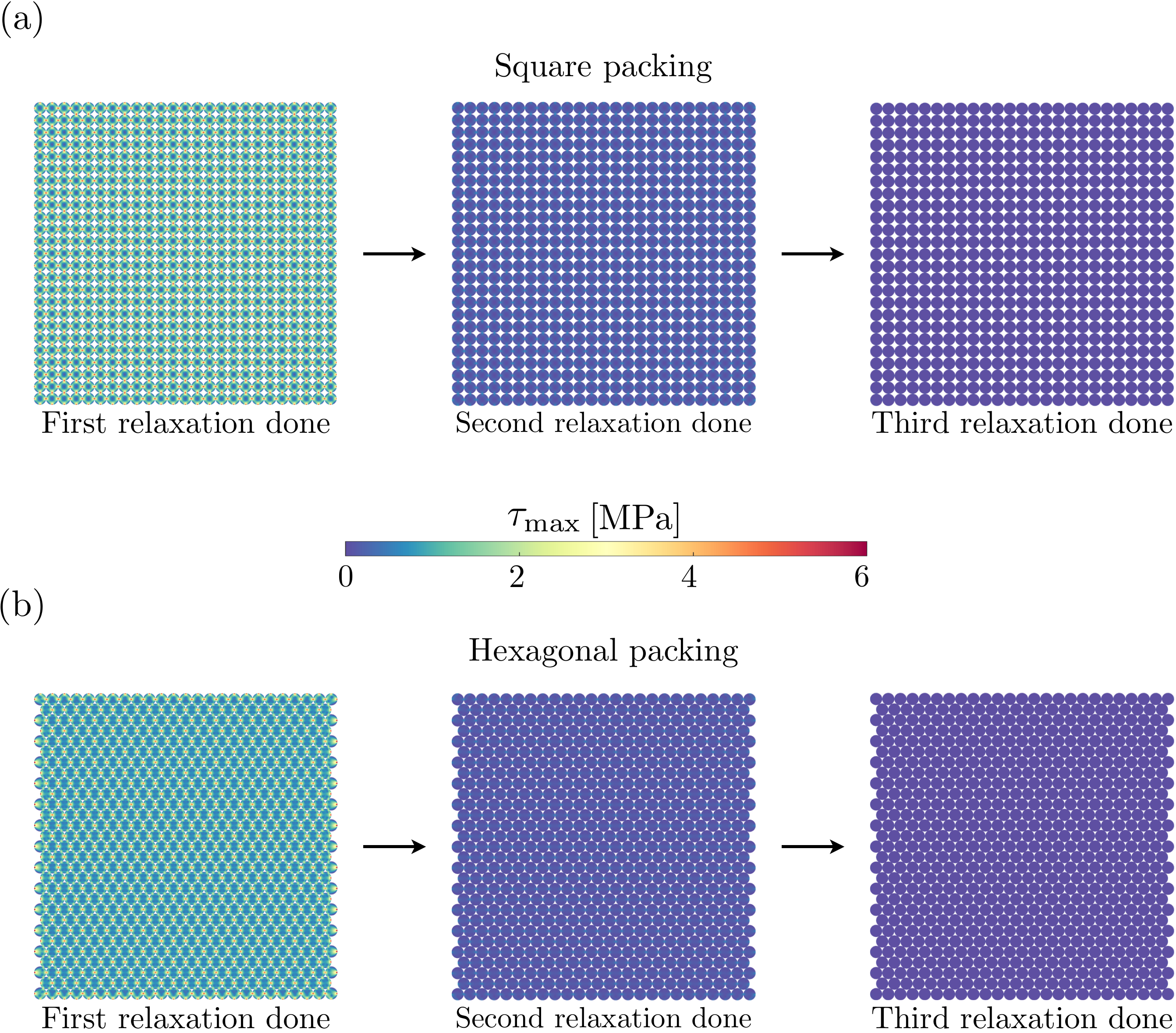}
\caption{The distribution of maximum shear stress $\tau_\text{max}$ within each disk for both the square packing (a) and the hexagonal packing (b) at each of three relaxation stages.}
\label{relaxationstress}
\end{figure}

For each packing lattice, similar to the procedure adopted in \cite{liu2020ils}, we prepare an initial configuration by stacking disks with slightly larger radii $5.01\,\,\text{mm}$ and relaxing the configuration with the four boundaries being held fixed. We apply this protocol to generate a very small overlap between two disks. Such overlap serves as a good initial guess for the Newton-Raphson solver, and it aids the convergence of our computations. We assign $E_\text{mean}$ to every disk and relax the configuration multiple times with $k_n=k_t=150\,\,\text{MPa/mm}$, to get sufficiently small inter-disk overlap. Multi-step relaxation can be realized by setting $\bm{U}^{a,z-1}_{\text{tot}} = \bm{0}$ after we finished the first $z-1$ relaxation steps, updated the configuration along the way and before we start the $z$th relaxation step. As expected, the inter-disk overlap becomes smaller and smaller as indicated by the distribution of maximum shear stress within each disk (Fig. \ref{relaxationstress}). For both packing lattices, we encounter convergence issues at the fourth relaxation step, which implies that the inter-disk overlap has become sufficiently small. Accordingly, for each packing lattice, we take the relaxed and updated configuration after the third relaxation as the initial configuration of our virtual experiments.
\subsection{Simulation results and discussions}
\subsubsection{Contact force heterogeneity: intensity}
\label{SectionProbabilityDistributionOfContactForces}
\begin{figure}[H]
\centering
\includegraphics[width=0.91\linewidth]{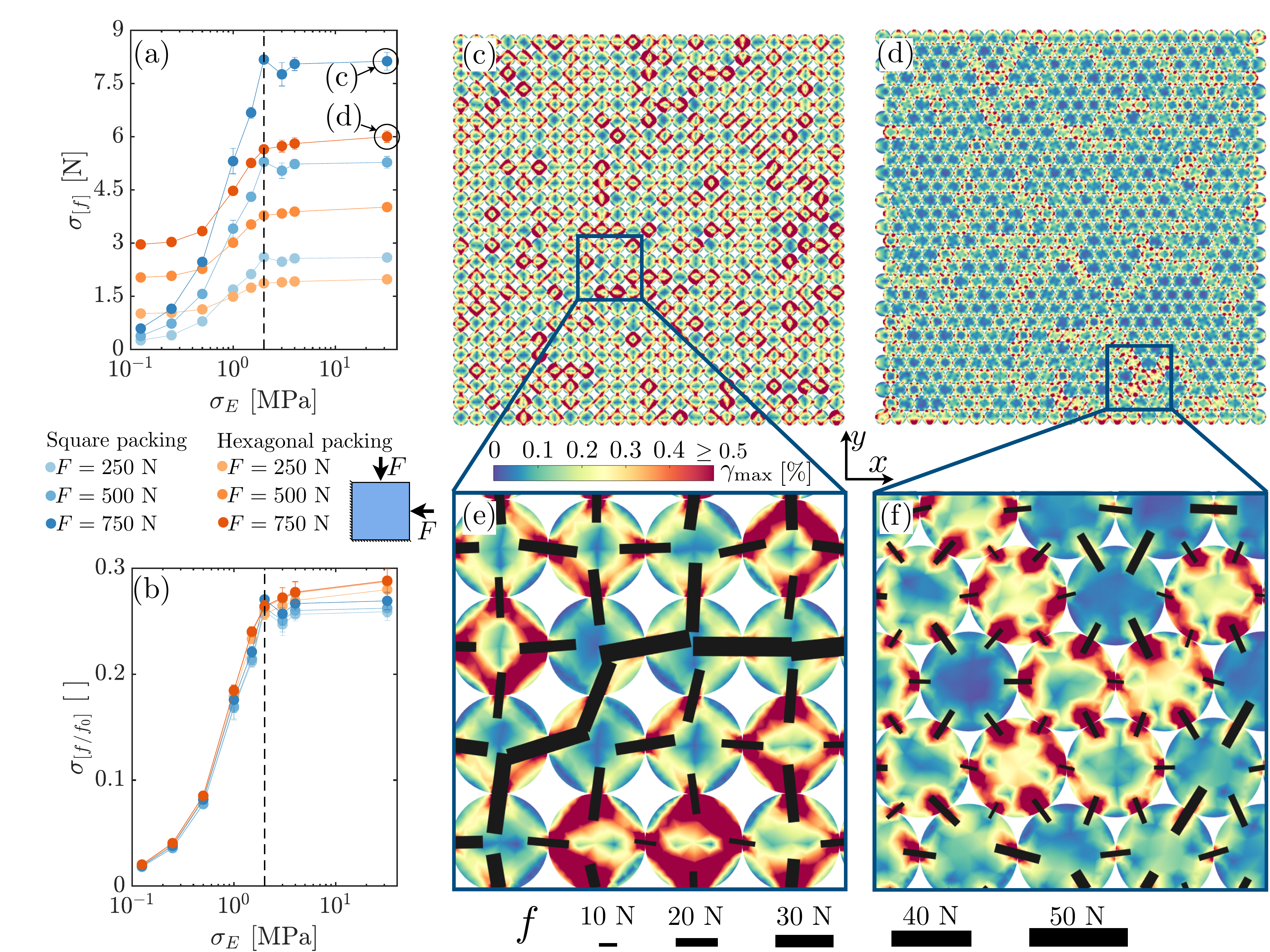}
\caption{(a) The standard deviation of contact force magnitudes, $\sigma_{[f]}$, as a function of $\sigma_E$, for both the square lattice (data in blue) and the hexagonal lattice (data in red), under three different compressive forces $F = 250\,\,\text{N},  500\,\,\text{N}$ and $750\,\,\text{N}$. Error bars indicating variations computed from ten independently sampled configurations using a single $\sigma_E$. The vertical dashed line indicates $\sigma_E = 2$ Mpa. (b) A similar figure to (a) but showing the normalized standard deviation of contact force magnitudes, $\sigma_[f/f_0]$. (c) The spatial distribution of maximum in-plane shear strain $\gamma_\text{max}$ with each disk for the square packing whose configuration is sampled from $\sigma_E = 32$. (d) A similar figure to (c) but showing results from the hexagonal packing. (e) A zoom-in figure showing the distribution of $\gamma_\text{max}$ overlaid with inter-particle forces whose magnitudes are represented by the length and thickness of solid black lines, and whose directions are aligned with the directions of those lines. (f) A similar figure to (e) but showing results from the hexagonal packing.}
\label{forcestrain}
\end{figure}

\begin{figure}[H]
\centering
\includegraphics[width=\linewidth]{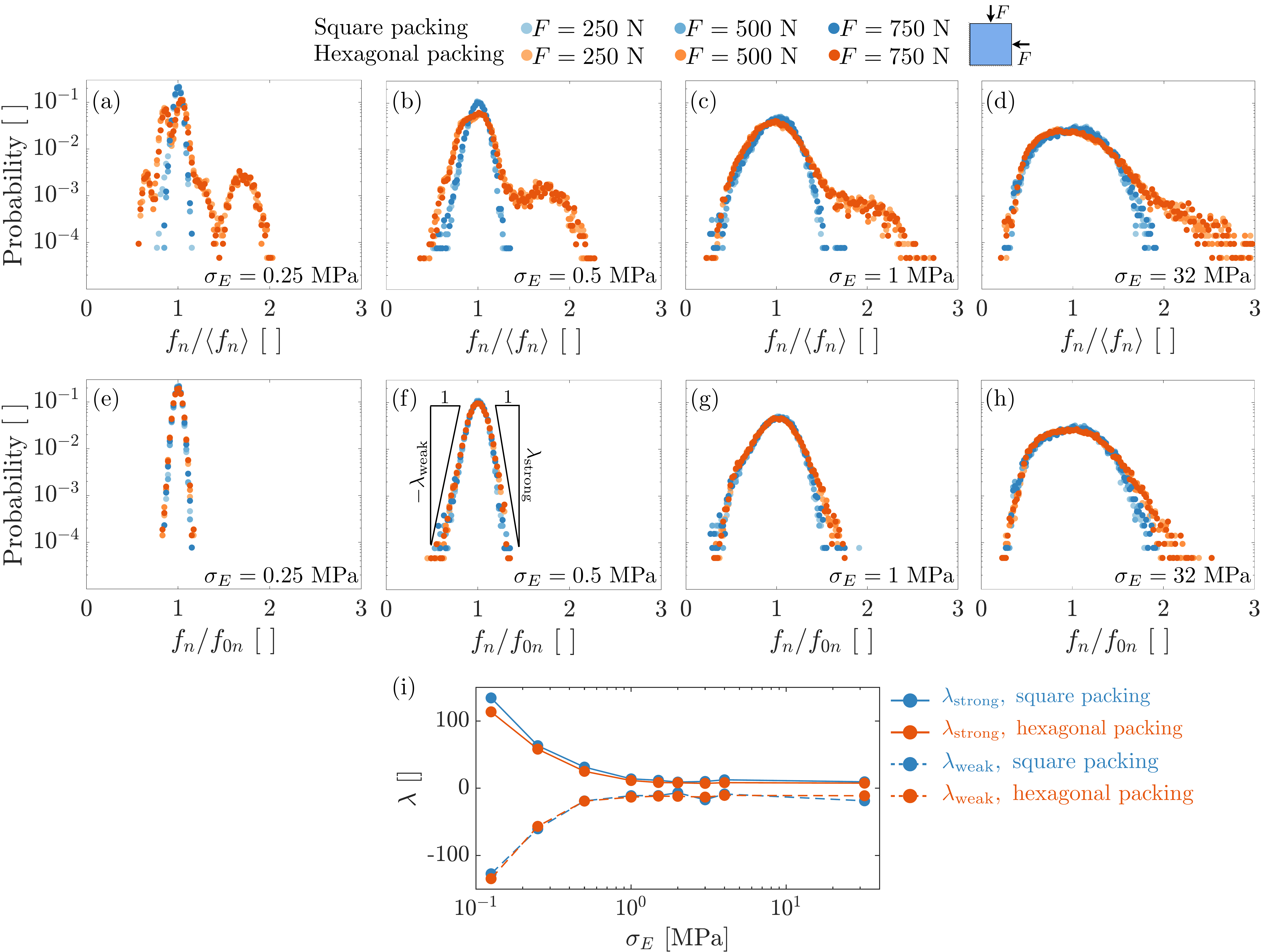}
\caption{The normalized probability distribution of normal contact forces shown in semi-log scale, for both the square packing (colored in blue) and the hexagonal packing (colored in red) using both $f_n/\langle f_n \rangle$ (first row) and  $f_n/ f_{0n}$ (second row). Results are presented as the average obtained from the ten independently sampled configurations from four different values of $\sigma_E$: $\sigma_E = 0.25$(a)(e), $\sigma_E = 0.5$(b)(f), $\sigma_E = 1$(c)(g), and $\sigma_E = 32$(d)(h). The fitted characteristic exponents $\lambda_\text{weak}$ and $\lambda_\text{strong}$ for both packings are shown in (i).}
 \label{normalforcedistribution}
\end{figure}

\begin{figure}[H]
\centering
\includegraphics[width=\linewidth]{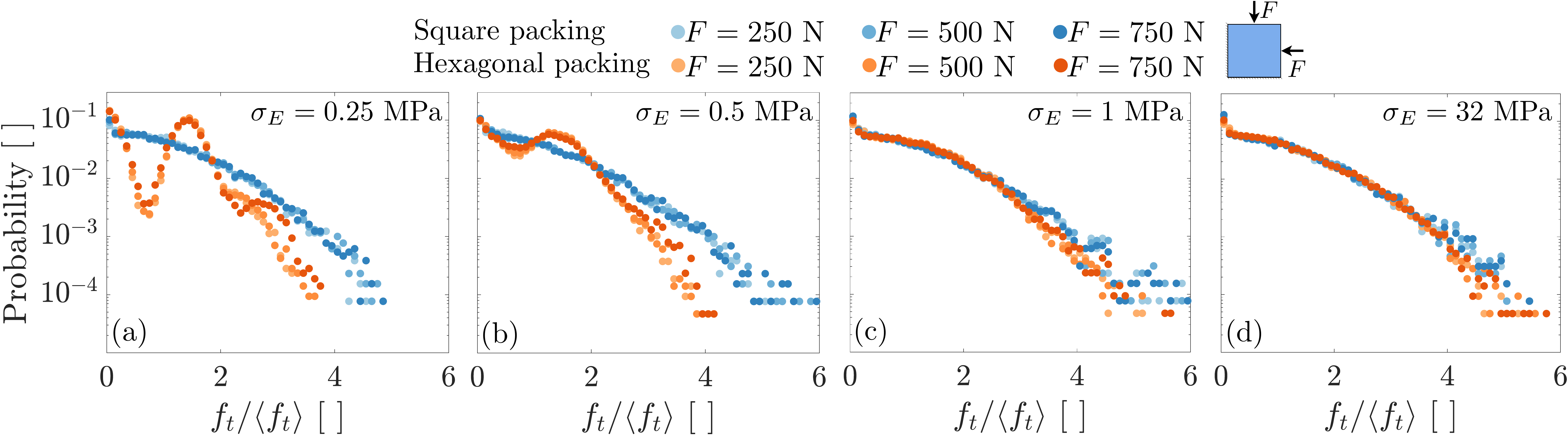}
\caption{The normalized probability distribution of tangential contact forces shown in semi-log scale, for both the square packing (colored in blue) and the hexagonal packing (colored in red) using $f_t/\langle f_t \rangle$. Results are presented as the average obtained from the ten independently sampled configurations from four different values of $\sigma_E$: $\sigma_E = 0.25$(a), $\sigma_E = 0.5$(b), $\sigma_E = 1$(c), and $\sigma_E = 32$(d).}
 \label{tangentialforcedistribution}
\end{figure}

We first investigate the system-level heterogeneity variations of contact forces in both packing lattices as $\sigma_E$ is varied. We first quantify such system-level heterogeneity by computing the standard deviation of all contact force magnitudes, terms as $\sigma_{[f]}$, in a simulated configuration. To obtain $[f]$, we compute $\bm{f}^{i\leftrightarrow j}$ (the vectorial form of $f^{i\leftrightarrow j}$) by $\bm{f}^{i\leftrightarrow j} = (\bm{f}^{i\leftrightarrow j}_i-\bm{f}^{i\leftrightarrow j}_j)/2$ between every two contacting disks, where $\bm{f}^{i\leftrightarrow j}_i$ and $\bm{f}^{i\leftrightarrow j}_j$ are the summation of forces on active nodes of solid $(i)$ with respect to solid $(j)$ and that on active nodes of solid $(j)$ with respect to solid $(i)$. We take a minus sign between these two quantities as they satisfy, in theory, $\bm{f}^{i\leftrightarrow j}_i+\bm{f}^{i\leftrightarrow j}_j = \bm{0}$ because of Newton’s third law. We take an average between these two quantities to reduce possible bias on evaluating $f^{i\leftrightarrow j}$ due to FEM discretization. Fig. \ref{forcestrain}(a) shows the variation of $\sigma_{[f]}$ with the variation of $\sigma_E$ for both packing lattices and for the three applied loads $F=250\,\,\text{N}$, $F=500\,\,\text{N}$ and $F=750\,\,\text{N}$, in a semi-log plot. Each data point shows the averaged value of $\sigma_{[f]}$ considering ten configurations independently sampled from a truncated normal distribution with a given $\sigma_E$, together with the error bar indicating the variation. For both packing lattices, $\sigma_{[f]}$ first increases with the increase of $\sigma_E$ and plateaus when $\sigma_E$ goes beyond 2 (the dashed vertical line). However, $\sigma_{[f]}$ from a square lattice increases much faster when $\sigma_E$ is smaller than two compared to that of a hexagonal lattice. In addition, the difference of $\sigma_{[f]}$ between the two packing lattices increases with the increase of external load $F$, with $\sigma_{[f]}$ of a square lattice being consistently greater than that of a hexagonal lattice for all $\sigma_E$. Alternatively, if we quantify the system-level heterogeneity using $\sigma_{[f/f_0]}$ instead of $\sigma_{[f]}$, as shown in Fig. \ref{forcestrain}(b), we find that contact forces in a hexagonal lattice is actually more heterogeneous than those in a square lattice. Here $\sigma_{[f/f_0]}$ indicates the standard deviation of $[f/f_0]$ where $[f_0]$ represents the collection of contact force magnitudes from the reference configuration in which all disks share the same $E$ ($\sigma_E \rightarrow 0$). This observation can be made more conclusively by looking at the probability distribution of normal contact force (which is discussed in more detail in the next paragraph), as shown in Figs. \ref{normalforcedistribution}(a)-(h), where the hexagonal packing shows a more widespread distribution than the square packing does. The apparent discrepancy between $\sigma_{[f]}$ and $\sigma_{[f/f_0]}$ may be explained by the fact that a hexagonal lattice allows for larger coordination number ($\langle Z \rangle = 6$) over a square lattice ($\langle Z \rangle = 4$). As a result, in an absolute term ($\sigma_{[f]}$), a contact in a square lattice sustains on average a larger contact force compared to a contact does in a hexagonal lattice, thus giving greater contact force heterogeneity under the same applied load. Additionally, a larger coordinate number allows for more options for contact forces to distribute in space, thus in a relative term ($\sigma_{[f/f_0]}$), enabling greater contact force heterogeneity. Moreover, we find that, once normalized by $[f_0]$, the relative contact force heterogeneity $\sigma_{[f/f_0]}$ become rather insensitive to the considered range of external loading, as shown by the collapse of curves in Fig. \ref{forcestrain}(b). Lastly, to aid visualization, we show the spatial distribution of maximum in-plane shear strain ($\gamma_\text{max}$) of one configuration sampled from $\sigma_E = 32$ and subjected to $F = 750\,\,\text{N}$ from both the square lattice (Fig. \ref{forcestrain}(c)) and the hexagonal lattice (Fig. \ref{forcestrain}(d)). We can clearly observe that some disks are under much larger shear deformation than others, in a similar way to how cylinders shine with different intensities in a photo-elastic experiment. However, we note that, unlike in a photo-elastic experiment where brighter cylinders indicate locations of larger contact forces, in our virtual experiments ``brighter" (in terms of shear strains) disks instead suggest locations of smaller contact forces. We present a zoomed-in plot of a small area of Fig. \ref{forcestrain}(c) and Fig. \ref{forcestrain}(d), respectively, as shown in Figs. \ref{forcestrain}(e)(f). It can be observed that larger contact forces (represented by longer and thicker solid black lines) take place generally between disks with smaller shear strains. In addition, the direction of each solid black line is aligned with the direction of the corresponding contact force. Its deviation from the direction of the branch vector (a vector connecting the center of mass of two contacting disks) suggests the existence of frictional forces that arise from non-zero contact friction.
We next move to analyze the probability distributions of contact forces from our virtual experiments and compare the results with the classical ones obtained from disordered packing of rigid disks (termed as DPRD hereafter for simplicity). We first compute the normal contact force magnitudes $[f_n]$ and tangential (frictional) contact forces magnitudes $[f_t]$ from $[f]$. Since we implement the contact law on a stress level instead of on a force level as in ordinary DEM \cite{cundall1979discrete}, for a contact between solid $(i)$ and solid $(j)$, we compute $f^{i\leftrightarrow j}_t$ and $f^{i\leftrightarrow j}_n$ in the following way: Suppose that we have $\bm{f}^{i\leftrightarrow j}$ (the vectorial form of $f^{i \leftrightarrow j}$) already computed following the procedure described in the preceding paragraph, and suppose that due to small deformation we can approximate the contact normal $\bm{n}^{i\leftrightarrow j}$ by the branch vector $\bm{l}^{i\leftrightarrow j}_c$ = $\bm{X}_c^{(i)} - \bm{X}_c^{(j)}$ using $\bm{n}^{i\leftrightarrow j} = \bm{l}^{i\leftrightarrow j}_c/||\bm{l}^{i\leftrightarrow j}_c||_2$, where $\bm{X}_c^{(i)}$ and $\bm{X}_c^{(j)}$ are the centroid positions of solid $(i)$ and solid $(j)$ in the undeformed configuration. With these quantities at hand, we can compute $f^{i\leftrightarrow j}_n = |\bm{f}^{i\leftrightarrow j}\cdot \bm{n}^{i\leftrightarrow j}|$ and $f^{i\leftrightarrow j}_t = \sqrt{\bm{f}^{i\leftrightarrow j}\cdot \bm{f}^{i\leftrightarrow j} - \left(f^{i\leftrightarrow j}_n\right)^2}$. Then, following the convention adopted in \cite{radjai1996force}, we plot the normalized probability distributions, $f_n/\langle f_n\rangle$ and $f_t/\langle f_t\rangle$, for both packing lattices subjected to all three loading steps and with $\sigma_E = 32$ (Fig. \ref{normalforcedistribution}(a) and Fig. \ref{tangentialforcedistribution}(a)), $\sigma_E = 1$ (Fig.\ref{normalforcedistribution}(b) and Fig. \ref{tangentialforcedistribution}(b)), $\sigma_E = 0.5$ (Fig. \ref{normalforcedistribution}(c) and Fig. \ref{tangentialforcedistribution}(c)) and $\sigma_E = 0.25$ (Fig. \ref{normalforcedistribution}(d) and Fig.\ref{tangentialforcedistribution}(d)). Here $\langle f_n\rangle$ and $\langle f_t\rangle$ are the mean normal contact force magnitude and the mean tangential (frictional) force magnitude, respectively. In addition, similar to data presented in Figs. \ref{forcestrain}(a)(b), data shown in every sub-figure of Fig. \ref{normalforcedistribution}  and Fig. \ref{tangentialforcedistribution} are results from the average of ten configurations independently sampled with a given $\sigma_E$.

First, let us focus on the normalized probability distribution of $f_n$. Generally speaking, the hexagonal packing shows a more widespread probability distribution of $f_n/\langle f_n\rangle$ over the square packing, indicating the distribution of contact force magnitudes for the hexagonal packing is more scattered. This general observation is consistent with the observation from Fig. \ref{forcestrain}(b) that $\sigma_{[f/f_0]}$ from the hexagonal packing is greater than that from the square packing. For the square packing, the probability distribution curve narrows toward the mean (where the probability peaks) as $\sigma_E$ decreases, indicating the convergence to a homogeneous contact force distribution where $f_n = \langle f_n \rangle $ for every contact. For the hexagonal packing, although the narrowing trend also appears, the probability distribution seems to converge to a different type of distribution as $\sigma_E$ decreases. This distribution has multiple peaks (Fig. \ref{normalforcedistribution}(a)) whose normal forces correspond to $f_{0n}$ (which is defined per contact) obtained from the reference configuration. Note that, in the reference configuration, unlike the square packing where $f_{0n}$ is the same for all contacts, $f_{0n}$ is not the same for all contacts for the hexagonal packing. This is due to the fact that, unlike the square lattice case, the loading direction (a square type) is not aligned with the lattice direction (a hexagonal type), thus leading to different contact forces among contacts. Interestingly, when we instead plot the probability distribution of $f_n/f_{0n}$, data from the hexagonal packing no longer show any peak when $\sigma_E$ us small, as indicated for example by Fig. \ref{normalforcedistribution}(e) in comparison to Fig. \ref{normalforcedistribution}(a) and by Fig. \ref{normalforcedistribution}(f) in comparison to Fig. \ref{normalforcedistribution}(b). Further, data from the hexagonal packing rapidly converge to data from the square packing as $\sigma_E$ decreases, and they both show exponential-like tails for normalized normal force above the mean ($f_n > f_{0n}$) and also for normalized normal force below the mean ($f_n < f_{0n}$), as shown in Figs. \ref{normalforcedistribution}(e), (f), (g) and (h). We further fit exponential functions of the form $\alpha \exp(-\lambda f_n/f_{0n})$ to those tails with characteristic exponents (Fig. \ref{normalforcedistribution}(f)) $\lambda_\text{weak}$ (for normalized normal forces below the mean) and $\lambda_\text{strong}$ (for normalized normal forces above the mean), and we plot the variation of $\lambda_\text{weak}$ and $\lambda_\text{strong}$ along $\sigma_E$ for both packing lattices, as shown in Fig. \ref{normalforcedistribution}(i). It can be observed that for both packing lattices, the magnitudes of both $\lambda_\text{weak}$ and $\lambda_\text{strong}$ rapidly decay as $\sigma_E$ increase and almost saturate when $\sigma_E$ goes beyond one. Interestingly, these exponential tails of the normal force distributions observed in our hexagonal but mechanically heterogenous packings are different from the Gaussian-like tails observed in hexagonally-arranged rigid frictionless packings \cite{van2007tail}. The apparent discrepancy observed in the normal force distribution for these two packings may be explained by different protocols from which forces are generated: In our work normal forces are generated from only one loading type (i.e., compressing uniformly along the x and y direction), while in \cite{van2007tail} normal forces are generated statistically from all force ensembles as long as the sampled forces satisfy overall stresses and balance each particle. It may be interesting to see whether the normal distribution will have a Gaussian-like tail if we further included normal forces generated from multiple other loading types (e.g., a uniaxial compression).
 
Next, we focus on discussing the normalized probability distribution of $f_t$.  In contrast to $f_n$, the hexagonal packing shows a less widespread probability distribution of $f_t$ than the square packing (Figs. \ref{tangentialforcedistribution}(a)(b)(c)). However, as $\sigma_E$ increases, results from the hexagonal packing gradually converge to those from the square packing, exhibiting a lattice-independent trend. In particular, when $\sigma_E$ is relatively large (i.e., $\sigma_E =32$), results from both packing lattices show nearly identical trend for $f_t$ above the mean (Figs. \ref{tangentialforcedistribution}(d)); this trend has a tail decaying faster than an exponential one which is commonly observed in DPRD. For $f_t$ below the mean the probability keeps increasing as $f_t$ approaches zero, which is in qualitative agreement with the power law scaling observed in DPRD for $f_t$ below the mean. 

\begin{figure}[H]
\centering
\includegraphics[width=\linewidth]{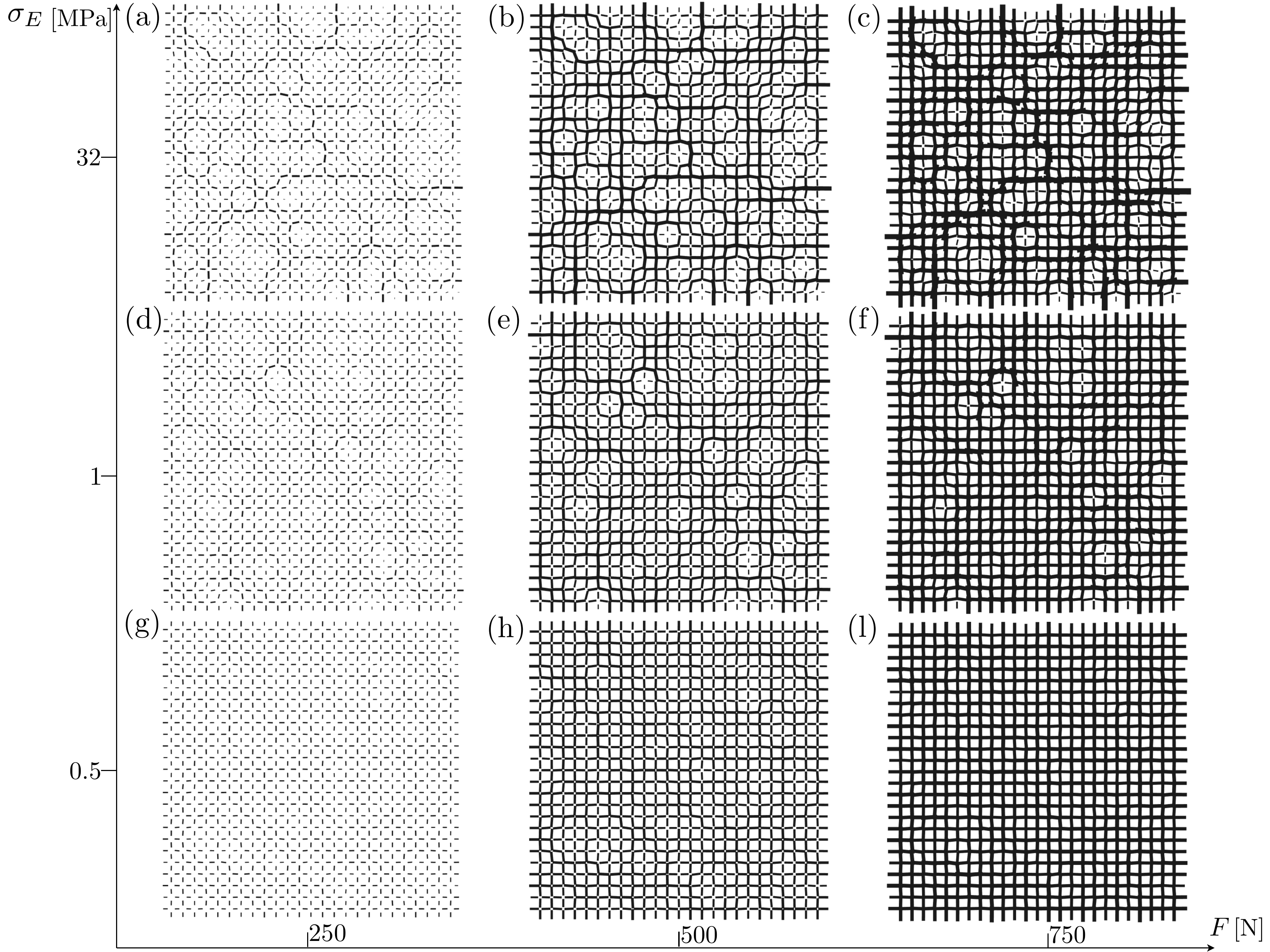}
\caption{Visualizations of the computed inter-particle forces in the square packing whose configurations are sampled from $\sigma_E = 32$(a)(b)(c), $\sigma_E = 1$(d)(e)(f) and $\sigma_E = 0.5$(g)(h)(l). Contact forces are represented by solid black lines whose length and thickness are scaled according to the magnitudes of contact forces.}
 \label{squarecontactforce}
\end{figure}

\begin{figure}[H]
\centering
\includegraphics[width=\linewidth]{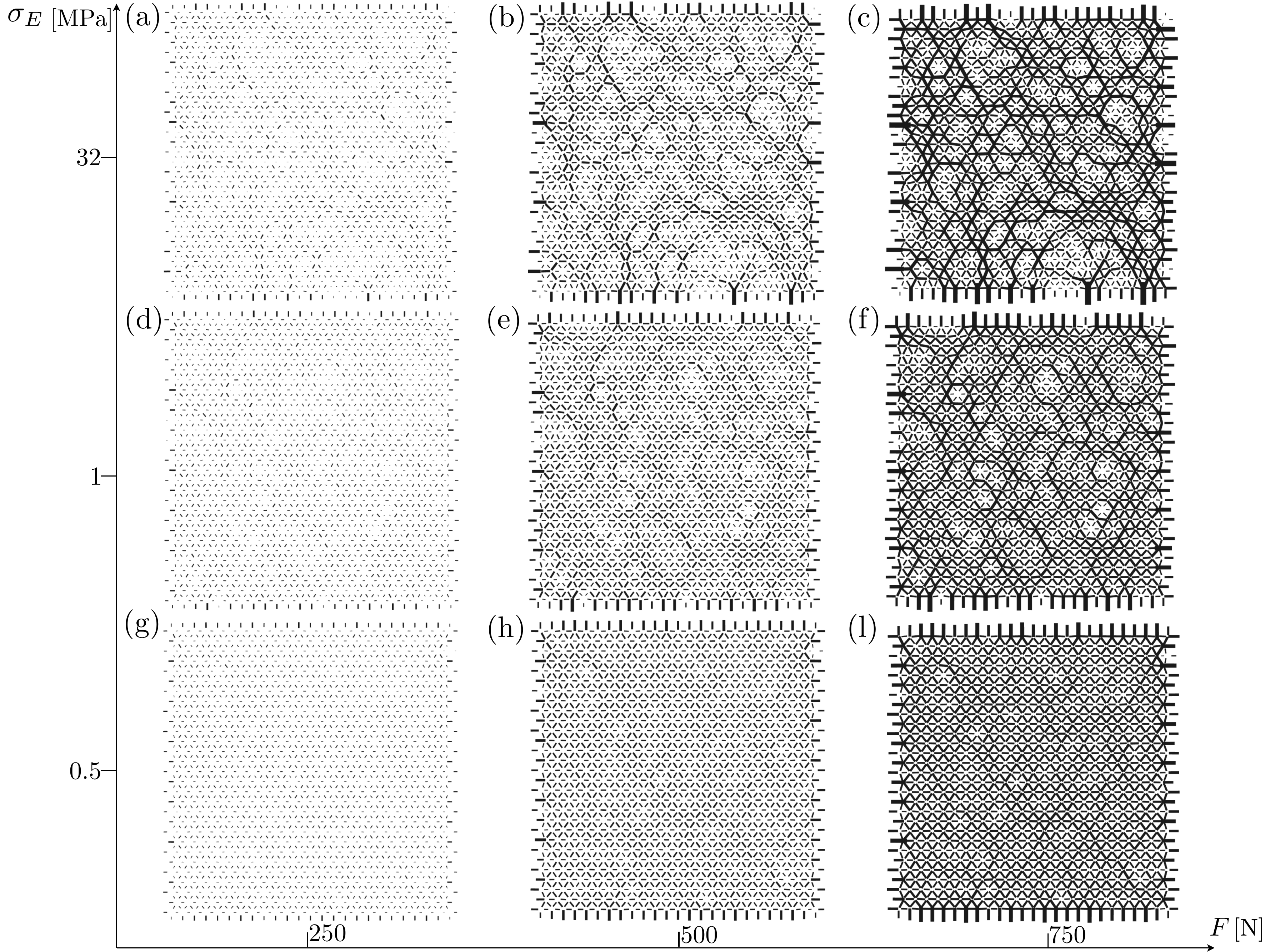}
\caption{A figure similar to Fig. \ref{squarecontactforce} but shows results from the hexagonal packing.}
 \label{hexacontactforce}
\end{figure}

In all, by comparing the probability distributions of $f_n$ and $f_t$ for both packing lattices, we conclude that, in the absence of particle reorganization, $f_n$ plays a more dominant role over $f_t$ in determining the degree of contact force heterogeneity of an ordered packing. We close this section by presenting two figures visualizing the contact force distributions for both the square lattice (Fig. \ref{squarecontactforce}) and the hexagonal lattice (Fig. \ref{hexacontactforce}) obtained from configurations sampled from different $\sigma_E$ and subjected to different $F$. Contact forces in both figures are scaled with the same constant. We can clearly observe the emergence of preferred locations of contact forces as $\sigma_E$ increases, in a way reminiscent of force chains observed in DPRD.

\subsubsection{Contact force heterogeneity: orientation}
In natural disordered granular media, it is well known that heterogeneity emerges not only as heterogeneous force intensities, but also as heterogeneous force orientations, and that both heterogeneities give rise to the ability of a natural disordered granular to resist an external loading. Accordingly, in this section, we shift our attention to analyze the contact force heterogeneity in terms of the orientation, instead of the intensity as discussed in section. \ref{SectionProbabilityDistributionOfContactForces}.

\begin{figure}[H]
\centering
\includegraphics[width=\linewidth]{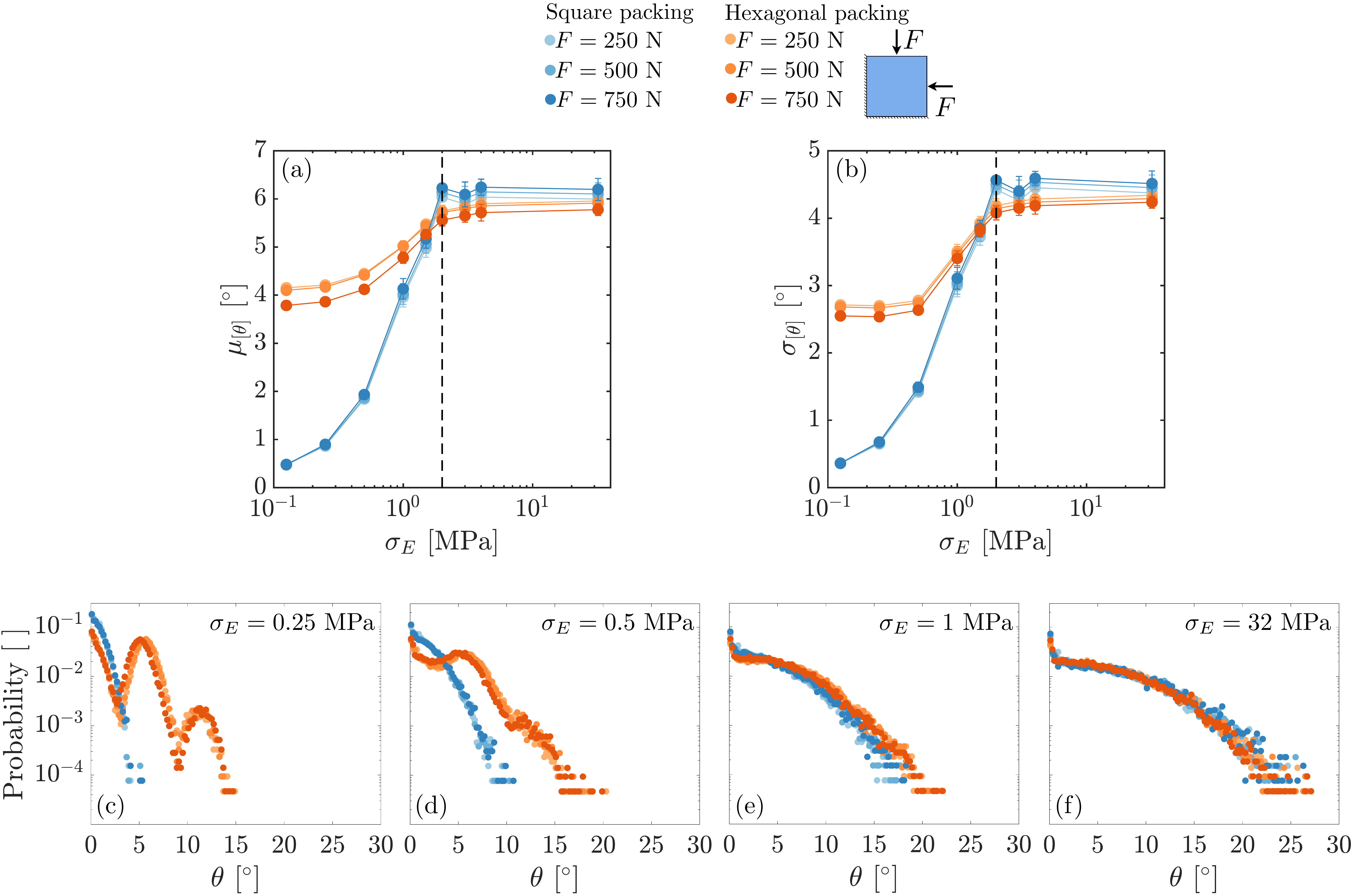}
\caption{(a) The average of the force orientation deviation, $\mu_{[\theta]}$, as a function of $\sigma_E$, for both the square lattice (data in blue) and the hexagonal lattice (data in red), under three different compressive forces $F = 250\,\,\text{N},  500\,\,\text{N}$ and $750\,\,\text{N}$. Error bars indicating variations computed from ten independently sampled configurations using a single $\sigma_E$. The vertical dashed line indicates $\sigma_E = 2$ MPa. (b) A similar figure to (a) but showing the standard deviation of the force orientation deviation, $\mu_{[\theta]}$. (c)-(f) The probability distribution of force orientation deviation,$\theta$, shown in semi-log scale, for both the square packing (colored in blue) and the hexagonal packing (colored in red). Results are presented as the average obtained from the ten independently sampled configurations from four different values of $\sigma_E$: $\sigma_E = 0.25$(c), $\sigma_E = 0.5$(d), $\sigma_E = 1$(e), and $\sigma_E = 32$(f).}
 \label{OrientationHeterogeneity}
\end{figure}

We propose a parameter $\theta$ to quantify the force orientation heterogeneity in a granular packing. We define $\theta$ as the deviation from the direction of a contact force, $\bm{f}$, to that of the contact force in the reference configuration, $\bm{f}_0$. In the reference configuration where every disk has the same mechanical property, $\bm{f}_0$ is either horizontal or vertical in a square packing, while in a hexagonal packing $\bm{f}_0$ can be horizontal, or $60^\circ$ from being horizontal, or $120^\circ$ from being horizontal. We compute both the average, $\mu_{[\theta]}$, and the standard deviation, $\sigma_{[\theta]}$, of $\theta$ for both packing lattices considering different values of $\sigma_E$, as shown in Fig. \ref{OrientationHeterogeneity}(a) and Fig. \ref{OrientationHeterogeneity}(b). Both results suggest that for both packing lattices the level of force orientation heterogeneity increases with the increase of mechanical heterogeneity (i.e., the increase of $\sigma_E$). In addition, when $\sigma_E$ is small, the force orientation heterogeneity of the square packing is weaker than that of the hexagonal packing, but it quicks converges to that of the hexagonal packing as $\sigma_E$ increases. These two observations can also be made by computing the probability distribution of $\theta$ at different values of $\sigma_E$, as shown from Fig. \ref{OrientationHeterogeneity}(c) to Fig.\ref{OrientationHeterogeneity}(f) where values of $\sigma_E$ are 0.25 MPa, 0.5 MPa, 1 MPa and 32 MPa, respectively. The observed generally stronger force orientation heterogeneity in the hexagonal packing may be explained by the larger coordination number the hexagonal packing possesses in comparison to the square packing, which give the hexagonal packing more possibilities for contact force orientation.

\subsubsection{Particle-scale gain of contact forces}
\label{SectionParticleScaleGain}
\begin{figure}[H]
\centering
\includegraphics[width=\linewidth]{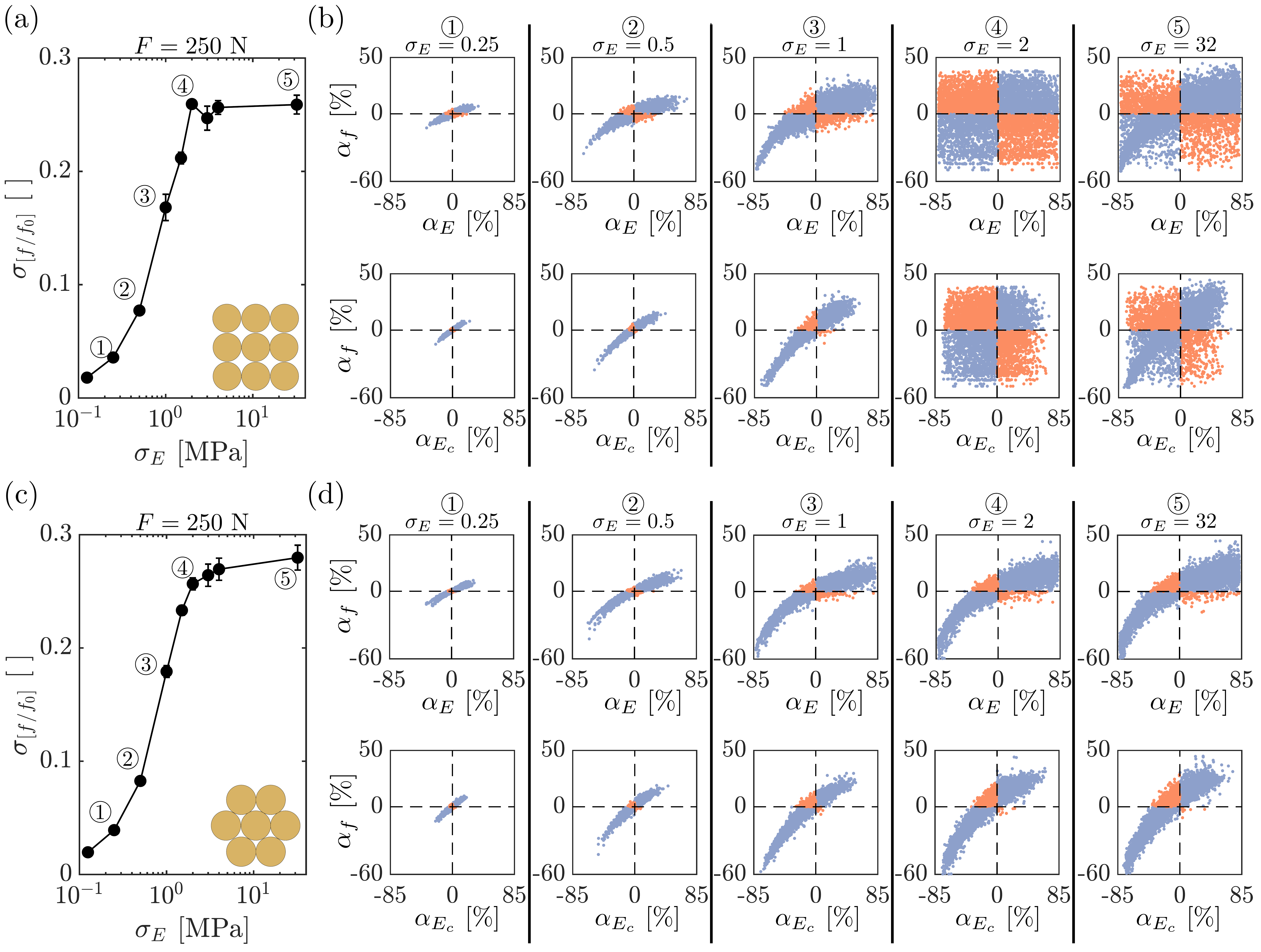}
\caption{(a) The variation of $\sigma_{[f/f_0]}$ against $\sigma_E$ for the square packing subjected to $F = 250 \,\,\text{N}$. We pick five $\sigma_E$ values and plot the correlation between $\alpha_E$ and $\alpha_f$ (top panel of (b)) and that between $\alpha_{E_c}$ and $\alpha_f$. Data are shown by overlaying results from the ten independently sampled configurations for each $\sigma_E$. Data colored in blue locating in the first and third quadrants, represent disks whose stiffness correlate positively to the contact force they sustain, i.e., $\alpha_E\alpha_f>0$ or $\alpha_{E_c}\alpha_f>0$. Data colored in red locating in the second and fourth quadrants, represent ``outliers" including softer disks ($\alpha_{E} < 0$ or $\alpha_{E_c} < 0$) gaining larger forces ($\alpha_f > 0$) and stiffer disks ($\alpha_E > 0$ or $\alpha_{E_c} > 0$) gaining smaller forces. (c) A similar figure to (a) but showing results from the hexagonal packing. (d) A similar figure to (b) but showing results from the hexagonal packing.}
\label{deviation}
\end{figure}

Observing the interesting patterns presented in Fig. \ref{squarecontactforce} and Fig. \ref{hexacontactforce}, it is then natural to wonder about a possible correlation between the mechanical property of a disk and the amount of contact force that disk sustains. Or in other words, does the mechanical property of a disk (in this work just the Young's modulus) play a role in determining the amount of contact force that disk receives? For disks being laterally confined into a one-dimensional chain, the answer is trivially that mechanical properties play no role, since geometrical constraints allows for a single path for contact forces to locate in space to balance the externally applied load regardless of how soft or how compressible a constituent disk is. However, when disks are arranged in 2D (and, of course, 3D) arrays the scenarios become much less straightforward, since there are multiple potential paths allowing for contact forces to form networks to balance the externally applied load. In particular, different mechanical properties of contacting disks lead to non-affine deformation from one disk to another (see for example Fig. \ref{forcestrain}(c)), which can cause stress redistribution (and subsequently contact force redistribution) that otherwise vanishes in disk packing with homogeneous mechanical properties. In this regard, we define the following two quantities to investigate the possible correlation between Young's modulus and the sustained contact force of a disk:
\begin{itemize}
\item $\alpha_E = (E/E_\text{mean}-1)\times 100\%$ defined as the relative ``stiffness" variation for a disk with respect to its reference state $E_\text{mean}$.
\item $\alpha_f = (f/f_0-1)\times100\%$ defined as the corresponding relative contact force gain/loss for the disk with respect to its referenced state $f_0$, where $f$ is the total contact force sustained by that disk with a given Young's modulus $E$, and $f_0$ is defined similarly but is obtained from the reference configuration in which $E = E_\text{mean}$ for every disk.
\end{itemize}
Solely from an energy point of view without any consideration of the potential spatial correlation among constituent disks, under the same external load, $\alpha_f$ should be positively correlated to $\alpha_E$, i.e., stiffer disks gain larger contact forces, which corresponds to less work done by the applied external load and consequently less strain energy stored by the disks\footnote{We assume that energy potentially dissipated through friction is not significant compared to the stored strain energy.}. First, we focus on discussing the correlation between $\alpha_E$ and $\alpha_f$ for the square packing. We pick five scenarios ($\sigma_E = 0.25, 0.5, 1, 2$ and $32$) under $F = 250\,\,\text{N}$ (Fig. \ref{deviation}(a)), and for each scenario, we plot all $\alpha_f$ and $\alpha_E$ data obtained from the ten independently sampled configurations (upper panel of Fig. \ref{deviation}(b)). We color data in blue if $\alpha_f$ and $\alpha_E$ appear in the first and third quadrants of each sub-figure, i.e., they are positively correlated as $\alpha_f\alpha_E > 0$, and we color data in red if otherwise ($\alpha_f\alpha_E < 0$ residing in the second and fourth quadrants). Two qualitative observations can be made. First, there indeed exist a positive correlation (data in blue) between $\alpha_E$ and $\alpha_f$ when $\sigma_E$ is relatively small ($\sigma_E < 2$). However, there are also ``outliers" (data in red) regardless of the particular value of $\sigma_E$, i.e., softer disks ($\alpha_E < 0$) ending up gaining larger forces ($\alpha_f > 0$) while stiffer disks ($\alpha_E > 0$) ending up gaining smaller forces ($\sigma_f < 0$). Second, as $\sigma_E$ increases, the positive correlation becomes weaker and the portion of ``outliers" become greater. The above two observations suggest the existence of nonlocal effect, i.e., how much force a disk sustains depends not only on how stiff that disk is, but also how stiff the surrounding disks are. We accordingly perform a first-order nonlocality check, labeling the stiffness of a disk by a new quantity $E_c$ that considers immediate contacting disks:
\begin{equation}
\label{effectivestiffness}
E_c = \frac{1}{N_c}\sum_{n=1}^{N_c}\left( 1/E_n+  1/E\right)^{-1},
\end{equation}
where $N_c$ is the number of contacting disks ($N_c = 4$ for a square packing), and $E_n$ is the corresponding Young's modulus of a contacting disk. For disks near the boundaries, we simply set $E_n$ to be infinity ($1/E_n \rightarrow 0$).  Each quantity associated with a contact inside the summation on the right-hand side of Eq. \eqref{effectivestiffness} can be viewed as an effective stiffness from two springs linked in a serial, each of whose stiffness equals the Young's modulus of a corresponding contacting disk. $E_c$ thus on average quantifies how ``stiff" a disk is by incorporating the stiffness of nearby disks. Accordingly, we use a new quantity $\alpha_{E_c} = (E_c/E_{c,\text{mean}}-1)\times 100\%$ and plot its correlation with $\alpha_f$ (shown in the second panel of Fig. \ref{deviation}(b)). Here $E_{c,\text{mean}}$ is simply the value of $E_c$ in the reference configuration where $E = E_\text{mean}$ for every disk. As shown in the second panel of Fig. \ref{deviation}(b)), compared to simply using $E$, $E_c$ appears to show better correlation with $\alpha_f$, especially when $\sigma_E$ is relatively small ($\sigma_E < 2$). More specifically, correlated data (colored in blue) are less scattered and the portion of ``outliers" (colored in red), particularly of those lying in the fourth quadrant ($E_c > E_{c,\text{mean}}$ but $f < f_0$), reduces considerably. However, when $\sigma_E$ further increases, the relevance of $E_c$ to $f$ decreases, though $E_c$ still performs slightly better than $E$. For the hexagonal packing, similarly, at the same five $\sigma_E$ values (Fig. \ref{deviation}(c)), we plot the correlation between $\alpha_E$ and $\alpha_f$ (top panel of Fig. \ref{deviation}(d)), and that between $\alpha_{E_c}$ and $\alpha_f$ (bottom panel of Fig. \ref{deviation}(d)). The general trends are very similar to those of the square packing case but the correlations are much stronger, especially when $\sigma_E$ is relatively large (e.g., $\sigma_E = 32$). The observed much stronger correlation in a hexagonal packing may be explained by its larger coordination number ($\langle Z \rangle= 6$) over that of a square packing ($\langle Z\rangle = 4 $), which statistically promotes energetically favored load-bearing paths formed by stiffer disks to balance the externally applied load, thereby implying a less pronounced nonlocal effect. For both packing lattices, however, we do observe the consistent existence of ``soft outliers" (those with $\alpha_{E_c} < 0$ but $\alpha_f > 0$) regardless of the use of $E_c$ or $E$. Such observations suggest that there is non-negligible spatial correlation among constituent disks, and that such correlation appears to be long-range extending beyond a first-order nonlocality.

\begin{figure}[htbp!]
\begin{subfigure}[t]{0.5\textwidth}
\includegraphics[scale=0.9, trim = -0.3in 0.1in 0 0]{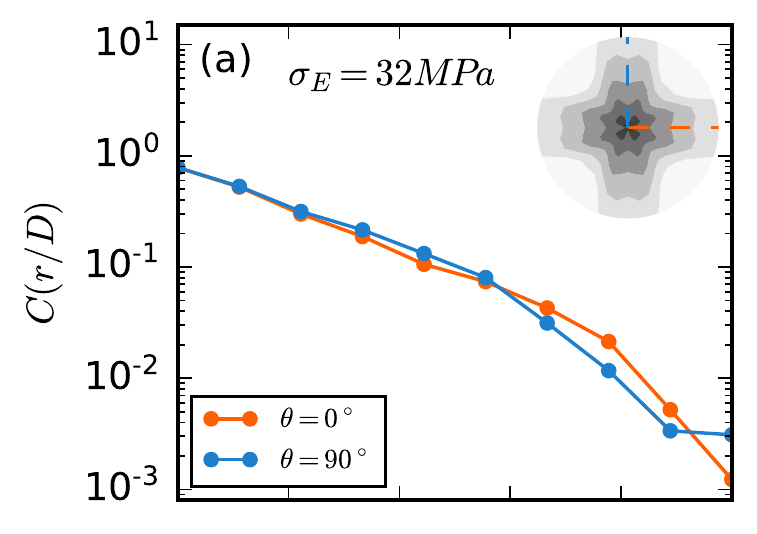}
\end{subfigure}
\begin{subfigure}[t]{0.5\textwidth}
\includegraphics[scale=0.9, trim = -0.3in 0.1in 0.2in 0]{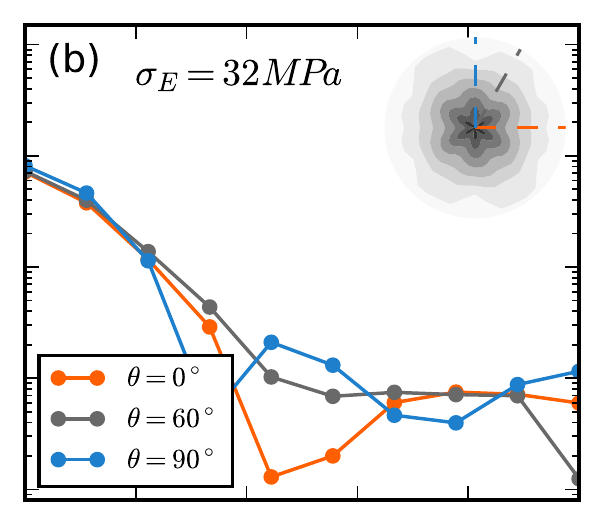}
\end{subfigure}
\begin{subfigure}[t]{0.5\textwidth}
\includegraphics[scale=0.9, trim = -0.3in 0.1in 0 0]{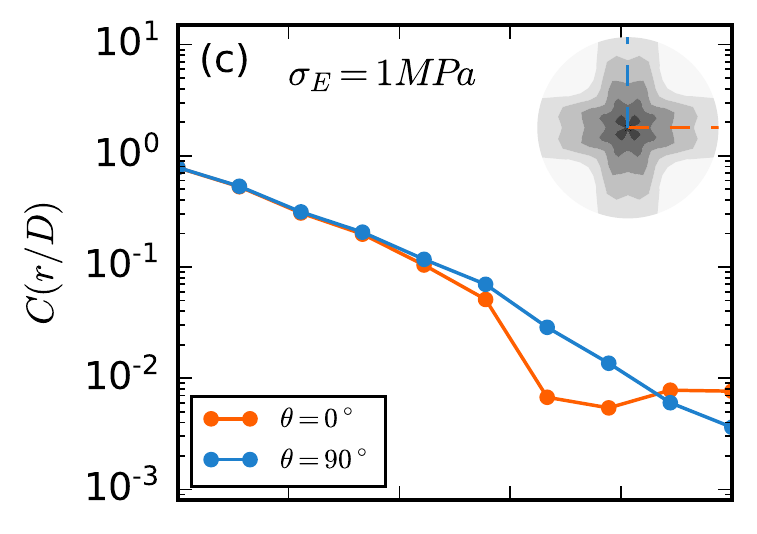}
\end{subfigure}
\begin{subfigure}[t]{0.5\textwidth}
\includegraphics[scale=0.9, trim = -0.3in 0.1in 0.2in 0]{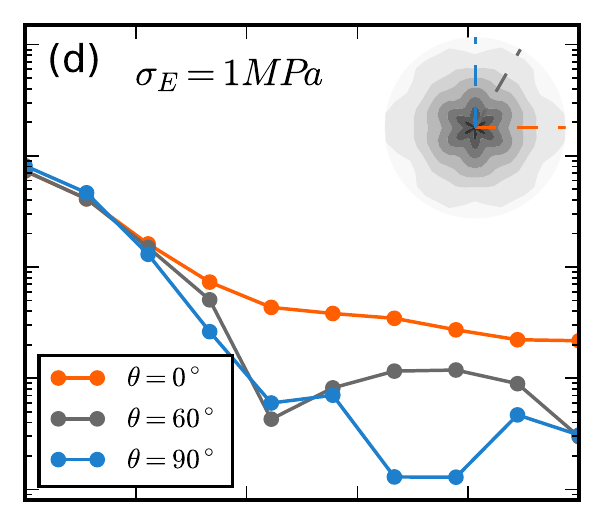}
\end{subfigure}
\begin{subfigure}[t]{0.5\textwidth}
\includegraphics[scale=0.9, trim = -0.3in 0.1in 0 0]{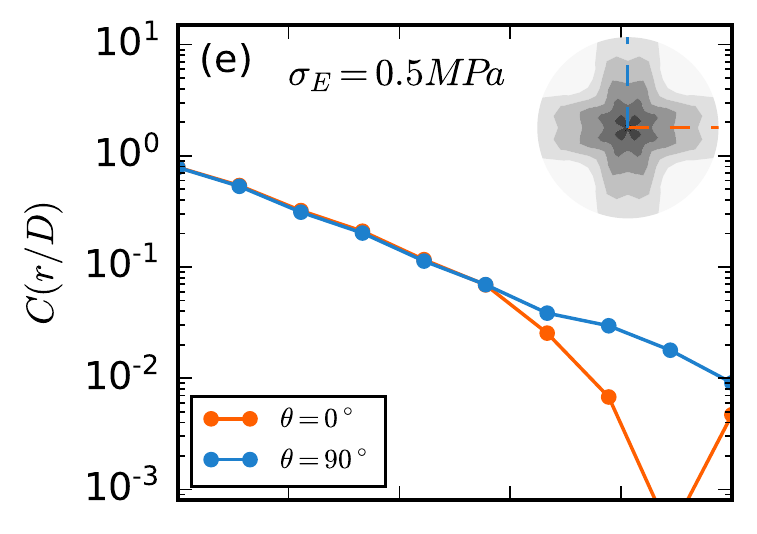}
\end{subfigure}
\begin{subfigure}[t]{0.5\textwidth}
\includegraphics[scale=0.9, trim = -0.3in 0.1in 0.2in 0]{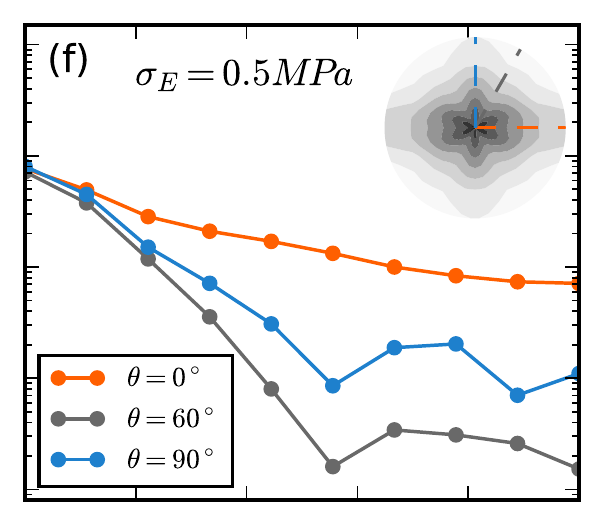}
\end{subfigure}
\begin{subfigure}[t]{0.5\textwidth}
\includegraphics[scale=0.9, trim = -0.3in 0.1in 0 0]{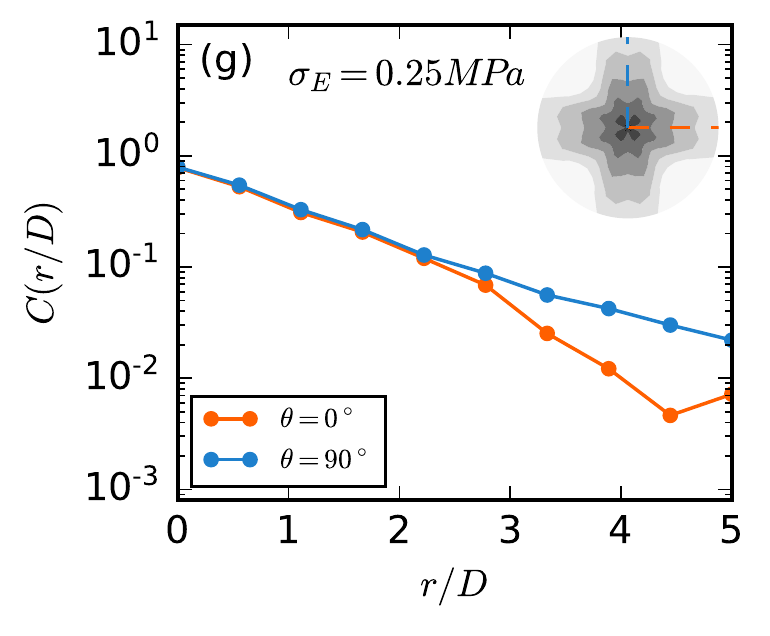}
\end{subfigure}
\begin{subfigure}[t]{0.5\textwidth}
\includegraphics[scale=0.9, trim = -0.25in 0.1in 0.2in 0]{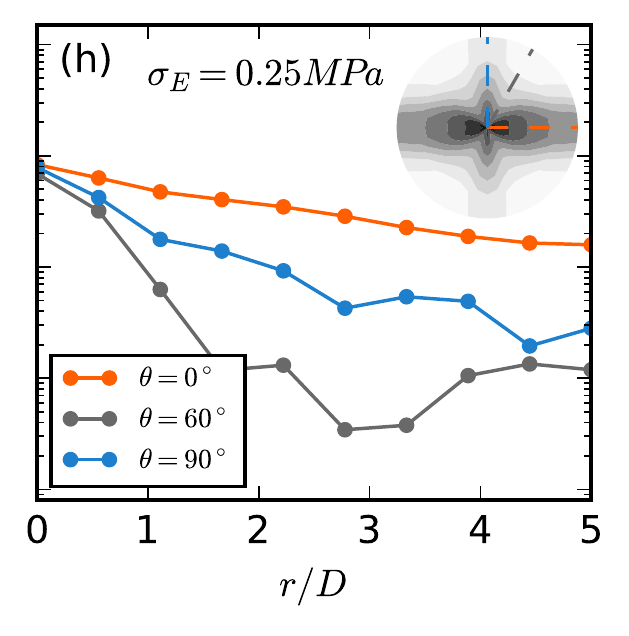}
\end{subfigure}
\caption{Spatial two-point correlation of the total magnitude of the force acting on a particle, evaluated along specific directions, and plotted as a function of the normalized distance. Plots (a), (c), (e), and (g) correspond to $\sigma_E=32,1,0.5,0.25$ MPa respectively for the square packing. Similarly, plots (b), (d), (f), and (h) correspond to $\sigma_E=32,1,0.5,0.25$ MPa respectively for the hexagonal packing. Insets show the polar contours of correlation for each case.}
\label{spatialCorrelationGraph}
\end{figure}

\begin{figure}[H]
\centering
\includegraphics[width=0.95\linewidth]{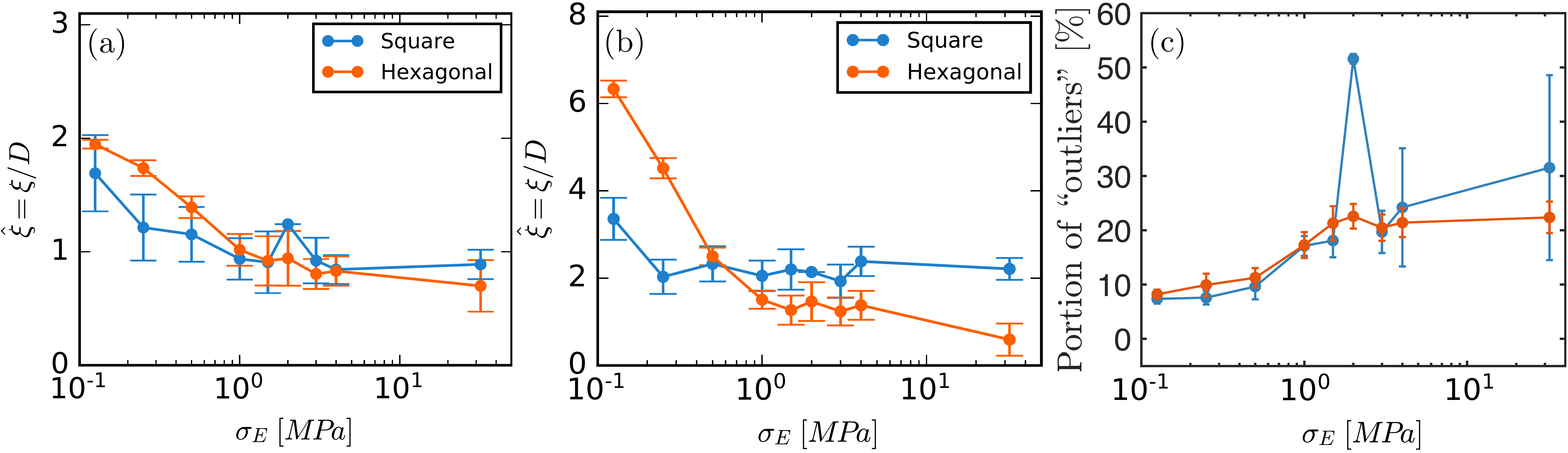}
\caption{(a) Angle-averaged normalized correlation length plotted against $\sigma_E$ for the square (colored blue) and the hexagonal (colored in red) packing. (b) A similar figure to (a) but for horizontally-oriented correlation length. (c) The portion of ``outliers" plotted against $\sigma_E$ for the square (blue) and the hexagonal (red) packing. Error bars indicate one standard deviation from the ten independently sampled configurations for each $\sigma_E$.}
\label{correlationLengthVsSigmaEGraph}
\end{figure}

To shed further light on nonlocality, we proceed to quantify the spatial force correlations within these ordered packings. Following \citep{majmudar2005contact}, we compute the two-point correlation function $C(\bm{r}) = \langle F(\bm{x}) F(\bm{x} + \bm{r}) \rangle_{\bm{x}}$, where $F$ is the sum
of the magnitudes of the contact forces on a particle\footnote{We also studied the spatial correlation of normal and tangential forces, which produced similar results, and are therefore not reported.}, and $\bm{x}$ is the position of its centroid. In order to investigate the correlation along different directions, we do not average over the angle. We pick the same four scenarios as earlier ($\sigma_E = 0.25, 0.5, 1$ and $32$) under $F = 250\,\,\text{N}$, and for each scenario, we plot the correlation (averaged over the ten independently sampled configurations) against the normalized radial distance along various angles (Fig.~\ref{spatialCorrelationGraph}). In particular, the left column (Figs~\ref{spatialCorrelationGraph} (a), (c), (e), and (g)) corresponds to the square packing, while the right column (Figs~\ref{spatialCorrelationGraph} (b), (d), (f), and (h)) corresponds to the hexagonal packing. In each plot, the insets reveal the polar contours of spatial correlation. The immediate observation is that, in both arrangements, the spatial force correlation indeed extends far beyond the first neighbors. This is consistent with the observation of force chain-like structures in Section~\ref{SectionProbabilityDistributionOfContactForces}. Despite the isotropic loading conditions, and in contrast to observations in DPRD, the correlation is highly anisotropic, and reflects the orientation of contacts in each arrangement. Interestingly, in the case of the square packing, the anisotropy appears to be independent of the degree of mechanical heterogeneity. On the contrary, the hexagonal packing with low mechanical heterogeneity exhibits a spatial distribution of forces whose correlation is more pronounced in the horizontal direction, while the same packing with large mechanical heterogeneity shows a spatial correlation that is equally pronounced along the six contact directions inherent to the arrangement. We postulate that this is due to the directional bias of the contact forces, inherent to the homogeneous hexagonal packing, which becomes less pronounced as heterogeneity increases. Finally, in Fig. \ref{correlationLengthVsSigmaEGraph}(a), we plot the evolution of the normalized correlation length $\hat{\xi} = \xi/D$ (where $D$ denotes the particle diameter) as a function of the material heterogeneity ($\sigma_E$). Note that the correlation length $\xi$ is obtained by fitting an exponential correlation kernel $C(r) = \alpha \exp(-r/\xi)$ to the angle-averaged correlation data, and is indicative of the characteristic size of particle chains and clusters that are responsible for force transmissions. In an isotropic (angle-averaged) sense, the hexagonal packing exhibits a slightly larger correlation length than the square packing in the low heterogeneity regime (e.g., $\sigma_E < 2$ MPa), while the opposite is true in the large heterogeneity regime (e.g., $\sigma_E \geq 2$ MPa). This phenomenon is more pronounced in the case where the correlation length is computed along the horizontal direction, as shown in Fig. \ref{correlationLengthVsSigmaEGraph}(b). These observations are in line with our expectations of the portion of ``outliers" (discussed in Section~\ref{SectionParticleScaleGain}) with respect to the total number of disks computed and shown in Fig. \ref{correlationLengthVsSigmaEGraph}(c), where the hexagonal packing shows a larger portion of ``outliers" over the square packing when $\sigma_E$ is relatively small (e.g., $\sigma_E < 2$ MPa), while the opposite is true when $\sigma_E$ goes beyond 2. Finally, as expected, both packings show an overall decaying correlation length upon increasing material heterogeneity.%

\subsubsection{Effect of contact friction}
We have also tried varying the contact friction $\mu_s$ to be either smaller or larger than 0.5, and we find $\mu_s$ has little effect on our findings so long as we are able to get converged solutions. In our case non-convergence happens when we reduce $\mu_s$, and it happens more frequently with the increase of $\sigma_E$. On the contrary, when $\mu_s > 0.5$ we are always able to find converged solutions. These non-converging scenarios imply reorganization events (e.g., large rotations of particles induced by frictional instabilities) which can also lead to loss of contacts\footnote{We rule out the possibility of numerical instabilities (e.g. those induced by poor mesh qualities) since our simulations converge for certain values of $\mu_s$.}. For instance, we find that for $\mu_s < 0.35$ we were unable to find converged solutions for hexagonal configurations sampled from $\sigma_E = 32$. It is possible in these scenarios that a relatively soft particle loses all contacts, while the surrounding particles are relatively stiffer and are connected in a way capable of sustaining the external load. When reorganizations happen under smaller $\mu_s$ they allow greater flexibility among particles to explore and form more energetically favored load-bearing paths, which may then lead to a more heterogeneous contact force distribution and a stronger correlation between $\alpha_{E_c}$ and $\alpha_f$. Extending our findings to these scenarios would either require an extension our current implementation to be dynamic, or experimentations with techniques \cite{hurley2014extracting,marteau2017novel} that can measure inter-particle forces among different types of materials.

\section{Summary and outlook}
\label{section:summary}
In this work, we explore the effect of mechanical heterogeneity on inter-particle forces in deformable granular media by means of numerical simulations using a FEM 2D contact mechanics algorithm. Specifically, we study two canonical packing lattices—a square lattice and a hexagonal packing lattice—under quasi-static isotropic compression. For both packing lattices, we show that heterogeneous inter-particle force distribution emerges as the Young's moduli of constituent disks gradually deviate from being homogeneous, despite disks being arranged orderly in the absence of geometric heterogeneity. 

At the system level, under the same level of mechanical heterogeneity, we observe that the hexagonal packing lattice shows a more heterogeneous inter-particle force distribution than the square packing lattice does. This observation may be explained by the larger coordination number of the hexagonal packing lattice that promotes more load-bearing paths. Correspondingly, we find that, on the one hand, for normal force well above the mean, the probability distribution from the hexagonal packing shows a longer tail than that from the square packing. For normal forces well below the mean, the probability distribution from both lattices show tails dipping toward zero. However, when the probability distribution is plotted by normalizing normal forces against their reference values (i.e., $f_{0n}$), it shows a much weaker lattice-dependent trend with both forces below and above the mean exhibiting exponential tails. On the other hand, tangential (frictional) forces well above the mean in both lattices show tails decaying faster than exponential ones that typically appear in DPRD. Finally, both studied systems exhibit long-range spatial force correlation, which is consistent with our observations of emerging force chain-like structures.

At the particle scale, both packing lattices show beyond-first-order spatial correlation (i.e., nonlocal) effect in the sense that the amount of contact force a disk sustains can be determined neither by considering the stiffness of that disk alone nor by further considering the stiffness of immediately contacting disks. Specifically, for both packing lattices we also observe the coexistence of ``outliers": softer disks gaining larger forces and stiffer disks gaining smaller forces. Additional analysis on spatial force correlation confirms that the spatial correlation effect is indeed beyond first order. Further, the analysis suggests that the hexagonal packing lattice shows strong nonlocal effect over the square packing when $\sigma_E$ is relatively small (i.e., $\sigma_E < 2$) and the opposite holds when $\sigma_E$ goes beyond 2, which is in line with the observation of the portion of ``outliers" in a hexagonal packing being larger than that of the square packing when $\sigma_E < 2$ and being smaller otherwise.

Concerning the effect of contact friction, we find that so long as no particle reorganization occurs, our findings are insensitive to the particular value of $\mu_s$. However, our simulations show that more mechanically heterogeneous packing requires a higher $\mu_s$ to prevent reorganizations from happening, suggesting a role of packing stabilization played by contact friction.

Looking forward, we present several potential future research directions. The first direction concerns a deeper understanding aiming at correlating the amount of contact force a disk sustains to the mechanical characterization of that disk. Our simulations show that even in simple ordered packing, it is challenging to account for the range of nonlocal effect. We believe that it is promising to tackle this challenge through a combination of network theory and machine learning. Specifically, network theory allows us to extract communities \cite{bassett2015extraction,karapiperis2021nonlocality}, thus implicitly taking into account the nonlocal effect. These communities serve as excellent sources from which one can extract multiple descriptors associated to a single disk in a way similar to \cite{cubuk2015identifying}. From these descriptors we could train machine learning (ML) algorithms to identify relevant descriptors, first as a labeling problem (i.e., correctly identifying ``soft" and ``stiff" disks) and later as a regression problem (i.e., quantitatively predicting the amount of contact forces). These trained ML models may be used to further explore possible finite size (boundary) effects. The second direction concerns extending our findings to regimes where particle reorganizations occur. Reorganization events allow granular media to explore more packing configurations, potentially leading to more energy-favored loading paths. It would be interesting to investigate how reorganization events change the inter-particle forces and their distribution. Such investigations can be achieved by extending our implementation to be fully dynamic (and possibly to account for finite kinematics), or through experiments using techniques capable of measuring inter-particle forces between particles made of different materials \cite{hurley2014extracting,marteau2017novel}. A final promising direction relates to the development of analytical models of force transmission in these mechanically heterogeneous packings. In this regard, it would be worthwhile studying how theories such as the q-model \citep{liu1995force} could be extended and adapted for these systems.

In all, by exploring the ``disordered" space in terms of particles’ mechanical properties, our work offers a fresh perspective to the classical understanding of inter-particle forces in granular media. It is in the hope of the authors that this work will promote further investigations along this direction.

\section*{Acknowledgement}
The authors would like to thank Dr. Ruobing Bai of Northeastern University and Dr. Siavash Monfared of Caltech for valuable comments on this paper. Part of the FEM contact mechanics implementation benefits from the computational mechanics course (AE/ME 108a) L.L. took in the fall of 2014 as a graduate student at Caltech. L.L. thanks the partial financial support provided by the Laboratory Directed Research and Development (LDRD) funding and the US Department of Energy (DOE), the Office of Nuclear Energy, Spent Fuel and Waste Disposition Campaign, under Contract No. DE-AC02-05CH11231 with Berkeley Lab. 
\newpage
\appendix

\section{Problem setting and FEM implementation}
 \label{subsection:FEM}
 \subsection{Governing equations}
 Let us consider a system consisting of $N$ finite solid bodies $\Omega^{(i)}$ together with their boundaries $\partial \Omega^{i}$, where $i \in \mathcal{I}\coloneqq\{1,2,...,N\}$ labels each solid body. For each solid boundary $\partial\Omega^{i}$, we may decompose it into a union as $\partial \Omega^{(i)} = \partial \Omega^{(i)}_{D} \cup \Omega^{(i)}_{N} \cup \Omega^{(i)}_{C} $ with $\partial \Omega^{(i)}_{D} \cap \partial\Omega^{(i)}_{N} = \varnothing, \partial \Omega^{(i)}_{C} \cap \partial\Omega^{(i)}_{N} = \varnothing$ and $\partial \Omega^{(i)}_{N} \cap \partial\Omega^{(i)}_{C} = \varnothing$, where $\partial \Omega^{(i)}_{D}, \partial \Omega^{(i)}_{N}$ and $\partial \Omega^{(i)}_{C}$ indicate the Dirichlet, Neumann and contact boundary conditions, respectively. We can further decompose $\partial \Omega^{(i)}_{C}$ into a union as:
 \begin{equation}
     \begin{aligned}
      \partial \Omega^{(i)}_{C} &= \bigcup\limits_{j \in\mathcal{I}^{(i)}_C} \partial \Omega^{(i\leftarrow j)}_C,\\
      \partial \Omega^{(i\leftarrow j)}_C \cap \partial \Omega^{(i\leftarrow k)}_C &= \varnothing \quad \forall \quad j,k \in \mathcal{I}^{(i)}_C \,\,\text{and}\,\, j\neq k,
      \end{aligned}
\end{equation}
where $\mathcal{I}^{(i)}_C$ is the set containing labels of all other solid bodies that are in contact with solid body $(i)$, and $\partial \Omega^{(i\leftarrow j)}_C$ means the contact boundary invoked by body $(j)$ onto $(i)$. Naturally, we must have $\partial \Omega^{(i\leftarrow j)}_C= \partial \Omega^{(j\leftarrow i)}_C$ under equilibrium. Note that, the contact boundary condition can be viewed as a special type of the Neumann boundary condition whose specificities are however unknown \textit{a priori}. Under quasi-statics and linearized kinematics conditions, we will need to solve, for every material point of every solid body, the following boundary volume problem (BVP), or the so-called strong form:
 \begin{equation}
    \label{strongform}
     \begin{aligned}
     \bm{\nabla_{x} \cdot \sigma} + \rho^{(i)}\bm{b}=\bm{0} \quad &\text{in}\quad \Omega^{(i)},\\
     \bm{u}^{(i)} = \hat{\bm{u}}^{(i)}\quad &\text{on} \quad \partial \Omega^{(i)}_D,\\
     \bm{\sigma}^{(i)}\cdot \bm{n}^{(i)}_N = \hat{\bm{t}}^{(i)}_N\quad &\text{on} \quad \partial \Omega^{(i)}_N,\\
     \bm{\sigma}^{(i)}\cdot \bm{n}^{(i)}_C = \hat{\bm{t}}^{(i)}_C\quad &\text{on} \quad \partial \Omega^{(i)}_C,
     \quad \forall \quad i \in \mathcal{I},\\
     \end{aligned}
\end{equation}
where $\bm{b}$ is the gravitational constant, and for each solid body $(i)$, $\bm{\sigma}$ is the Cauchy stress tensor, $\rho$ is the material density, $\bm{\hat{u}}$ is the imposed displacement field along $\partial \Omega_D$, $\bm{n}_N$ and $\hat{\bm{t}}_N$ are the boundary outward normal and imposed external traction along $\partial \Omega_D$ respectively, $\bm{n}_C$ and $\hat{\bm{t}}_C$ are the contact boundary outward normal and the contact traction along $\partial \Omega_D$ respectively. Due to the presence of contacts that are unknown \textit{a priori} (i.e., depending on the deformed configuration $\bm{x}$), here we evaluate every quantity in the deformed configuration, although for quantities not involved in contact it makes little difference to evaluate them instead in the undeformed configuration $\bm{X}$, thanks to the linearized kinematics (small deformation) assumption. For example, $\bm{\nabla_{X} \cdot \sigma} \simeq \bm{ {\nabla_{x} \cdot \sigma}}$ and $\bm{n}_N(\bm{x}) \simeq \bm{n}_N(\bm{X})$. Using the same coordinate frame for $\bm{X}$ and $\bm{x}$, we may want to solve for the displacement field $\bm{u}(\bm{x}) \simeq \bm{u}(\bm{X})$ that both the contact traction $\hat{\bm{t}}_C$  and the boundaries $\partial \Omega_C$ depend on. For quantities involved in contact between any two solid bodies $(i)$ and $(j)$, compatibility and Newton’s third law further impose the following constraints:
 \begin{equation}
    \label{contactconstraint}
     \begin{aligned} 
     \hat{\bm{t}}^{(i\leftarrow j)}_C &= -\hat{\bm{t}}^{(j\leftarrow i)}_C,\\
     \hat{\bm{n}}^{(i\leftarrow j)}_C &= -\hat{\bm{n}}^{(j\leftarrow i)}_C, \quad \forall \quad \bm{x} \in  \Omega^{(i\leftarrow j)}_C( = \Omega^{(j\leftarrow i)}_C).
     \end{aligned}
 \end{equation}
 \subsection{Variational formulations}
Now, with the strong form Eq. \eqref{strongform} and the contact constraints Eq. \eqref{contactconstraint} being defined, we construct the corresponding weak form through variational formulation for each solid body:
\begin{equation}
   \label{weakform}
    \begin{aligned}
    &\text{find}\,\, \bm{u^}{(i)} \in \mathcal{U}^{(i)} = \big\{\bm{u}|\bm{u} \in \mathcal{H}^{1}(\Omega^{(i)}), \bm{u}=\hat{\bm{u}}^{(i)} \,\,\text{on}\,\, \partial\Omega^{(i)}_D \big\} \\
    &\text{s.t.} \quad \int_{\Omega^{(i)}} \bm{\sigma}(\bm{u}^{(i)})\bm{\nabla_x}\bm{v}dV = \int_{\Omega^{(i)}}\rho^{(i)}\bm{b}\bm{v}dV + \int_{\partial \Omega^{(i)}_N} \hat{\bm{t}}^{(i)}_N\bm{v}dS+ \int_{\partial \Omega^{(i)}_c} \hat{\bm{t}}^{(i)}_C\bm{v}dS\\
 &\forall \quad \bm{v} \in \mathcal{V}^{(i)} = \big\{\bm{v}|\bm{v} \in \mathcal{H}^{1}(\Omega^{(i)}), \bm{v}=\bm{0}  \,\,\text{on}\,\, \partial\Omega^{(i)}_D\big\},\\
    \end{aligned}
\end{equation}
where 
\begin{equation}
  \label{contactterm}
   \int_{\partial \Omega^{(i)}_c} \hat{\bm{t}}^{(i)}_C\bm{v}dS = 
    \sum_{j\in\mathcal{I}^{(i)}_C}\int_{\partial \Omega^{(i\leftarrow j)}_C} \hat{\bm{t}}^{(i\leftarrow j)}_C\bm{v}dS.
\end{equation}
We can see that, due to the presence of contacts among solid bodies as imposed by Eq. \eqref{contactterm}, Eq. \eqref{weakform} is coupled across different solid bodies. Further, Eq. \eqref{weakform} must be solved iteratively as quantities presented in Eq. \eqref{contactterm} depend on the solution $\bm{u}^{(i)}$ of each solid body $(i)$ which is unknown \textit{a priori}. For example, $\hat{\bm{t}}^{(i\leftarrow j)}$ and $\partial \Omega^{(i\leftarrow j)}_C$ depend on both $\bm{u}^{(i)}$ and $\bm{u}^{(j)}$ through a specific choice of contact law. To this end, let us close Eq. \eqref{weakform} with a constitutive law governing the behavior of the solids. For simplicity, we consider isotropic linear elasticity with $\bm{\sigma} = \bm{\mathcal{C}} :\bm{\epsilon}$, where $\bm{\epsilon} = [\bm{\nabla u}+(\bm{\nabla {u}})^{T}]/2$ is the infinitesimal strain tensor and $\bm{\mathcal{C}}$ is the fourth-order material stiffness tensor. 
\subsubsection{FEM implementations}
Now with the weak form Eq. \eqref{weakform} being defined, we then discretize it over a triangulation $\mathcal{T}^{h,(i)}$ for each solid body $\Omega^{(i)}$. For simplicity, we choose $\mathcal{P}_1(\mathcal{T}^{h,(i)})$ the continuous piecewise linear function space on $\mathcal{T}^{h,(i)}$, a subspace of the Sobolev space $\mathcal{H}^1(\Omega^{(i)})$, to discretize both $\bm{u}$ and $\bm{v}$ following the Bubnov-Galerkin approximation. We finally arrive at solving the following weak form which is the discretized version of Eq. \eqref{weakform}:
\begin{equation}
\label{FEMweak}
  \begin{aligned}
    &\text{find}\,\, \bm{u^}{h,(i)} \in \mathcal{U}^{h,(i)} = \big\{\bm{u}|\bm{u} \in \mathcal{P}_1(\mathcal{T}^{h,(i)})), \bm{u}=\hat{\bm{u}}^{(i)} \,\,\text{on}\,\, \partial\mathcal{T}^{h,(i)}_D \big\} \\
    &\text{s.t.} \quad \int_{\Omega^{(i)}} \bm{\sigma}(\bm{u}^{h,(i)})\bm{\nabla_x}\bm{v}^hdV = \int_{\Omega^{(i)}}\rho^{(i)}\bm{b}\bm{v}^hdV + \int_{\partial \Omega^{(i)}_N} \hat{\bm{t}}^{(i)}_N\bm{v}^hdS+ \int_{\partial \Omega^{(i)}_c} \hat{\bm{t}}^{(i)}_C\bm{v}^hdS\\
   &\forall \quad \bm{v}^h \in \mathcal{V}^{h,(i)} = \big\{\bm{v}|\bm{v} \in \mathcal{P}_1(\mathcal{T}^{h,(i)})), \bm{v}=\bm{0}  \,\,\text{on}\,\, \partial\mathcal{T}^{h,(i)}_D\big\}.\\
   \end{aligned}
\end{equation}
Now, due to the arbitrariness of $\bm{v}^h$, Eq. \eqref{FEMweak} further implies the following system of algebraic equations to hold:
\begin{equation}
\label{compact}
\bm{K}\bm{U}^a = \bm{F}_{\text{ext}} + \bm{F}_{\text{contact}}(\bm{U}^a),
\end{equation}
with
\begin{align}
\label{globalstiffness}
\bm{K} &= \text{diag}\big\{\bm{K}^{(1)},\bm{K}^{(2)},...,\bm{K}^{(N)}\big\},\\
\label{globalexternal}
\bm{U}^a &= [\bm{u}^{a,(1)},\bm{u}^{a,(2)},...,\bm{u}^{a,(N)}\big]^{T},\\
\bm{F}_{\text{ext}} &= [ \bm{F}^{(1)}_{\text{ext}},\bm{F}^{(2)}_{\text{ext}},...,\bm{F}^{(N)}_{\text{ext}} ]^{T},\\
\bm{F}_{\text{contact}} &= [ \bm{F}^{(1)}_{\text{contact}}(\bm{U}^a),\bm{F}^{(2)}_{\text{contact}}(\bm{U}^a),...,\bm{F}^{(N)}_{\text{contact}}\bm{U}^a) ]^{T},
\end{align}
where for each solid body $(i)$, $\bm{K}^{(i)}$ is the stiffness matrix, $\bm{u}^{a,(i)}$ is the nodal displacements in need of solving, $\bm{F}^{(i)}_{\text{ext}}$ accounts for the ordinary Neumann boundary condition and the body force term, and $\bm{F}_{\text{contact}}$ accounts for the contact forces incurred through interaction with neighboring solid bodies. Since $\bm{F}_{\text{contact}}$ depends on the solution $\bm{U}^a$, Eq. \eqref{compact} is nonlinear in $\bm{U}^a$ and must be solved iteratively. Equivalently, we want to solve the following root-finding problem:
\begin{equation}
\label{globalresidue}
\begin{aligned}
\text{Find}\,\, \bm{U}^a, \,\,\text{s.t.}\,\, \bm{R}(\bm{U}^a) = \bm{K}\bm{U}^a - \bm{F}_{\text{ext}} - \bm{F}_{\text{contact}}(\bm{U}^a)=\bm{0}.
\end{aligned}
\end{equation}
Suppose that we know $\bm{U}^{a,k}$ at the $k$th iteration, then at the $k+1$-th iteration, the displacement field can be found through Newton-Raphson as the following:
\begin{equation}
\label{newtonraph}
\begin{aligned}
\bm{U}^{a,k+1} &= \bm{U}^{a,k} -(\bm{J}^{k})^{-1}\bm{R}(\bm{U}^{a,k}), \\
\text{with} \quad \bm{J}^k &= \frac{\partial \bm{R}}{\partial \bm{U}^a}\bigg\rvert_{\bm{U}^a = \bm{U}^{a,k}} = \bm{K} - \bm{J}^k_c,\,\,\text{and}\,\, \bm{J}^k_c = \frac{\partial \bm{F}_{\text{contact}}}{\partial \bm{U}^a}\bigg\rvert_{\bm{U}^a= \bm{U}^{a,k}},
\end{aligned}
\end{equation}
where $\bm{J}^k$ is the global jacobian which takes a contribution from the contact jacobian $\bm{J}^k_c$. The above procedure is iterated until the difference between $\bm{U}^{a,k}$ and $\bm{U}^{a,k+1}$ is sufficiently small. Note that the form of $\bm{J}^k$ shown in Eq. \eqref{newtonraph} is deduced assuming linearized kinematics and linear elasticity, as values of $\bm{K}$ and $\bm{F}_{\text{ext}}$ change negligibly after deformation and can thus be treated as being independent of $\bm{U}^{a}$. However, for finite kinematics and nonlinear materials this is no longer true and their gradients with respect to $\bm{U}^a$ will need to be evaluated at each iteration step as well. 

In general, deriving the analytical expression of $\bm{J}^k_c$ is very hard since it depends not only on the specific local geometry within the contact region between any two solid bodies, but also the specific choice of contact law. Therefore, without loss of generality, we choose to compute $\bm{J}^k_c$ numerically via finite difference:
\begin{equation}
\label{contactjacobian}
    \begin{aligned}
    \bm{J}_c^k(:,m) \simeq \frac{\bm{F}_{\text{contact}}(\bm{U}^{a,k}+\bm{e}^{m,k})-\bm{F}_{\text{contact}}(\bm{U}^{a,k})}{h},
    \end{aligned}
\end{equation}
where $\bm{J}_c^k(:,m)$ means the $m$-th column of $\bm{J}_c^k$, and $m$ represents the global indexing of a node of a solid body along either the $x$ or the $y$ degree of freedom (in 2D). $\bm{e}^{m,k}$ is a unit vector with all entries being zero except the $m$-th one which has a value of $h$ with $h$ being a small value that takes the following form similar to \cite{liu2020ils}:
\begin{equation}
\label{perturbationvector}
\bm{e}^{m,h} = [0,...,0,\underbrace{h}_{m\text{-th entry}},0,...,0]^{T}, \quad h = \text{max}(\epsilon,\epsilon|U^{a,k}_m|).
\end{equation}
Here $U^{a,k}_m$ means, at the $k$-th iteration, the nodal displacement of the solid body’s node whose global index is $m$ along either the $x$ or the $y$ degree of freedom (in 2D). $\epsilon$ is a user-defined small value and may be related to the specific machine precision of the executing computer. We can think of the $m$-th column of $\bm{J}^k_c$ as the resulting contact forces acting on each solid body when the $m$-th degree of freedom of the whole system, which corresponds to a certain degree of freedom of one node of one solid body, is being slightly perturbed. From this perspective, we can view $\bm{J}_c^k$ as an incremental contact stiffness matrix at the $k$-th iteration. In light of this, at each iteration $k$, we do not need to perturb every single node of every solid body, but instead we only need to perturb nodes of each solid body that are on the boundary and are ``active". By ``active" we refer to nodes near the contact region that, when whose displacements are perturbed, can induce changes on $\bm{F}_{\text{contact}}$. For contact between any two solids $(i)$ and $(j)$, we carry out the contact force computation twice by interchanging the role of $(i)$ and $(j)$ for the master and the slave and averaging the results. We take this role-interchanging step as being beneficial to minimize the potential bias \cite{laursen2013computational,sauer2015unbiased,mlika2017unbiased} that can arise in the computed contact forces. Alternatively, one can perform contact computations on a so-called neutral contact surface \cite{chi2015level,leichner2019contact,liu2020ils} without computing contact forces twice. Finally, our implementation is outlined by Algorithm 1 (see Appendix) where we leverage the sparse representations and solvers available in the open-source library Eigen \cite{eigenweb} to efficiently store $\bm{K}$ and $\bm{J}^k_c$, and to solve for $\bm{U}^{a,k}$. We design the geometries of solid bodies and generate the triangulation/mesh using Gmsh \cite{geuzaine2009gmsh}.

\section{Contact contribution computation}
 \label{subsection:contactcomputation}
 In this subsection, we discuss in detail how we compute the contact forces $\bm{F}_{\text{contact}}$ and contact jacobians $\bm{J}_c$ that appear in Algorithm 1. For simplicity, we focus on 2D cases, while the 3D version of the introduced methodology can be deduced analogously.
 \subsection{Contact law regularization}
\begin{figure}[H]
\centering
\includegraphics[width=0.5\linewidth]{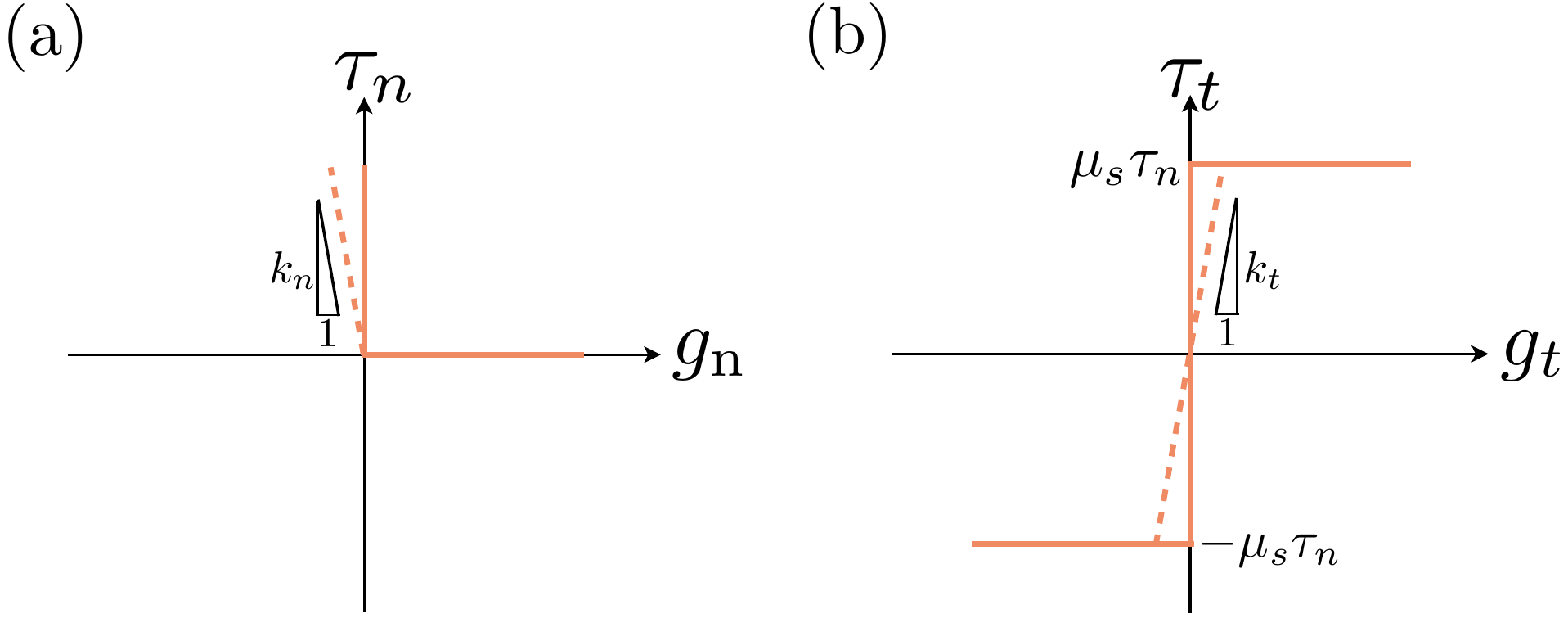}
\caption{Graphical representations of the contact laws. (a) Contact law in the normal direction shown in its original form (solid red lines) and regularized (penalized) form (red dashed lines), where $k_n$ is a user-defined regularization parameter termed as the normal contact stiffness. (b) Contact law in the tangential direction (the Coulomb friction law) shown in its original form (solid red lines) and regularized (penalized) form (red dashed lines), where $k_t$ is a user-defined regularization parameter termed as the tangential contact stiffness.}
\label{contactlawgraph}
\end{figure}

 As the object of this work is to study dry granular material, we consider contact interactions between solid bodies to be purely repulsive in the normal direction, and we assume the tangential interaction to obey the macro-scale Coulomb friction law. Further, since we focus in the quasi-static regime, we simply use a single constant friction coefficient $\mu_s$, making no distinction between static and dynamic friction coefficients. We acknowledge that, although the assumption of $\mu_s$ being constant is widely used in modeling frictional granular media, it is a simplification of the reality where $\mu_s$ can be highly stochastic and depends spatially on local geometric properties of a contact surface \cite{albertini2021stochastic}. To this end, to better illustrate the chosen contact law, let us focus on a simply case of contacts between two solid bodies $(i)$ and $(j)$ and with solid body $(i)$ being the master and solid body $(j)$ being the slave. Suppose that, in the deformed configuration, we pick a point $i_n$ with position $\bm{x}^{(i)}_{i_n}$ on the boundary of solid body $(i)$, and we find its closest projection material point $\bar{i}_n$ with position $\bm{x}^{(j)}_{\bar{i}_n}$ on the boundary of solid body $(j)$ together with the associated outward surface normal $\bm{n}$ (being parallel to $\bm{x}^{(i)}_{i_n}-\bm{x}^{(j)}_{\bar{i}_n}$ but pointing outside of solid body $(j)$) and tangential direction $\bm{t}$ satisfying $\bm{t}\cdot \bm{n} = 0$. With these quantities being defined, we further introduce two quantities associated with $i_n$:
 \begin{equation}
 \label{contactgap}
 \begin{aligned}
 \bm{g}^{i_n}_n &= g_n\bm{n}\,\,\text{with}\,\, g_n = (\bm{x}^{(i)}_{i_n}-\bm{x}^{(j)}_{\bar{i}_n})\cdot\bm{n},\\
  \bm{g}^{i_n}_t &= g_t\bm{t}\,\,\text{with}\,\, g_t = (\bm{u}^{(i)}_{i_n}-\bm{u}^{(j)}_{\bar{i}_n})\cdot\bm{t},
  \end{aligned}
 \end{equation}
where $\bm{u}^{(i)}_{i_n}$ and $\bm{u}^{(j)}_{\bar{i}_n}$ are the displacements of $i_n$ and $\bar{i}_n$, respectively. They are linked to the undeformed configuration through $\bm{x}^{(i)}_{i_n} =  \bm{X}^{(i)}_{i_n}+\bm{u}^{(i)}_{i_n}$ and $\bm{x}^{(j)}_{\bar{i}_n} =  \bm{X}^{(j)}_{\bar{i}_n}+\bm{u}^{(j)}_{\bar{i}_n}$, where $\bm{X}^{(i)}_{i_n}$ and $\bm{X}^{(j)}_{\bar{i}_n}$ are the positions of $i_n$ and $\bar{i}_n$ in the undeformed configuration, respectively. Physically, non-penetrability requires that $g_n \geq 0$. Then we can describe the contact law by the following two conditions (known as the Signorini conditions \cite{jean1999non}):
 \begin{equation}
 \label{contactlawperp}
 \begin{aligned}
   g_n = 0 \,\,\text{with}\,\, \tau_n > 0 &\perp g_n > 0 \,\,\text{with}\,\, \tau_n = 0,\\
   g_t = 0 \,\,\text{with}\,\, |\tau_t| \leq \mu_s\tau_n &\perp g_t > 0 \,\,\text{with}\,\, |\tau_t| = \mu_s\tau_n,\\
   \bm{\tau}_n &= \tau_n\bm{n},\\
   \bm{\tau}_t &=  -\text{sign}({g_t})|\tau_t|\bm{t},
 \end{aligned}
 \end{equation}
where $\tau_n$ and $\tau_t$ are the (signed) magnitude of normal and tangential traction acting on $\bm{p}$. The negative sign in the right-hand side of the tangential traction term $\bm{\tau}_t$ merely states that friction is always in the opposite direction of the relative motion. The notation ``$\perp$" represents that only one condition on either side of it can hold at a given material point. Graphically, Eq. \eqref{contactlawperp} can be described by the solid red lines shown in Fig. \ref{contactlawgraph}(a) and Fig. \ref{contactlawgraph}(b).
 
Enforcing the constraints shown in Eq. \eqref{contactlawperp} may be accomplished exactly through the Lagrangian multiplier method or the Augmented Lagrangian method, or approximately through the penalty/regularization method \cite{laursen2013computational}. Each method has its advantages and disadvantages, and one can choose a specific method based on the specific problem in need of analyzing. For modeling granular materials, in the opinion of the authors the penalty/regularization formulation (red dashed lines in Fig. \ref{contactlawgraph}(a) and Fig. \ref{contactlawgraph}(b)) is the most suitable route for the following two reasons: (1) it allows for much easier contact detections between solid particles, and (2) it allows for more flexible control of balancing the trade-off between the convergence of a computation and the capture of the contact physics. In short, the penalty formulation allows for a finite normal penetration through a normal contact stiffness $k_n$ (red dashed lines in Fig. \ref{contactlawgraph}(a)), and it allows for a finite tangential slip before frictional yielding through a tangential contact stiffness $k_t$ (red dashed lines in Fig. \ref{contactlawgraph}(b)). Self-evidently, the exact contact law is enforced when values of $k_n$ and $k_t$ go to infinity. We emphasize here that one can program the penalty formulation to behave in a very similar way to the Augmented Lagrangian method, by programming into an algorithm the variation of $k_n$ and $k_t$ during the Newton-Raphson iterations. More specifically, one can increase the values of $k_n$ and $k_t$ (e.g., double the values) after the algorithm converges under smaller values of $k_n$ and $k_t$, then continue the computation, and repeat the procedure until the algorithm converges at a desired value of $k_n$ and $k_t$ that can capture reasonably well the contact behaviors. Within the penalty formulation, we can express the regularized version of Eq. \eqref{contactlawperp} in its vectorial form as the following:
\begin{equation}
\label{contactpenalized}
 \begin{aligned}
  \bm{\tau}_n &= k_n|\text{min}(g_n,0)|\bm{n},\\
  \bm{\tau}_t &= -\text{min}(k_t|g_t|,\mu_s||\bm{\tau}_n||_2)\text{sign}(g_t)\bm{t}.\\
 \end{aligned}
\end{equation}
It then follows that, in the deformed configuration, we enforce Eq. \eqref{contactpenalized} instead of Eq. \eqref{contactlawperp} to every $i_n$ on the boundary of the master solid body $(i)$ near the contact region. Vice versa, if we interchange the role of master and slave for $(i)$ and $(j)$, we enforce Eq. \eqref{contactpenalized} to every $j_n$ on the boundary of $(j)$ near the contact region. It then follows that, when the Newton-Raphson converges under large enough $k_n$ and $k_t$, the true contact traction $\hat{\bm{t}}_C$ is well approximated as $\hat{\bm{t}}_C \simeq \bm{\tau}_n+\bm{\tau}_t$ along $\partial \Omega_C$ of a solid body.
 \subsubsection{Contact force and jacobian computation}
\begin{figure}[H]
\centering
\includegraphics[width=0.9\linewidth]{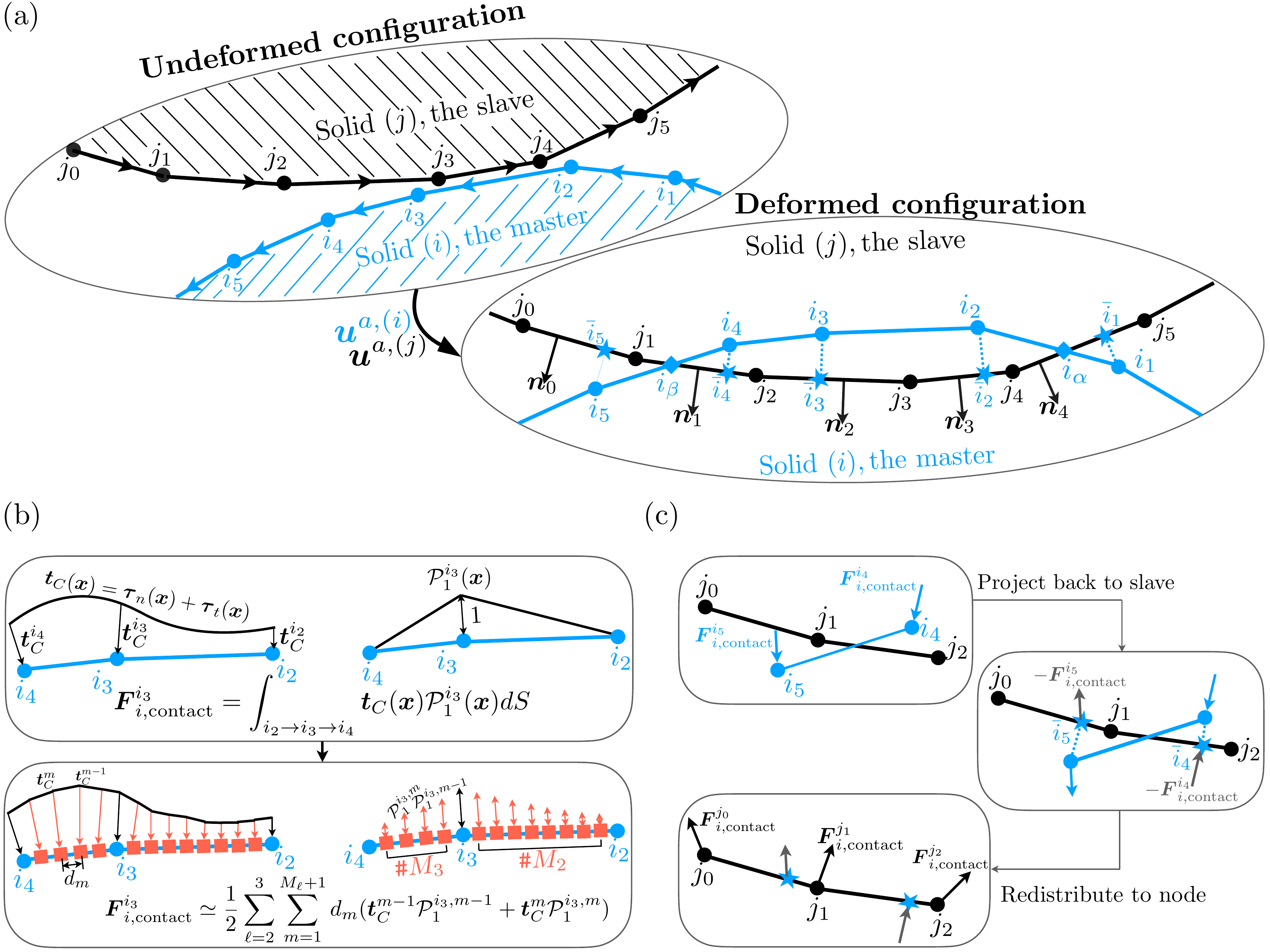}
\caption{A schematic illustration of computing contact traction between two solid bodies with given mesh conditions. (a) A generic case where three boundary nodes ($i_2, i_3$ and $i_4$) of solid $(i)$ penetrates into solid $(j)$ after a set of guessed displacement field $\bm{u}^{a,(i)}$ and $\bm{u}^{a,(j)}$. $i_\alpha$ and $i_\beta$ represent two intersection points representing, respectively, the starting point and ending point of the contact boundary $\Omega^{(i\leftarrow j)}_C$. $\bar{i}_1$, $\bar{i}_1$, $\bar{i}_2$, $\bar{i}_3$, $\bar{i}_4$, and $\bar{i}_5$ are the closest projections of nodes $i_1$ to $i_5$ on the boundary of solid $(j)$. $\bm{n}_0$, $\bm{n}_1$, $\bm{n}_2$, $\bm{n}_3$ and $\bm{n}_3$ is the outward surface normals of solid body $(j)$ in terms of its boundary connections $j_0 \rightarrow j_1$, $j_1 \rightarrow j_2$, $j_2 \rightarrow j_3$, $j_3 \rightarrow j_4$, and $j_4 \rightarrow j_5$. (b) An example of computing the nodal contact force $\bm{F}_{i,\text{contact}}^{i_3}$ at node $i_3$ in our work. A number of $M_2$ and $M_3$ material points are selected from $i_2 \rightarrow i_3$ and $i_3 \rightarrow i_4$ respectively. Contact traction $\bm{t}_C$ and values of the test function $\mathcal{P}^{i_3}$ are computed at these points to estimate $\bm{F}^{i_3}_\text{contact}$ at node $i_3$. (c) An example of projecting the computed nodal contact forces on solid $(i)$, the master, back to the boundary of solid $(j)$, the slave, following the Newton’s third law. Taking nodes $i_4$ and $i_5$ as an example. First, the computed nodal contact forces $\bm{F}^{i_4}_{i,\text{contact}}$ and $\bm{F}^{i_5}_{i,\text{contact}}$ are projected onto their closest projections $\bar{i}_4$ and $\bar{i}_5$; Second, these two projected forces, expressed as  $-\bm{F}^{i_4}_{i,\text{contact}}$ and $-\bm{F}^{i_5}_{i,\text{contact}}$, are treated as contact traction with only nonzero values at $\bar{i}_4$ and $\bar{i}_5$ (i.e., as Direc-delta distributions) and are redistributed onto the neighboring boundary nodes ($j_0$, $j_1$ and $j_2$) of solid $(j)$ to get nodal contact forces $-\bm{F}^{j_0}_{i,\text{contact}}$, $-\bm{F}^{j_1}_{i,\text{contact}}$ and $-\bm{F}^{j_2}_{i,\text{contact}}$.}
\label{contactgraph}
\end{figure}

\begin{figure}[H]
\centering
\includegraphics[width=0.9\linewidth]{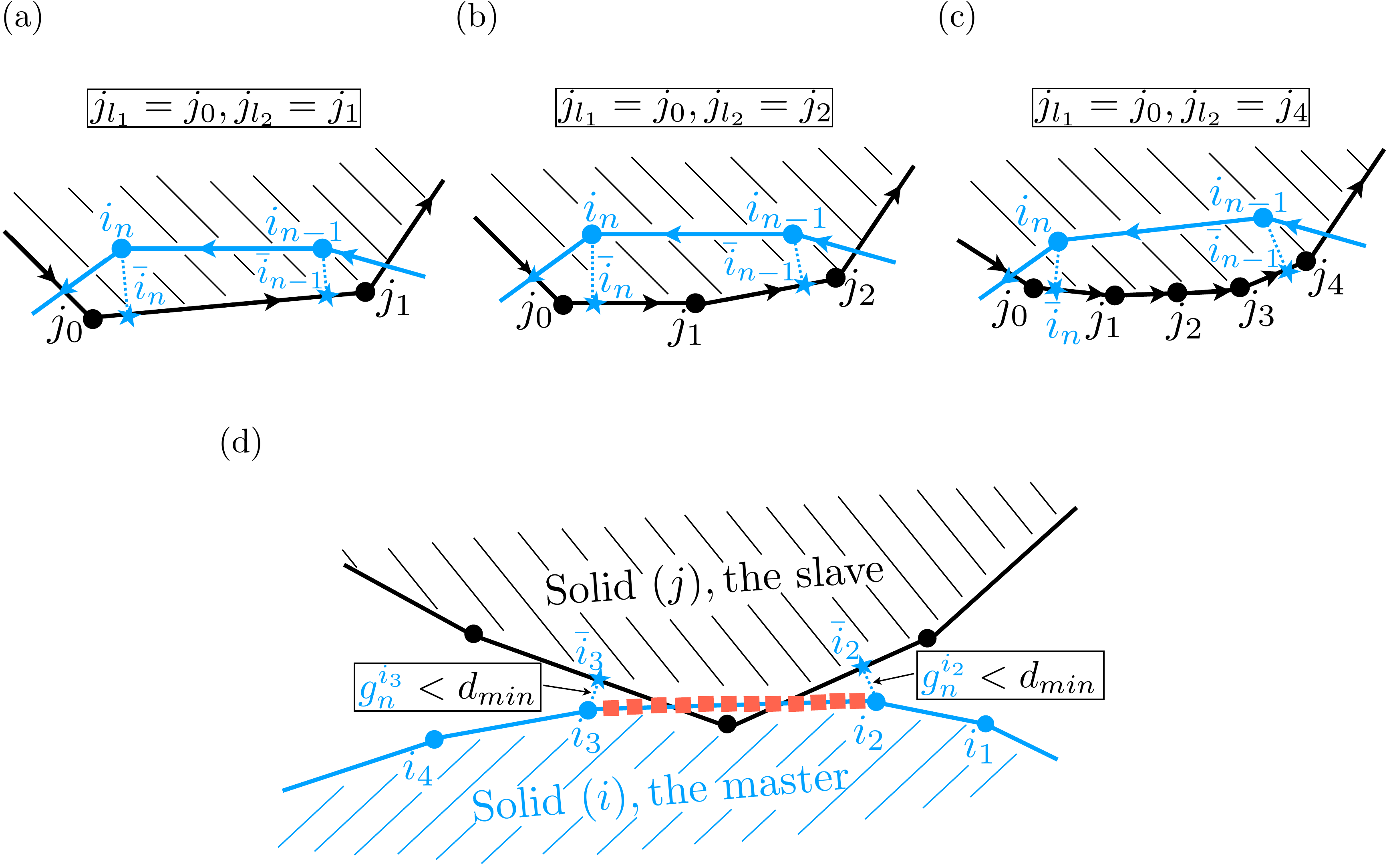}
\caption{(a) An example of the shortest path $j_{l_1}\rightarrow ...\rightarrow j_{l_2}$ having size one with $j_{l_1} = j_0$ and $j_{l_2} = j_1$. (b) An example of the shortest path $j_{l_1}\rightarrow ...\rightarrow j_{l_2}$ having size two with $j_{l_1} = j_0$ and $j_{l_2} = j_2$. (c) An example of the shortest path $j_{l_1}\rightarrow ...\rightarrow j_{l_2}$ having size four with $j_{l_1} = j_0$ and $j_{l_2} = j_4$. (d) An example where nodal contact forces being nonzero even though the nodes ($i_2$ and $i_3$) are geometrically outside the slave. These nodes are considered as long as the corresponding $g_n$ is smaller than a user-defined threshold $d_\text{min}$.}
\label{contactlawgraphspecial}
\end{figure}

In this subsection, we discuss how we compute the contact forces and jacobians using the regularized contact law Eq. \eqref{contactpenalized} in FEM where geometries of solid bodies are discretized into meshes. Again, for simplicity, here we focus on the simplest case involving only two solid bodies in contact. Consider a generic case where in the undeformed configuration $\bm{X}$ solid body $(i)$ (taken as the master) and solid body $(j)$ (taken as the slave) are free of contact, then after the application of a guessed set of nodal displacement fields $\bm{u}^{a,(i)}$ and $\bm{u}^{a,(j)}$, the two bodies come into contact in the deformed configuration $\bm{x}$, as shown in Fig. \ref{contactgraph}(a). Without loss of generality while at the same time aiding clarity, we specify also the local mesh condition near the contact region of the two contacting solid bodies.
We label the boundary FEM nodes near the contact region and arrange them counter-clockwise to represent the boundary connections as  $j_0\rightarrow j_1,j_1 \rightarrow j_2, j_2 \rightarrow j_3, j_3\rightarrow j_4$, and $j_4 \rightarrow j_5$ (for solid body $(j)$), and $i_1\rightarrow i_2, i_2\rightarrow i_3, i_3\rightarrow i_4$ and $i_4\rightarrow i_5$ (for solid body $i$). In the deformed configuration, we can identify $\partial \Omega^{i \leftarrow j}_C$ as $\partial \Omega^{i \leftarrow j} _C = i_{\alpha}\rightarrow i_2\rightarrow i_3 \rightarrow i_4 \rightarrow i_\beta$, where $i_\alpha$ and $i_\beta$ label the intersection point between line $i_1\rightarrow i_2$ and line $j_4 \rightarrow j_5$, and that between line $i_4 \rightarrow i_5$ and line $j_1 \rightarrow j_2$, respectively. The next step is to obtain $\bm{t}_C(\bm{x})$ along $\partial \Omega^{i \leftarrow j}_C$ and compute the nodal contact force at each ``active" node accordingly. These nodal contact forces are then assembled to obtain $\bm{F}^{i \leftarrow j}_{i,\text{contact}}$ shown in Algorithm 1. Taking node $i_3$ as an example (see also the top panel of Fig. \ref{contactgraph}(b)), the corresponding nodal contact force $\bm{F}^{i_3}_{i,\text{contact}}$ is given by:
 \begin{equation}
 \label{nodalforce}
 \begin{aligned}
 \bm{F}^{i_3}_{i,\text{contact}} &= \int_{\partial \Omega^{i \leftarrow j}_C}\bm{t}_C(\bm{x})\mathcal{P}_1^{i_3}(\bm{x})dS =  \int_{i_2\rightarrow i_3\rightarrow i_4}\bm{t}_C(\bm{x})\mathcal{P}_1^{i_3}(\bm{x})dS.
 \end{aligned}
 \end{equation} 
We only need to consider $i_2\rightarrow i_3\rightarrow i_4$ instead of the entire $\partial \Omega^{i \leftarrow j}_C$, because $\mathcal{P}^{i_3}_1$ is constructed as a continuous piece-wise linear function with value $1$ at node $i_3$ and value $0$ at every other node. Contact forces on other boundary nodes $i_1, i_2, i_4$ and $i_5$ can be computed analogously. We emphasize that, although geometrically nodes $i_5$ and $i_1$ are not in contact with solid $(j)$, but they are also ``active" nodes because one of their neighboring nodes $i_4$ and $i_2$ is ``active", leading to nonzero nodal contact forces at nodes $i_5$ and $i_1$. For instance, taking node $i_5$ as an example,$\bm{F}^{i_5}_{i,\text{contact}}$ is given by:
\begin{equation}
\label{nodelforcei5}
\begin{aligned}
 \bm{F}^{i_5}_{i,\text{contact}} &= \int_{\partial \Omega^{i \leftarrow j}_C}\bm{t}_C(\bm{x})\mathcal{P}_1^{i_5}(\bm{x})dS =  \int_{i_4\rightarrow i_\beta}\bm{t}_C(\bm{x})\mathcal{P}_1^{i_5}(\bm{x})dS.
\end{aligned}
\end{equation} 
As we can see, unlike ordinary Neumann boundary conditions where $\hat{\bm{t}}_N$ is often prescribed explicitly with an analytical expression, a contact traction $\bm{t}_C$ distribution depend implicitly not only on the specific geometry of the slave solid but also the specific contact law, making an analytical evaluation of Eq. \eqref{nodalforce} virtually impossible. Furthermore, the need to identify intersection points explicitly like $i_\beta$ and $i_\alpha$, which also depends on the specific mesh conditions, adds additional complications against getting the analytical evaluation. In light this, we propose using the trapezoidal rule to evaluate Eq. \eqref{nodalforce}(and Eq. \eqref{nodelforcei5}) numerically. The basic idea is to pick a finite number of points for each boundary connections, to compute $\bm{t}_C$ at each of these points, and to approximate the integration as a summation of trapezoids. By doing so, we make no explicit computation to locate those intersections points (those like $i_\alpha$ and $i_\beta$), but rather, we directly perform the trapezoidal summation over an entire connection (for instance, evaluating along $i_4 \rightarrow i_5$ instead of along $i_4 \rightarrow i_\beta$ in Eqn.\eqref{nodelforcei5}). As the number of points goes larger, we can expect more accurate results in resolving the actual integration region $i_4 \rightarrow i_\beta$. Again, let us taking node $i_3$ as an example. As a starting point, we can readily compute $\bm{t}^{i_2}_C,\bm{t}^{i_3}_C$ and $\bm{t}^{i_4}_C$ located at nodes $i_2$, $i_3$ and $i_4$ according to Eq. \eqref{contactgap} and Eq. \eqref{contactpenalized}, once we find these nodes' closest projections (points $\bar{i}_2,\bar{i}_3$ and $\bar{i}_4$ shown in Fig. \ref{contactgraph}(a)) on the boundary of the slave solid. For getting $\bm{t}_C(\bm{x})$ between node $i_2$ and $i_3$, and between nodes $i_3$ and $i_4$, we consider the following. First, in the un-deformed configuration, we select a finite number (taking as M) material points equally spaced along each boundary connections. For instance, for boundary connection $i_3 \rightarrow i_4$,  let $m=0$ represent $i_3$, $m=M+1$ represent $i_4$ and $0<m<M+1$ represent those selected in between $i_3$ and $i_4$. Then, as we know the guessed nodal displacements $\bm{u}^{a,(i)}$,  we can locate these points (shown as red squares in the bottom panel of Fig. \ref{contactgraph}) in the deformed configuration through interpolation. Next, we can compute $d_m$ which is defined as the distance between the $m-1$-th point and the $m$-th point (see the bottom panel of Fig. \ref{contactgraph}(b)):
\begin{equation}
\label{distance}
d_m = ||\bm{x}_{m-1}-\bm{x}_m||_2,
\end{equation}
where $\bm{x}_{m}$($\bm{x}_{m-1}$) is the position of the $m$-th ($m-1$-th point) in the deformed configuration. Specifically, we have $\bm{x}_{i_3} = \bm{x}_0$ and $\bm{x}_{i_4} = \bm{x}_{M+1}$. In the meantime, we can compute $\bm{t}_C = \bm{\tau}_n+\bm{\tau}_t$ at each of these points (see the bottom panel of Fig. \ref{contactgraph}(b)) following the same procedure as how $\bm{t}_C$ is computed at $i_2, i_3$ and $i_4$. Lastly, we can easily find the value of $\mathcal{P}_1^{i_3,m}$ at these points (see the bottom panel of Fig. \ref{contactgraph}(b)) since we know their positions in the deformed configuration. Here $\mathcal{P}_1^{i_3,m}$ means the value of $\mathcal{P}_1^{i_3}$ at the deformed position of the $m$-th point. Similarly, we have $\mathcal{P}_1^{i_3,0} = \mathcal{P}_1^{i_3,i_3}$ and $\mathcal{P}_1^{i_3,M+1} = \mathcal{P}_1^{i_3,i_4}$. With these ingredients at hand, $\bm{F}^{i_3}_{i,\text{contact}}$ shown in Eq. \eqref{nodalforce}  and $\bm{F}^{i_5}_{i,\text{contact}}$ shown in Eq. \eqref{nodelforcei5} are then approximated by:
\begin{equation}
\label{discretesum}
\begin{aligned}
\boldsymbol{F}_{i,\text{contact}}^{i_3} &\simeq \frac{1}{2}\sum_{\ell=2}^3\sum_{m=1}^{M_\ell+1}d_m(\boldsymbol{t}^{m-1}_C\mathcal{P}_1^{i_3,m-1}+\boldsymbol{t}^{m}_C\mathcal{P}_1^{i_3,m}),\\
\boldsymbol{F}_{i,\text{contact}}^{i_5} &\simeq \frac{1}{2}\sum_{m=1}^{M_4+1}d_m(\boldsymbol{t}^{m-1}_C\mathcal{P}_1^{i_5,m-1}+\boldsymbol{t}^{m}_C\mathcal{P}_1^{i_5,m}),
\end{aligned}
\end{equation}
where $M_2$ ($M_3, M_4$) is the number of selected points on boundary connection $i_2 \rightarrow i_3$ ($i_3 \rightarrow i_4, i_4 \rightarrow i_5$).  We note that $M$ does not have to be uniform across every boundary connection, and the selected points do not have to be equally spaced in the undeformed configuration. In our implementation we choose constant $M$ and equally spaced points for simplicity. Certainly, it may be desirable to use larger $M$ near the contact region and use very small $M$ away from the contact region. Once we get all nodal contact forces acting on solid $(i)$ taking solid $(j)$ as the slave, the next step is to project back these nodal contact forces onto the boundary of solid $(j)$ according to Newton’s third law. The idea is to project the nodal forces at ``active" nodes as point loads (with the same magnitude but in opposite direction) onto the closest corresponding projection points. For example, as shown in Fig. \ref{contactgraph}(c), nodal forces $\bm{F}^{i_4}_{i,\text{contact}}$ and $\bm{F}^{i_5}_{i,\text{contact}}$ are projected back onto the boundary of solid $(j)$ as $-\bm{F}^{i_4}_{i,\text{contact}}$ located at point $\bar{i}_4$ and $-\bm{F}^{i_5}_{i,\text{contact}}$ located at point $\bar{i}_5$. After that, these two point loads are redistributed onto the closest nodes of solid $(j)$ following the same procedure as presented in Eq. \eqref{nodalforce}:
\begin{equation}
\label{redistribution}
\begin{aligned}
\bm{F}^{j_0}_{i,\text{contact}} &= \int_{j_0\rightarrow j_1}-\bm{F}^{i_5}_{i,\text{contact}}\delta(\bm{x}-\bm{{x}}_{\bar{i_5}})\mathcal{P}_1^{j_0}(\bm{x})dS\\
 &= -\bm{F}^{i_5}_{i,\text{contact}}\mathcal{P}_1^{j_0,\bar{i}_5},\\
\bm{F}^{j_1}_{i,\text{contact}} &= \int_{j_0\rightarrow j_1\rightarrow j_2}-[\bm{F}^{i_5}_{i,\text{contact}}\delta(\bm{x}-\bm{{x}}_{\bar{i_5}})+\bm{F}^{i_4}_{i,\text{contact}}\delta(\bm{x}-\bm{{x}}_{\bar{i_4}})]\mathcal{P}_1^{j_1}(\bm{x})dS\\
&= -\bm{F}^{i_5}_{i,\text{contact}}\mathcal{P}_1^{j_1,\bar{i}_5}-\bm{F}^{i_4}_{i,\text{contact}}\mathcal{P}_1^{j_1,\bar{i}_4},\\
\bm{F}^{j_2}_{i,\text{contact}} &= \int_{j_1\rightarrow j_2}-\bm{F}^{i_4}_{i,\text{contact}}\delta(\bm{x}-\bm{{x}}_{\bar{i_4}})\mathcal{P}_1^{j_2}(\bm{x})dS\\
 &= -\bm{F}^{i_4}_{i,\text{contact}}\mathcal{P}_1^{j_2,\bar{i}_4},
\end{aligned}
\end{equation}
where $\delta(\cdot)$ is the Dirac delta function. Lastly, these redistributed nodal forces are assembled to get $\bm{F}_{i,\text{contact}}^{j\leftarrow i}$ shown in Alrogithm 1. With contact forces being computed, the corresponding contact jacobian can be computed following the above procedure by perturbing the displacement field as shown in Eq. \eqref{contactjacobian} and Eq. \eqref{perturbationvector}. We close this subsection by presenting Algorithm 2 (see Appendix for details) which outlines our implementation of computing $\bm{F}^{i\leftarrow j,k}_{i,\text{contact}}$ and $\bm{F}^{j\leftarrow i,k}_{i,\text{contact}}$ at a given iteration $k$. During the same iteration, Algorithm 2 is then repeated to get $\bm{F}^{j\leftarrow i,k}_{j,\text{contact}}$ and $\bm{F}^{i\leftarrow j,k}_{j,\text{contact}}$ by taking solid $(j)$ as the master. Algorithm 2 can be described by three parts: the first part computes the nodal traction $\bm{t}^{i_n}_C$ at each node $i_n$, the second part computes the contact traction $\bm{t}^{m}_C$ at every material point on a connection $i_{n-1} \rightarrow i_{n}$, and the last part computes the nodal contact forces using relevant $\bm{t}^{i_n}_C$ and $\bm{t}^{m}_C$. For the second part, we construct the shortest path $j_{l_1}\rightarrow ... \rightarrow j_{l_2}$ on the slave that starts from the connection where $\bar{i}_{n}$ lives and ends at the connection where $\bar{i}_{n-1}$ lives. By doing so, there is no need looping through every boundary connection of the slave to compute contact traction $\bm{t}^{m}_C$. The size of the path $j_{l_1}\rightarrow ... \rightarrow j_{l_2}$ depends on the mesh size near the contact region of both the master and the slave. It can be one (a single connection $j_0\rightarrow j_1$, see Fig. \ref{contactlawgraphspecial}(a)), two (two connections $j_0\rightarrow j_1\rightarrow j_2$, see Fig. \ref{contactlawgraphspecial}(b)) or more ($j_0\rightarrow ...\rightarrow  j_4$, see Fig. \ref{contactlawgraphspecial}(c)). 
Normally, the size of the path is mostly one or two, as we often ensure similar mesh size for the master and the slave near the contact region for better convergence. For the last part computing nodal contact forces, we consider boundary nodes whose $g_n$ are smaller than a threshold value $d_{min}$. This is to include the possible scenario where, although both nodes of a boundary connection do not penetrate into the slave, the connection itself does. As an example, as shown in Fig. \ref{contactlawgraphspecial}(d), for connection $i_2 \rightarrow i_3$ where $g_n^{i_2} < d_{min}$ and $g_n^{i_3} < d_{min}$, contact force computation is also performed. One may pick $d_\text{min}$ to be close to the local element size near the contact region. Algorithm 2 involves three more algorithms aiming at computing and correcting (when needed) the contact traction. The need for correction often happens when the projection of a node $i_n$can be made on more than one boundary connection of the slave. Such corrections are especially useful for simulating contact problems involving solid bodies with sharp corners. We leave the detailed discussions to the next section.

 \subsection{Closest projection and geometry-informed correction}
\begin{figure}[H]
\centering
\includegraphics[width=\linewidth]{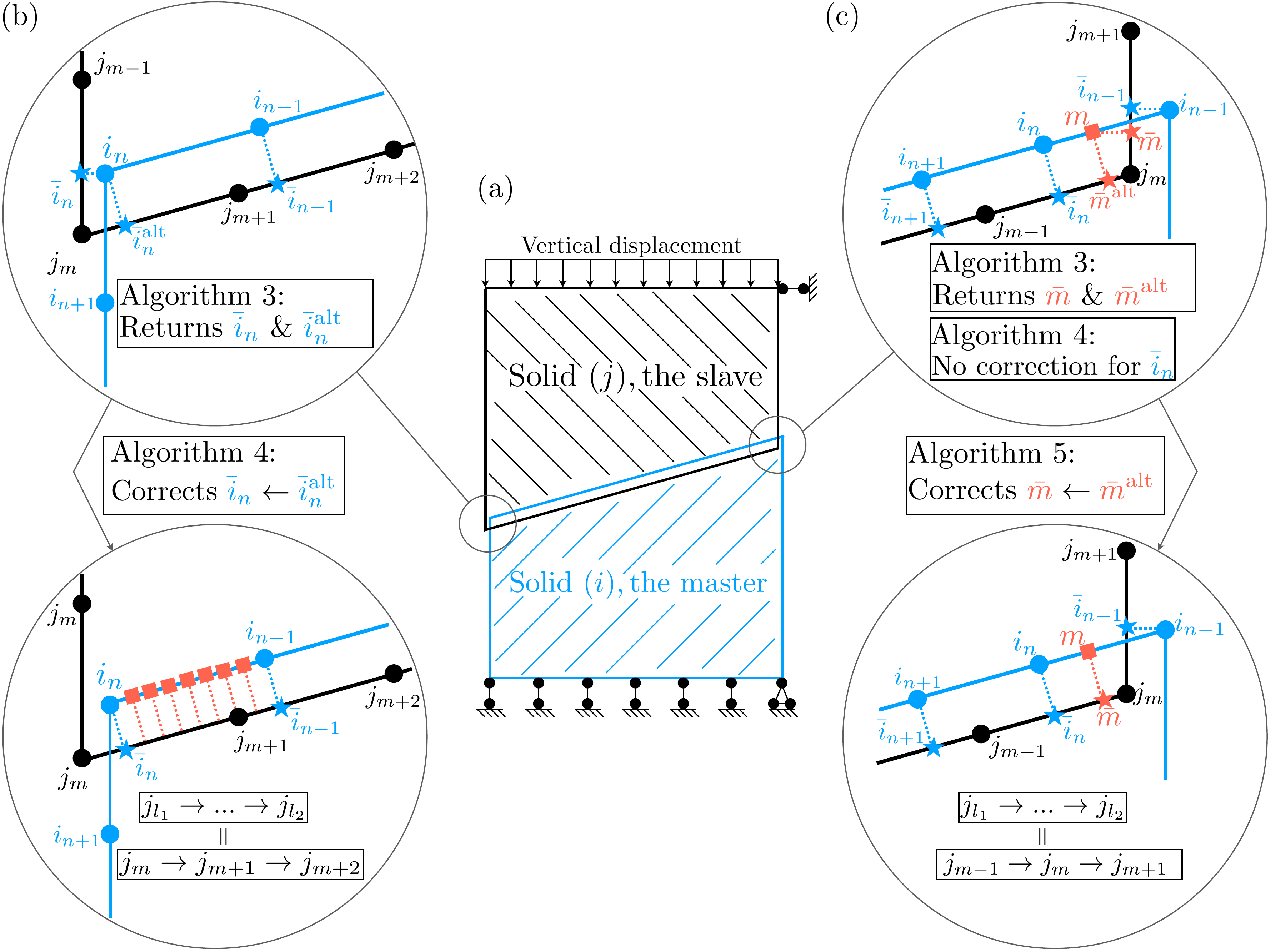}
\caption{An illustrative example of two contacting trapezoids to explain Algorithm 3, 4 and 5 aiming at ensuring correct contact traction computations. (a) The configuration of the two considered contact trapezoids. (b) A scenario where node $i_n$'s closest projection $\bar{i}_n$ needs to be corrected to its alternative $\bar{i}^{\text{alt}}_n$. Algorithm 3 returns contact information at both $\bar{i}_n$ and $\bar{i}^{\text{alt}}_n$, and Algorithm 4 checks and performs correction if needed. Once the contact information is corrected at node $i_n$, the shortest sub-path $j_{l_1}\rightarrow ... \rightarrow j_{l_2}$ is also corrected from $j_{m-1}\rightarrow j_m \rightarrow j_{m+1}$ to $j_{m}\rightarrow j_{m+1} \rightarrow j_{m+2}$, ensuring subsequent correct contact tractions of the material points on $i_{n-1}\rightarrow i_n$. (c) A scenario where Algorithm 4 performs no projection correction on node $i_n$ and thus no correction on the shortest sub-path $j_{l_1}\rightarrow ... \rightarrow j_{l_2}$ (which is $j_{m-1}\rightarrow j_m \rightarrow j_{m+1}$); Algorithm 5 accordingly checks and corrects the projections of each material point $m$ on $i_{n-1} \rightarrow i_n$ if needed.} 
\label{contactcorrection}
\end{figure}

In some cases, finding the ``correct" projection $\bar{i}_n$ for a node $i_n$ can be challenging. This is particularly true for simulating contacting solid bodies with sharp corners. To better illustrate the concept, we discuss the following sliding block geometry shown in Fig. \ref{contactcorrection}(a) as an illustrative example, where there is a very small penetration between the top solid (the master) and the bottom solid (the slave). Concerning the boundary node $i_n$ of the master as shown in Fig. \ref{contactcorrection}(b), in this particular configuration, the closest projection $\bar{i}_n$ is on $j_{m-1} \rightarrow j_m$, while the desired projection $\bar{i}^{\text{alt}}_n$ is on $j_m \rightarrow j_{m+1}$. If we proceed a simulation considering only the closest projection, the simulation is only able to give correct result when $i_n$ is on $j_m \rightarrow j_{m+1}$. This implies that our implementation would become highly unstable and sensitive to the specific geometries and positions of the solid bodies.\\

To overcome such a limitation, we construct Algorithm 3, 4 and 5 (see Appendix for details) to make sure that we locate the correct projection. Generally speaking, Algorithm 3 returns the closest projection of a node and additionally the alternative projection if there is one, provided that the node is geometrically inside the slave; Algorithm 4 check every boundary node $i_n$ which has an alternative projection, and decide whether to correct the closest projection by the alternative projection; Algorithm 5 works essentially the same as Algorithm 4, but it deals with the material points on a connection $i_{n-1} \rightarrow i_n$. 

Taking this sliding block configuration as an example, we first use Algorithm 3 which returns contact information ($\bm{g}^{i_n}_n$, $\bm{g}^{i_n,\text{alt}}_n$, $\bm{t}^{i_n}_C$ and $\bm{t}^{i_n,\text{alt}}_C$) computed from $\bar{i}_n$ and $\bar{i}^{\text{alt}}_n$ (Fig. \ref{contactcorrection}(b)). We then store these contact information and use Algorithm 4 to decide whether to make the correction $\bar{i}_n \leftarrow \bar{i}^{\text{alt}}_n$ and $\bm{t}^{i_n}_C \leftarrow \bm{t}^{i_n,\text{alt}}_C$. In Algorithm 4, we only consider projection correction for ``susceptible" nodes. We deem a node $i_n$ to be ``susceptible" if it satisfies all the following three requirements: (1) the node itself is geometrically inside the slave solid, (2) its two immediate neighboring nodes ($i_{n-1}$ and $i_{n+1}$) must not at the same time be geometrically outside or inside the slave solid, or equivalently, one immediate neighboring node must be outside and the other one must be inside, and (3) if $i_{n-1}$ is inside then $i_{n-2}$ must also be inside, or if  $i_{n+1}$ is inside then $i_{n+2}$ must also be inside. Once we deem a node $i_n$ to be susceptible and if the node has an alternative projection, we check whether $\bm{g}^{i_n,\text{alt}}_n$ instead of $\bm{g}^{i_n}_n$ is more aligned with $\bm{g}^{i_{n-1}}_n$ (let us say $i_{n-1}$ is inside). If so, we perform the projection correction $\bar{i}_n \leftarrow \bar{i}^{\text{alt}}_n$ and $\bm{t}^{i_n}_C \leftarrow \bm{t}^{i_n,\text{alt}}_C$. As an example, as shown in Fig. \ref{contactcorrection}(b), $i_n$ is ``susceptible" as $i_{n+1}$ is outside the slave while $i_{n-1}$ and $i_{n-2}$ (not shown) are inside the slave. Because $i_n$ has an alternative projection and $\bm{g}^{i_n,\text{alt}}_n$ is more aligned with $\bm{g}^{i_{n-1}}_n$ than $\bm{g}^{i_n}_n$ does, Algorithm 4 corrects the projection for node $i_n$. This correction also corrects the path $j_{l_1}\rightarrow ...\rightarrow j_{l_2}$ on which we compute the projections (using Algorithm 3) of all materials points along a boundary connection that contains the ``susceptible" node. For instance, as shown in Fig. \ref{contactcorrection}(b), for computing traction on material points on connection $i_{n-1} \rightarrow i_{n}$, the path $j_{l_1}\rightarrow ...\rightarrow j_{l_2}$ is corrected from $j_{m-1}\rightarrow j_m\rightarrow j_{m+1} \rightarrow j_{m+2}$ to $j_{m}\rightarrow j_{m+1}\rightarrow j_{m+2}$. Therefore, once Algorithm 4 corrects the projection of a ``susceptible" node, there is no longer need to consider correcting projections for material points near that node. In all, Algorithm 4 performs a nonlocal check around a ``susceptible" node by going through the projections of its neighboring nodes. Of course, the range of the nonlocality can be larger than one. One may revise the third requirement on checking a ``susceptible" node to include more nodes. For instance, let us say $i_{n+1}$ is inside the slave, then a nonlocality range of two means that both $i_{n+2}$ and $i_{n+3}$ must be inside the slave as well. 

However, it is possible that, even though a node is ``susceptible", its closest projection is the desired one and Algorithm 4 performs no correction. In these cases we will further need to correct potential erroneous projections on those material points around that node. We accordingly construct Algorithm 5 to deal with these cases.

An example is shown in Fig. \ref{contactcorrection}(c), where node $i_n$ is ``susceptible" by our definition, but it does not need correction, since $\bm{g}^{i_n}_n$ (which is associated with $\bar{i}_n$) is more aligned with $\bm{g}^{i_{n+1}}_n$ than $\bm{g}^{i_n,\text{alt}}_n$ does (which is associated with $\bar{i}^{\text{alt}}_n$, not shown but residing on $j_m \rightarrow j_{m+1}$). In this example, with $\bm{g}^{m}_n$ and $\bm{g}^{m,\text{alt}}_n$ computed (using Algorithm 3) for each material point $m$ on $i_{n-1} \rightarrow i_{n}$ and inside the slave, Algorithm 5 checks if $\bm{g}^{m,\text{alt}}_n$ instead of $\bm{g}^{m}_n$ is more aligned with $\bm{g}^{i_n}_n$. if so (as is the case in this particular example), Algorithm 5 makes the correction $\bar{m} \leftarrow \bar{m}^{\text{alt}}$ and $\bm{t}^{m}_C \leftarrow \bm{t}^{m,\text{alt}}_C$. Lastly, we point out that for scenarios where the path $j_{l_1}\rightarrow ...\rightarrow j_{l_2}$ has size one (for example, see Fig. \ref{contactlawgraphspecial}(a)), there are no alternative projections and thus no need for correction. Algorithm 5 is skipped in these scenarios. In all, with Algorithm 3, 4 and 5 assembled into Algorithm 2, and with Algorithm 2 assembled into Algorithm 1, Algorithm 1 is completed and we verify the implementation using several benchmark tests. We leave the verifications to section. \ref{subsection:verification}.
 \subsection{Extension to scenarios with incremental loading}
Combining section \ref{subsection:contactcomputation} with section \ref{subsection:FEM}, Algorithm 1 present a complete implementation for a single loading step, i.e., with a single set of Dirichlet and Neumann boundary condition. However, in many scenarios, we may wish to perform simulations with incremental loadings, i.e., with multiple sets of Dirichlet and Neumann boundary conditions. In fact, for multi-body contact mechanics problems, an implementation capable of simulating incremental loading is always preferred, since it is always better to solve the non-linear Eq. \eqref{globalresidue} through Newton-Raphson with an initial guess that is not too far from the solution. In this regard, we extend Algorithm 1 to consider displacement solutions obtained from preceding loading steps. Specifically, suppose that we already know the displacement solutions $\bm{U}^{a,1}, \bm{U}^{a,2}, ..., \bm{U}^{a,z-1}$ at each of the first $z-1$ loading steps of a system, and we have updated the configuration of the system along the way. Then, at the $z-$th loading step starting from a configuration which has been updated by all previous displacements, we revise Eq. \eqref{globalresidue} to solve the following:
 \begin{equation}
 \label{residuetotal}
 \begin{aligned}
\text{Find}\,\, \bm{U}^{a,z},\,\,\text{s.t.}\,\, \bm{R}(\bm{U}^{a,z}_{\text{tot}}) = \bm{K}(\bm{U}^{a,n}_{\text{tot}}) - \bm{F}^z_{\text{ext}} - \bm{F}_{\text{contact}}(\bm{U}^{a,z}_{\text{tot}})=\bm{0},
\end{aligned}
 \end{equation}
where $\bm{U}^{a,n}_{\text{tot}} = \sum_{l=1}^{n}\bm{U}^{a,z}$ is the total displacement after the $n$-th loading step. It then follows that $\bm{U}^{a,z} = \bm{U}^{a,z}_{\text{tot}} - \bm{U}^{a,z-1}_{\text{tot}}$, where $\bm{U}^{a,z-1}_{\text{tot}}$ is known. Since we assume small deformation, there is no need to update $\bm{K}$ when we update the configuration at the conclusion of each loading step. Of course one may still choose to update $\bm{K}$ but the results will make little difference so long as the assumption of small deformation still holds. $\bm{F}^z_{\text{ext}}$ represents a ``total" Neumann boundary condition at the $z$-th loading step. For instance, if we would like to impose an incremental pressure $\bm{p}(z) = \bm{p}_0z$ where $\bm{p}_0$ is constant and $z$ is the loading step, $\bm{F}^z_{\text{ext}}$ is computed from $\bm{p}(z)$ not $\bm{p}_0$. On the contrary, since at each loading step the configuration has been updated by previous loading steps, the Dirichlet boundary condition is instead enforced incrementally. For instance, suppose that we would like to impose a vertical displacement field $\bm{u}(z) =\bm{u}_0(z)$ where  $\bm{u}_0$ is constant and $z$ is the loading step, the Dirichlet boundary condition is enforced by $\bm{u}_0$ not $\bm{u}(z)$. W note that the Neumann and the Dirichlet boundary conditions do not need to be monotonous, or in another word, we can use Eq. \eqref{residuetotal} to simulate ``loading-unloading" contact problems. Lastly, for contact force computations, normal traction depends only on $\bm{U}^{a,z}$ since the configuration is already updated by $\bm{U}^{a,z-1}_{\text{tot}}$. However, the tangential (frictional) traction depends on $\bm{U}^{a,z}_{\text{tot}}$. Specifically, they are computed by replacing $\bm{u}^{(i)}_{i_n}$ and $\bm{u}^{(j)}_{\bar{i}_n}$ with $\sum_{\ell=1}^{z}\bm{u}^{(i),\ell}_{i_n}$ and $\sum_{\ell=1}^{z}\bm{u}^{(j),\ell}_{\bar{i}_n}$ in Eq. \eqref{contactgap}.
\subsection{Verification tests}
\label{subsection:verification}
In this section, we verify our implementation using three benchmark tests: a Brazilian disk compression test, a frictional contact test between two cylinders, and a sliding block test whose geometry has already been shortly discussed in section 2.2.3. Additionally, we consider plane-strain condition and neglect body forces. From an experimental point of view, we may consider these simulations as virtual tabletop experiments with confinements along the out-of-plane direction. Lastly, we take $k_n = k_t$ for all simulations, and we determine their values on the specific Young's modulus used in each simulation. 
\subsubsection{A Brazilian compression test}
\begin{figure}[H]
\centering
\includegraphics[width=0.93\linewidth]{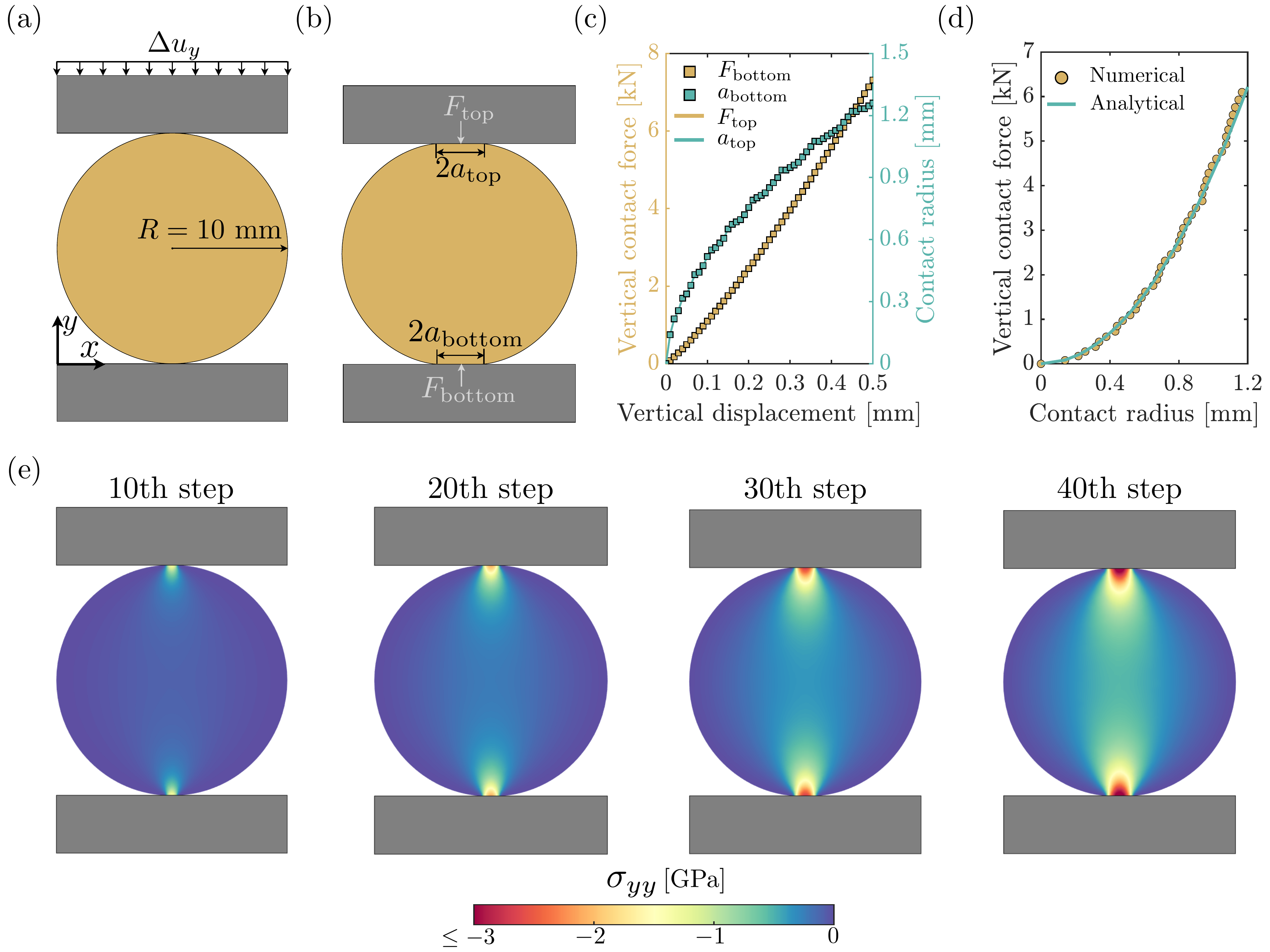}
\caption{Setup and simulation results. (a) The configuration of the Brazilian disc compression test, where the top rigid plate is displacement controlled and the bottom rigid plate is fixed. (b) An illustration of the contact radii ($a_\text{top}$, $a_\text{bottom}$) and vertical contact forces ($F_\text{top}$, $F_\text{bottom}$) at the top plate and bottom plate, respectively. (c) Simulation results of the evolutions of $a_\text{top}$, $a_\text{bottom}$ (colored in yellow) and of $F_\text{top}$, $F_\text{bottom}$ (colored green) with the imposed vertical displacement. Both contact radius and contact force measured at the top (solid lines) and bottom plate (filled squares) show excellent agreement with each other. (d) The relationship between contact radius and vertical contact force. The simulation result (filled yellow circles) shows excellent agreement with the analytical prediction (the solid green line) using Eq. \eqref{brazilianequation}. (e) Four snapshots visualizing the spatial distribution of $\sigma_{yy}$ inside the disc at the $10$th, $20$th, $30$th and $40$th loading steps.}
 \label{brazilian}
\end{figure}

We consider a disk (radius $R=10\,\,\text{mm}$, Young’s modulus E = 50 MPa and Poisson’s ratio $\nu = 0.3$) being confined between two rigid plates along the $y$ direction, see Fig. \ref{brazilian}(a). We fix the bottom plate and move the top plate downward at each quasi-static loading step with a constant vertical displacement $\Delta u_y = 0.01\,\,\text{mm}$. We simulate a total of 50 loading steps. At each loading step, after convergence we compute the contact radii ($a_\text{top}$, $a_\text{bottom}$) and vertical contact forces ($F_\text{top}$, $F_\text{bottom}$) at both the top and the bottom plate, see Fig. \ref{brazilian}(b)). We assign a small friction coefficient $\mu_s = 0.05$ to prevent potential rigid body motions, and we take $k_n = k_t = 1000\,\,\text{MPa/mm}$. As expected, due to symmetry, at equilibrium of each loading step, we have $F_\text{top} = F_\text{bottom}$ and $a_\text{top} = a_\text{bottom}$, see Fig. \ref{brazilian}(c). Further, our simulation shows excellent agreement (see Fig. \ref{brazilian}(d)) with the analytical prediction \cite{barber2002elasticity} of the vertical contact force using the measured contact radius, according to the following equation:
\begin{equation}
\label{brazilianequation}
a = 2\sqrt{\frac{F(1-\nu^2)R}{\pi E}},
\end{equation}
where we compute the contact force $F$ using the contact radius $a$ measured from the simulation. Lastly, we visualize the distribution of $\sigma_{yy}$ inside the disk (see Fig. \ref{brazilian}(e)) at four different loading steps, from which the evolution of $\sigma_{yy}$ with the loading step can be clearly observed.
 \subsubsection{A frictional contact test between two cylinders}
\begin{figure}[H]
\centering
\includegraphics[width=0.93\linewidth]{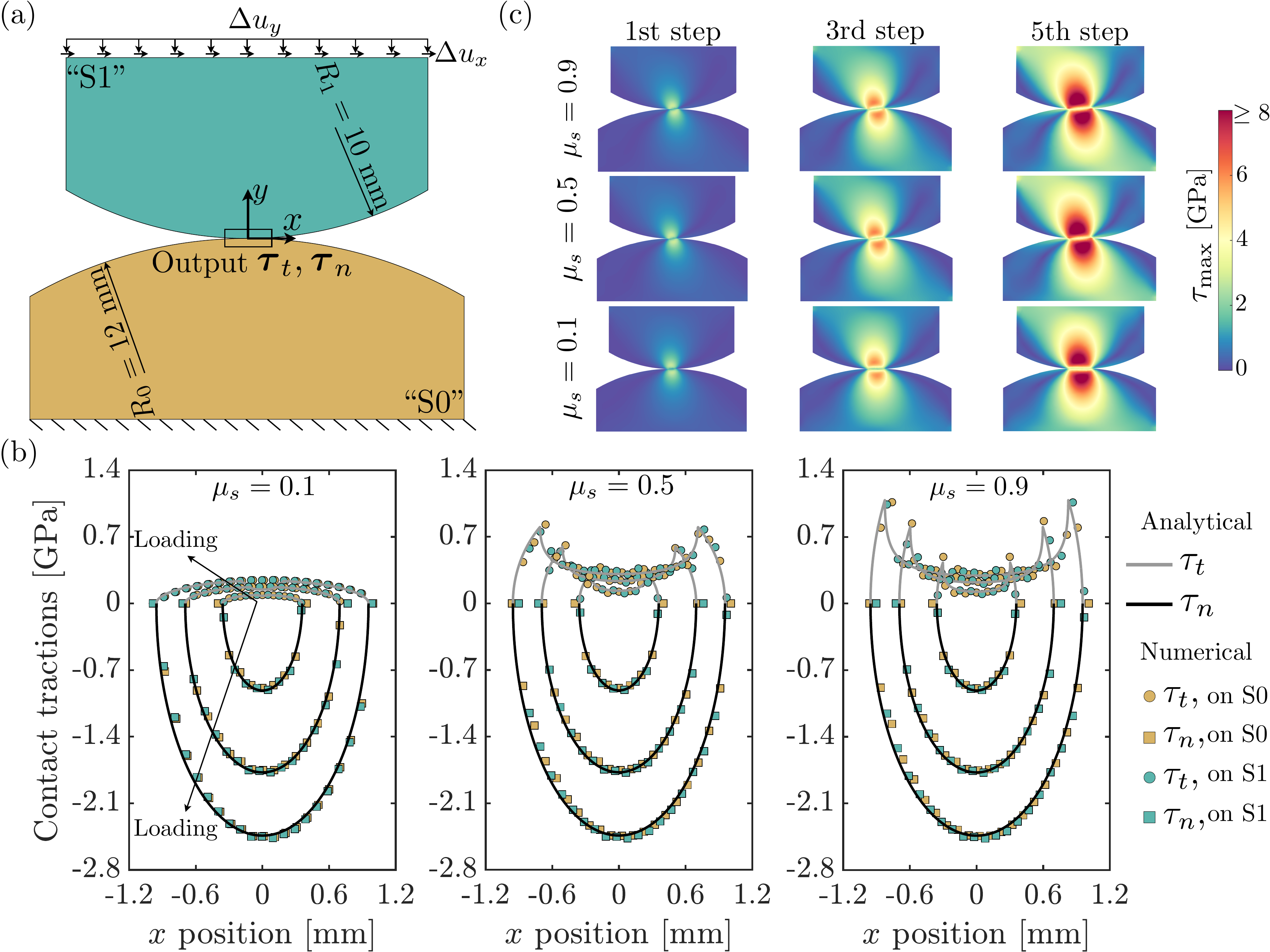}
\caption{Setup and simulation results. (a) The configuration of the frictional contact test between two cylinders (S0 and S1), where the bottom edge of S0 is fixed and the top edge of S1 is subjected to a displacement couple $(\Delta u_x, \Delta u_y)$. (b) Distributions of contact traction magnitudes $\tau_t$ (filled circles) and $\tau_n$ (filled squares) along the $x$ direction of the contact surfaces of both S0 (colored in yellow) and S1 (colored in green), under $\mu_s = 0.1$ (left figure), $\mu_s = 0.5$ (middle figure) and $\mu_s = 0.9$ (right figure) for the first, third and fifth loading steps. Analytical predictions of $\tau_t$ (solid gray lines) and $\tau_n$ (solid black lines) computed using Eq. \eqref{hertzianequation} show excellent agreement with those directly output from the simulations. (c) Visualizations of the distributions of maximum shear stress $\tau_\text{max}$ of both S0 and S1 under $\mu_s = 0.1$ (bottom row), $\mu_s = 0.5$ (middle row) and $\mu_s = 0.9$ (top row) for the first (left column), third (middle column) and fifth (right column) loading steps.}
 \label{hertzian}
\end{figure}

We consider frictional contacts between two cylindrical surfaces. The bottom solid body (termed as ``S0" hereafter) has a radius $R_0 = 12\,\,\text{mm}$ , a Young's modulus $E_0$ = 50 MPa and a Poisson's ratio $\nu_0 = 0.3$, while the top solid body (termed as ``S1" hereafter) has a radius $R_1 = 10\,\,\text{mm}$, a Young's modulus $E_1$ = 50 MPa and a Poisson's ratio $\nu_1 = 0.3$. We further fix the bottom edge of S0 and apply a displacement couple $(\Delta u_x, \Delta u_y)$ to the top edge of S1 at each loading step, see Fig. \ref{hertzian}(a), where $\Delta u_x$ points to the right while $\Delta u_y$ points downward. We simulate a total of five steps and we output both the normal ($\bm{\tau}_n$) and tangential ($\bm{\tau}_t$) traction along the contact surface (see Fig. \ref{hertzian}(a))of each solid body at each load step. In addition, we can compute the reaction forces (the tangential force $F_t$ and the normal force $F_n$) along the top edge of S1 and similarly along the bottom edge of S0. It then follows that we can compute $\tau_t$ and $\tau_n$ (the magnitudes of $\bm{\tau}_t$ and $\bm{\tau}_n$) using these reaction forces ($F_t$ and $F_n$) via the analytical solutions available in the literature \cite{barber2002elasticity}:
 \begin{equation}
 \label{hertzianequation}
 \begin{aligned}
 \tau_n(x) &= \frac{2F_n\sqrt{a^2-x^2}}{\pi a^2},\,\,\tau_t(x) = \frac{2\mu_sF_n}{\pi a^2}\left[\sqrt{a^2-x^2}-H(c^2-x^2)\sqrt{c^2-x^2}\right],\,\, -a<x<a,\\
 \text{with}\,\, a &= \left[\frac{4F_nR_0R_1}{\pi(R_0+R_1)}\left(\frac{1-\nu_0^2}{E_0}+\frac{1-\nu_1^2}{E_1}\right)\right]^{\frac{1}{2}}\,\,\text{and}\,\,c = a\left(1-\frac{F_t}{\mu_sF_n}\right)^{\frac{1}{2}},
 \end{aligned}
 \end{equation}
where $H(\cdot)$ is the Heaviside function. We compare this analytically computed traction to those directly output from the simulations. Note that, force balance implies that the reaction forces ($F_t$ and $F_n$) along the two Dirichlet boundaries of the solid bodies sum up to zero at each loading step. We take the contact penalization parameters $k_n = k_t = 1000\,\,\text{MPa/mm}$ remaining the same as in the previous example. We vary the contact friction coefficient $\mu_s = 0.1, 0.5$ and $0.9$ to see how contact traction varies with the variation of $\mu_s$. As shown in Fig. \ref{hertzian}(b), our simulation captures well the distribution of $\tau_t$ and $\tau_n$ along the contact surface belonging to both S0 and S1. We flip $\tau_n$ to be negative for a clear comparison with $\tau_t$. The three curves in each subfigure of Fig. \ref{hertzian}(b) correspond to the first, the third and the fifth loading step. We also visualize the distribution of maximum shear stress $\tau_\text{max}$ in Fig. \ref{hertzian}(c) for the three loading steps considered in Fig. \ref{hertzian}(b). As expected, smaller friction ($\mu_s = 0.1$) gives more symmetric distribution of $\tau_\text{max}$ (with respect to the $y$ direction) than larger frictions do ($\mu_s = 0.5$ and $\mu_s = 0.9$).

 \subsection{A frictional sliding block test}
\begin{figure}[H]
\centering
\includegraphics[width=0.93\linewidth]{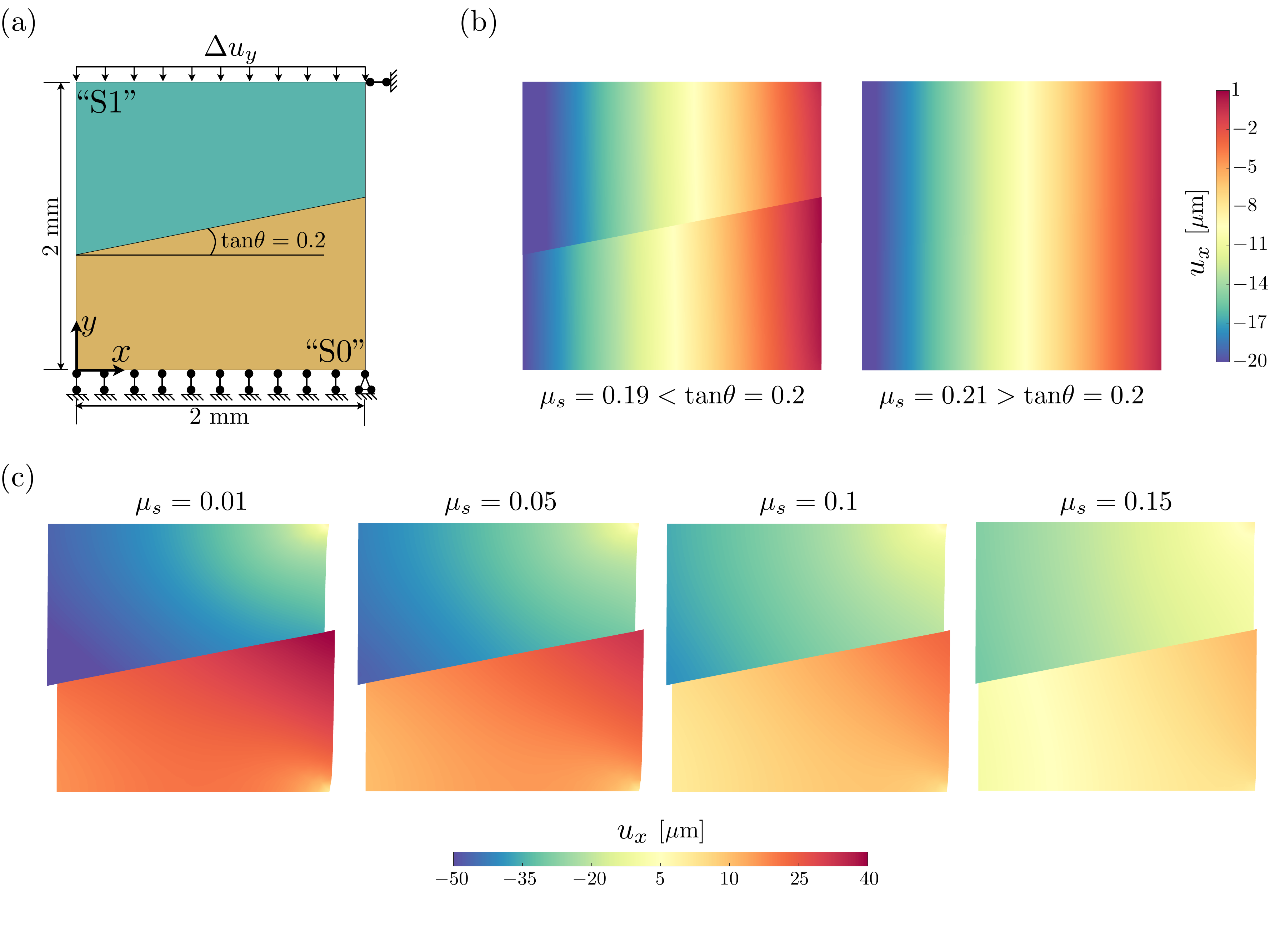}
\caption{Setup and simulation results. (a) The configuration of the two contacting trapezoids (S0 and S1) forming an interface with inclination $\text{tan}\theta = 0.2$. The bottom edge of S0 is fixed along the $y$ direction with only the right bottom corner being also fixed along the $x$ direction, and the top edge of S1 is subjected to a displacement $\Delta u_y$ along the $y$ direction with only the right top corner being fixed along the $x$ direction. (b) Simulation results showing the spatial distribution of $u_x$ over the deformed configuration under a contact friction coefficient $\mu_s = 0.19$ (left figure) and a contact friction coefficient $\mu_s = 0.21$ (right figure). (c) Similar simulation results comparing to (b) but with four different contact friction coefficients starting from the left: $\mu_s = 0.01$, $\mu_s = 0.05$, $\mu_s = 0.1$ and $\mu_s = 0.15$.}
 \label{sliding}
\end{figure}

 We consider the example presented in \cite{annavarapu2014nitsche}, where two trapezoids are vertically stacked together with an interface inclination $\text{tan}\theta = 0.2$, see Fig. \ref{sliding}(a). Again, we term the bottom trapezoid as ``S0" and the top trapezoid as ``S1", and we apply the same boundary condition as in \cite{annavarapu2014nitsche} (see also Fig. \ref{sliding}(a)), where the top edge of S0 is subjected to a vertical displacement $\Delta u_y = 0.05\,\,\text{mm}$ pointing downward. Such configuration allows for frictional sliding to initiate between S0 and S1 when the contact friction $\mu_s$ is chosen appropriately. More specifically, force balance implies that when $\mu_s < 0.2$ sliding occurs, while when $\mu_s \geq 0.2$ no sliding occurs. Same as the previous two simulations, both trapezoids have a Young's modulus E = 50 MPa and a Poisson's ratio $\nu = 0.3$, while contact penalization parameters $k_n = k_t = 1000\,\,\text{MPa/mm}$. We first assign $\mu_s = 0.19$ and $\mu_s = 0.21$ to test the implementation. As shown in Fig. \ref{sliding}(b) by the distribution of the horizontal displacement $u_x$, sliding indeed occurs when $\mu_s = 0.19$ and sticking indeed occurs when $\mu_s = 0.21$. Further, our implementation also captures relatively large sliding scenarios that can occur with even smaller contact friction coefficients, as shown in Fig. \ref{sliding}(c) that reports results with $\mu_s = 0.01, 0.05, 0.1$ and $0.15$. As expected, the relative sliding between S0 and S1 decreases when $\mu_s$ increases.

\section{Algorithm 1 outlines our multi-body contact mechanics implementation}
In Algorithm 1, the notation ``$[:]_i$"(``$[:]_j$") appearing in a matrix means all the row entries of that matrix which correspond to all nodes of solid $(i)$ (solid $(j)$).\\
\begin{algorithm}[H]
\SetAlgoLined
\KwData{Number of solid bodies $N$, undeformed configuration $\bm{X}$, $\bm{K}$ as shown in Eq. \eqref{globalstiffness}, $\bm{F}_{\text{ext}}$ as shown in Eq. \eqref{globalexternal} computed from the undeformed configuration, initial guess $\bm{U}^{a,k}$, initial residue $\bm{R^k}$ picked arbitrarily s.t. $||\bm{R}^k||_2 >$ threshold, with the iteration index $k=0$.}
\KwResult{Converged solution $\bm{U}^{a,\text{conv}}$.}
\While{$||\bm{R}^k||_2 \geq$ threshold} {
 Initialize global $\bm{F}^k_{\text{contact}} = \bm{0}$;\\
 Initialize global $\bm{J}^k_{\text{c}}$ to be all zeros;\\
 Update the deformed configuration as $\bm{x}^k = \bm{X}+\bm{U}^{a,k}$;\\
  \For{Solid i=1; i $\leq$ N; i++} {
    \For{Solid j=i+1; j $\leq$ N; j++ }{
      \If{Solid i is in contact with Solid j under $\bm{x}^k$} {
           Initialize local $\bm{J}^{i \leftrightarrow j,k}_c$ between $i$ and $j$ with all zeros;\\
           \textcolor{blue}{Compute local $\bm{F}^{i \leftarrow j,k}_{i,\text{contact}}$ and $\bm{F}^{j \leftarrow i,k}_{i,\text{contact}}$ taking $i$ as master, $j$ as slave}; \textcolor{blue}{[Alg.2]}\\       
           \textcolor{blue}{Compute local $\bm{F}^{j \leftarrow i,k}_{j,\text{contact}}$ and $\bm{F}^{i \leftarrow j,k}_{j,\text{contact}}$ taking $j$ as master, $i$ as slave}; \textcolor{blue}{[Alg.2]}\\ 
           Compute local $\bm{F}^{i \leftarrow j,k}_{\text{contact}}$ = $(\bm{F}^{i \leftarrow j,k}_{i,\text{contact}}+\bm{F}^{i \leftarrow j,k}_{j,\text{contact}})/2$;\\
           Compute local $\bm{F}^{j \leftarrow i,k}_{\text{contact}}$ = $(\bm{F}^{j \leftarrow i,k}_{i,\text{contact}}+\bm{F}^{j \leftarrow i,k}_{j,\text{contact}})/2$;\\
           Assemble $\bm{F}^{i \leftarrow j,k}_{\text{contact}}$ and $\bm{F}^{j \leftarrow i,k}_{\text{contact}}$ into the global $\bm{F}^k_{\text{contact}}$;\\
           \For {Each active node $n_i$ of Solid i w.r.t Solid j}{
             \For {Each degree of freedom of $n_i$ taking as $d_i$} {
                    Compute $h$ and $\bm{e}^{n_i+d_i,k}$ via Eq. \eqref{perturbationvector};\\
                    Repeat lines 9-12 to get $\bm{F}^{i \leftarrow j,k}_{\text{perturbed},i}$ and $\bm{F}^{j \leftarrow i,k}_{\text{perturbed},i}$;\\
                    Populate $\bm{J}^{i \leftrightarrow j,k}_c([:]_i,n_i+d_i)$ = $(\bm{F}^{i \leftarrow j,k}_{\text{perturbed},i} - \bm{F}^{i \leftarrow j,k}_{\text{contact}})/h$;\\
                    Populate $\bm{J}^{i \leftrightarrow j,k}_c([:]_j,n_i+d_i)$ = $(\bm{F}^{j \leftarrow i,k}_{\text{perturbed},i} - \bm{F}^{j \leftarrow i,k}_{\text{contact}})/h$;\\                                       
             }
           }
           \For {Each active node $n_j$ of Solid j w.r.t Solid i}{
             \For {Each degree of freedom of $n_j$ taking as $d_j$} {
                    Compute $h$ and $\bm{e}^{n_j+d_j,k}$ via Eq. \eqref{perturbationvector};\\
                    Repeat lines 9-12 to get $\bm{F}^{i \leftarrow j,k}_{\text{perturbed},j}$ and $\bm{F}^{j \leftarrow i,k}_{\text{perturbed},j}$;\\
                    Populate $\bm{J}^{i \leftrightarrow j,k}_c([:]_i,n_j+d_j)$ = $(\bm{F}^{i \leftarrow j,k}_{\text{perturbed},j} - \bm{F}^{i \leftarrow j,k}_{\text{contact}})/h$;\\
                    Populate $\bm{J}^{i \leftrightarrow j,k}_c([:]_j,n_j+d_j)$ = $(\bm{F}^{j \leftarrow i,k}_{\text{perturbed},j} - \bm{F}^{j \leftarrow i,k}_{\text{contact}})/h$;\\                                       
             }
           }
           Assemble $\bm{J}^{i \leftrightarrow j,k}_c$ into the global $\bm{J}^k_{\text{c}}$;
       }
    }
  }
  Compute $\bm{R}^k(\bm{U}^{a,k}) $ and $\bm{J}^k $ according to Eqs.\eqref{newtonraph} and \eqref{globalresidue};\\
  Enforce Dirichlet boundary conditions (if any) to $\bm{J}^k$ and $\bm{R}^k$;\\
  Update $\bm{U}^{a,k} \leftarrow  \bm{U}^{a,k}-(\bm{J}^{k})^{-1}\bm{R}^k$;\\
  Update $k \leftarrow  k+1$;
}
Set $\bm{U}^{a,\text{conv}} = \bm{U}^{a,k}$.
\caption{A FEM implementation for solving multi-body contact mechanics problem under quasi-static loading condition}
\end{algorithm}
\section{Algorithm 2 computing contact forces between two solid bodies}
\begin{algorithm}[H]
\SetAlgoLined
\KwData{The undeformed configuration $\bm{X}$ and displacement field $\bm{u}^{a,k}$ of solids $(i)$ and $(j)$, a user-defined threshold $d_\text{min}$ to initiate contact computation, contact regularization parameters $k_n$ and $k_t$, and Coulomb friction coefficient $\mu$.}
\KwResult{Nodal contact forces $\bm{F}^{i\leftarrow j,k}_{i,\text{contact}}$, $\bm{F}^{j\leftarrow i,k}_{i,\text{contact}}$, and ``active" node list $\mathcal{I}_{i,A}$.}
Initialize the deformed configuration for both solids: $\bm{x}^k = \bm{X}+\bm{u}^{a,k}$;\\
Initialize a nodal checklist $\mathcal{I}_{i,C}$ to be empty;\\
Initialize a projection checklist $\mathcal{I}_{\bar{i},C}$ and an alternative one $\mathcal{I}^{'}_{\bar{i},C}$ both to be empty;\\
Initialize a nodal traction list $\mathcal{F}_{i,C}$ and an alternative one $\mathcal{F}^{'}_{i,C}$ both to be empty;\\
Initialize a nodal gap distance list $\mathcal{G}_{i,C}$ and an alternative one $\mathcal{G}^{'}_{i,C}$ both to be empty;\\
Initialize nodal contact forces $\bm{F}^{i\leftarrow j,k}_{i,\text{contact}} = \bm{0}$ and $\bm{F}^{j\leftarrow i,k}_{i,\text{contact}}=\bm{0}$;\\
\For{every boundary node $i_n$ of solid (i)} {
    Identify the boundary node $j_m$ of solid $(j)$ whose distance to $i_n$ in $\bm{x}^k$ is the smallest;\\
    \If{the smallest distance $<d_\text{min}$}{
       Identify the closest two boundary connections $j_{m-1}\rightarrow j_m \rightarrow j_{m+1}$;\\
       \textcolor{blue}{Compute $\bar{i}_n$, $\bm{g}^{i_n}_n$,$\bm{t}_C^{i_n}$ and their alternatives $\bar{i}_n^{\text{alt}}$, $\bm{g}^{i_n,\text{alt}}_n$ and $\bm{t}_C^{i_n,\text{alt}}$}; \textcolor{blue}{[Alg.3]}\\
       \If{$||\bm{g}_n^{i_n}||_2 < d_{min}$}{
          Insert $i_n$ into $\mathcal{I}_{\bar{i},C}$, $\bar{i}_n$ into $\mathcal{I}_{\bar{i},C}$ and  $\bar{i}_n^{\text{alt}}$ into $\mathcal{I}^{'}_{\bar{i},C}$;\\
          Insert $\bm{t}_C^{i_n}$ into $\mathcal{F}_{i,C}$, $\bm{t}_C^{i_n,\text{alt}}$ into $\mathcal{F}^{'}_{i,C}$, $\bm{g}^{i_n}_n$ into $\mathcal{G}_{i,C}$ and $\bm{g}^{i_n,\text{alt}}_n$ into $\mathcal{G}^{'}_{i,C}$;\\
          Initialize $\bm{F}^{i_{n}}_{i,\text{contact}} = \bm{0}$;\\
       }
    }
}
\textcolor{blue}{Correct projections $\bar{i}_n \leftarrow \bar{i}^{\text{alt}}_n$, $\bm{g}^{i_n}_n \leftarrow \bm{g}^{i_n,\text{alt}}_n$, and $\bm{t}_C^{i_n} \leftarrow \bm{t}_C^{i_n,\text{alt}}$ if needed}; \textcolor{blue}{[Alg.4]}\\
\For{every boundary connection $i_{n-1}\rightarrow i_n$ of solid $(i)$}{
  \If{$i_{n-1} \in \mathcal{I}_{i,C}$ or $i_n \in \mathcal{I}_{i,C}$}{
    Locate $\bm{t}^{i_{n-1}}_C$ and $\bm{t}^{i_n}_C$ in $\mathcal{F}_{i,C}$ and let $\bm{t}^{0}_C = \bm{t}^{i_{n-1}}_C, \bm{t}^{M+1}_C = \bm{t}^{i_n}_C$;\\
    Compute $\mathcal{P}_1^{i_{n-1},0},\mathcal{P}_1^{i_{n-1},M+1},\mathcal{P}_1^{i_{n},0}$ and $\mathcal{P}_1^{i_{n},M+1}$;\\
    Locate $\bar{i}_{n-1}$ and $\bar{i}_n$ in $\mathcal{I}_{\bar{i},C}$ corresponding to $i_{n-1}$ and $i_{n}$;\\
    Determine the shortest path $j_{l_1}\rightarrow ...\rightarrow j_{l_2}$ from $\bar{i}_{n-1}$ to $\bar{i}_n$; [see Figs. \ref{contactlawgraphspecial}(a)-(c)]\\
     \For{every material point $\bm{x}_m$ with $0<m<M+1$ on $i_{n-1} \rightarrow i_{n}$}{
       Compute the corresponding $\mathcal{P}_1^{i_{n-1},m}$, $\mathcal{P}_1^{i_{n},m}$, and $d_m$ according to Eq. \eqref{distance};\\
       Repeat line 7 and 9 using $j_{l_1} \rightarrow ...\rightarrow j_{l_2}$ to get a sub-path $j_{s-1} \rightarrow j_{s} \rightarrow j_{s+1}$; \\
       \textcolor{blue}{Compute $\bar{m}$, $\bm{g}^{m}_n$,$\bm{t}_C^{m}$ and their alternatives $\bar{m}^{\text{alt}}$, $\bm{g}^{m,\text{alt}}_n$ and $\bm{t}_C^{m,\text{alt}}$}; \textcolor{blue}{[Alg.3]}\\
       \textcolor{blue}{Correct projections $\bar{m} \leftarrow \bar{m}^{\text{alt}}$, $\bm{g}^{m}_n \leftarrow \bm{g}^{m,\text{alt}}_n$, and $\bm{t}_C^{m} \leftarrow \bm{t}_C^{m,\text{alt}}$ if needed}; \textcolor{blue}{[Alg.5]}\\     
     }
     Update $\bm{F}^{i_{n-1}}_{i,\text{contact}}$ and $\bm{F}^{i_{n}}_{i,\text{contact}}$ with contributions from $i_{n-1}\rightarrow i_n$ using Eq. \eqref{discretesum};\\
  }
  Populate $\bm{F}^{i_{n-1}}_{i,\text{contact}}$ and $\bm{F}^{i_{n}}_{i,\text{contact}}$ into $\bm{F}^{i\leftarrow j,k}_{i,\text{contact}}$;\\
  Project back onto solid $(j)$ through $\bm{F}^{\bar{i}_{n-1}}_{i,\text{contact}} = -\bm{F}^{i_{n-1}}_{i,\text{contact}}$ and $\bm{F}^{\bar{i}_{n}}_{i,\text{contact}} = -\bm{F}^{i_{n}}_{i,\text{contact}}$;\\
  Redistribute to get nodal forces following Eq. \eqref{redistribution} and populate them into $\bm{F}^{j\leftarrow i,k}_{i,\text{contact}}$;\\
}
Insert every $i_n$ into $\mathcal{I}_{i,A}$ as long as $||\bm{F}^{i_{n}}_{i,\text{contact}}||_2 > 0$;\\
Return $\bm{F}^{i\leftarrow j,k}_{i,\text{contact}}$, $\bm{F}^{j\leftarrow i,k}_{i,\text{contact}}$, and $\mathcal{I}_{i,A}$.
\caption{Compute the contact forces $\bm{F}^{i\leftarrow j,k}_{i,\text{contact}}$ and $\bm{F}^{j\leftarrow i,k}_{i,\text{contact}}$, taking solid $(i)$ as the master and solid $(j)$ as the slave at the $k$-th iteration.}
\end{algorithm}
\section{Algorithm 3 computing contact projection for one boundary node of the master solid}
\begin{figure}[H]
\centering
\includegraphics[width=0.91\linewidth]{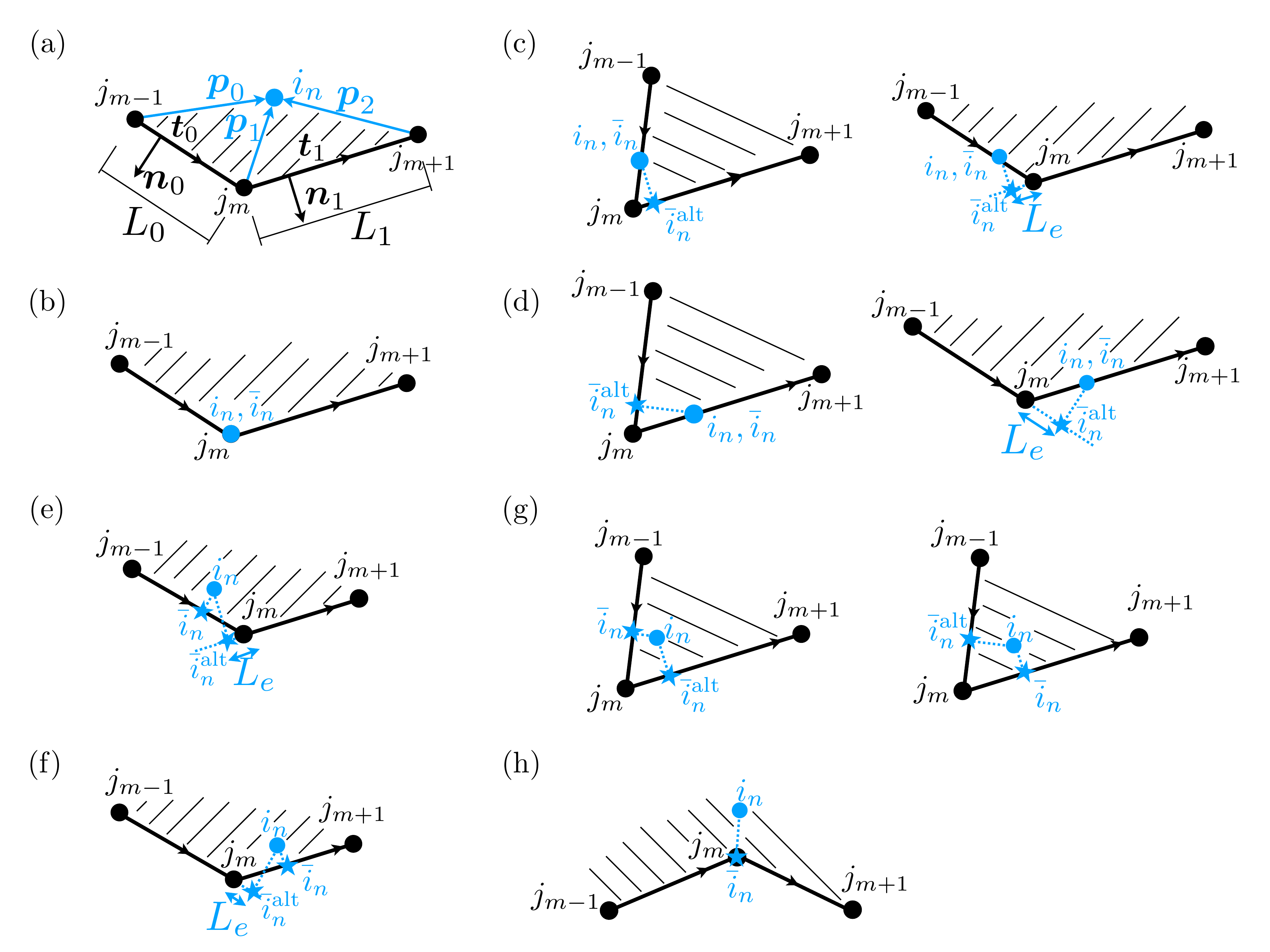}
\caption{A configuration of one node $i_n$ and its several possible relative positions with respect to its nearest two boundary connections $j_{m-1} \rightarrow j_m\rightarrow j_{m+1}$. (a) An example of $i_n$ is inside solid body $(j)$ determined from its relative position from $j_0 \rightarrow j_1\rightarrow j_2$. $\bm{p}_0$, $\bm{p}_1$, and $\bm{p}_2$ are vectors pointing from nodes $j_0$, $j_1$ and $j_2$ to node $i_n$, respectively; $\bm{t}_0, \bm{n}_0$ and $\bm{t}_1, \bm{n}_1$ are the unit tangent and outward normal vectors associated with $j_0\rightarrow j_1$ and $j_1\rightarrow j_2$, respectively; $L_0$ and $L_1$ are the lengths of $j_{m-1}\rightarrow j_m$ and $j_m\rightarrow j_{m+1}$, respectively. (b) $i_n$ is exactly on $j_m$. (c) $i_n$ is exactly on $j_{m-1}\rightarrow j_m$ and can be projected onto $j_m\rightarrow j_{m+1}$ (left figure) or weakly projected onto $j_m\rightarrow j_{m+1}$ if $L_e \leq e_s L_1$ where $e_s$ is a user-defined small constant. (d) A similar scenario to (c) but $i_n$ is exactly on $j_m\rightarrow j_{m+1}$. (e) $i_n$ is inside solid body $(j)$ and can be projected onto $j_{m-1}\rightarrow j_m$ and weakly projected onto $j_{m}\rightarrow j_{m+1}$. (f) A similar scenario to (e) but $i_n$ can be projected onto $j_{m}\rightarrow j_{m+1}$ and weakly projected onto $j_{m-1}\rightarrow j_{m}$. (g) $i_n$ is inside solid body $(j)$ and can be projected onto both $j_{m-1}\rightarrow j_{m}$ (where $\bar{i}_n$ resides, the left figure) and $j_{m}\rightarrow j_{m+1}$ (where $\bar{i}_n$ resides, the right figure). (h) $i_n$ is inside solid body $(j)$ but can be projected onto neither $j_{m-1}\rightarrow j_{m}$ nor $j_{m}\rightarrow j_{m+1}$, $i_n$ is projected onto $j_m$, i.e., $\bar{i}_n$ overlaps with $j_m$.}
\label{contactdetect}
\end{figure}

To better present Algorithm 3, we first introduce the following notations and shorthand expressions. In addition, to aid with visualization, we introduce Fig. \ref{contactdetect} containing several graph examples that are referenced in Algorithm 3.
 \begin{itemize}
\item $\bm{p}_0 = \bm{x}_{i_n}-\bm{x}_{j_{m-1}}$, $\bm{p}_1 = \bm{x}_{i_n}-\bm{x}_{j_{m}}$ and $\bm{p}_2 = \bm{x}_{i_n}-\bm{x}_{j_{m+1}}$ satisfying $||\bm{p}_1||_2 \leq ||\bm{p}_0||_2$ and  $||\bm{p}_1||_2 \leq ||\bm{p}_2||_2$  via the selection done at line 9 of Algorithm 2, see Fig. \ref{contactdetect}(a);
\item $\bm{t}_0 = \frac{\bm{x}_{j_m}-\bm{x}_{j_{m-1}}}{||\bm{x}_{j_m}-\bm{x}_{j_{m-1}}||_2}$ and $\bm{t}_1 =  \frac{\bm{x}_{j_{m+1}}-\bm{x}_{j_{m}}}{||\bm{x}_{j_{m+1}}-\bm{x}_{j_{m}}||_2}$ the unit tangents of $j_{m-1}\rightarrow j_m$ and $j_{m}\rightarrow j_{m+1}$ respectively, see Fig. \ref{contactdetect}(a);
\item $\bm{n}_0$ and $\bm{n}_1$ the outward normals of $j_{m-1}\rightarrow j_m$ and $j_{m}\rightarrow j_{m+1}$ satisfying $\bm{n}_0\cdot\bm{t}_0=0$ and $\bm{n}_1\cdot\bm{t}_1=0$ respectively, see Fig. \ref{contactdetect}(a);
\item $L_0 = ||\bm{x}_{j_m}-\bm{x}_{j_{m-1}}||_2$ and $L_1 = ||\bm{x}_{j_{m+1}}-\bm{x}_{j_{m}}||_2$ the length of $j_{m-1}\rightarrow j_m$ and $j_{m}\rightarrow j_{m+1}$ respectively, see Fig. \ref{contactdetect}(a);
\item ``Proj0" means that $i_n$ can be projected onto $j_{m-1}\rightarrow j_m$: $(\bm{p}_0\cdot\bm{t}_0)(\bm{p}_1\cdot\bm{t}_0) \leq 0$;
\item ``WProj0" means that $i_n$ can be weakly projected onto $j_{m-1}\rightarrow j_m$: $(\bm{p}_0\cdot\bm{t}_0)(\bm{p}_1\cdot\bm{t}_0) > 0$ and $L_e \leq e_s L_0$ where $L_e = (\bm{x}_{i_n}-\bm{x}_{j_m})\cdot \bm{t}_0$, where $e_s$ is a user-defined small constant, see the second figure in Fig. \ref{contactdetect}(d) and Fig. \ref{contactdetect}(f);
\item ``Proj1" means that $i_n$ can be projected onto $j_{m}\rightarrow j_{m+1}$: $(\bm{p}_1\cdot\bm{t}_1)(\bm{p}_2\cdot\bm{t}_1) \leq 0$;
\item ``WProj1" means that $i_n$ can be weakly projected onto $j_{m}\rightarrow j_{m+1}$: $(\bm{p}_1\cdot\bm{t}_1)(\bm{p}_2\cdot\bm{t}_1) > 0$ and $L_e \leq e_s L_1$ where $L_e = (\bm{x}_{i_n}-\bm{x}_{j_m})\cdot \bm{t}_0$, see the second figure in Fig. \ref{contactdetect}(c) and \ref{contactdetect}(e) for example;
\item ``In0" means that $i_n$ is inside the slave in terms of $j_{m-1}\rightarrow j_m$: $\bm{p}_0\cdot \bm{n}_0 < 0$;
\item ``In1" means that $i_n$ is inside the slave in terms of $j_{m}\rightarrow j_{m+1}$: $\bm{p}_1\cdot \bm{n}_1 < 0$.
\end{itemize}
\,
\begin{algorithm}[H]
\SetAlgoLined
\KwData{$i_n$, $\bm{X}_{i_n}$, $\bm{u}^{a,k}_{i_n}$, $j_{m-1}\rightarrow j_m \rightarrow j_{m+1}$, $\bm{X}_{j_{m-1}}$, $\bm{X}_{j_m}$, $\bm{X}_{j_{m+1}}$, $\bm{u}^{a,k}_{j_{m-1}}$, $\bm{u}^{a,k}_{j_m}$, and $\bm{u}^{a,k}_{j_{m+1}}$.}
\KwResult{$\bar{i}_n$, $\bar{i}^{\text{alt}}_n$, $\bm{g}^{i_n}_n$, $\bm{g}^{i_n,\text{alt}}_n$, $\bm{t}^{i_n}_C$, and $\bm{t}^{i_n,\text{alt}}_C$.}
Set $\bm{x}_{i_n}$ = $\bm{X}_{i_n}+\bm{u}^{a,k}_{i_n}$ and initialize $\bar{i}_n$, $\bar{i}^{\text{alt}}_n$, $\bm{g}^{i_n}_n$, $\bm{g}^{i_n,\text{alt}}_n$, $\bm{t}^{i_n}_C$, and $\bm{t}^{i_n,\text{alt}}_C$;\\
Set $\bm{x}_{j_{m-1}}$ = $\bm{X}_{j_{m-1}}+\bm{u}^{a,k}_{j_{m-1}}$, $\bm{x}_{j_{m1}}$ = $\bm{X}_{j_{m}}+\bm{u}^{a,k}_{j_{m}}$ and $\bm{x}_{j_{m+1}}$ = $\bm{X}_{j_{m+1}}+\bm{u}^{a,k}_{j_{m+1}}$;\\
\uIf{$i_n$ overlaps with $j_m$}{
   Let $\bar{i}_n = j_m$, $\bm{g}^{i_n} = \bm{0}$ and set $\bm{t}^{i_n}_C = \bm{0}$; [see Fig. \ref{contactdetect}(b)]\\
}
\uElseIf{$i_n$ is on $j_{m-1}\rightarrow j_m$}{
 Let $\bar{i}_n$ = $i_n$, $\bm{g}^{i_n}_n = \bm{0}$,$\bm{t}^{i_n}_C = \bm{0}$;\\
  \If{(``In1" and ``Proj1") or (``In1" and ``WProj1")}{
     Compute $\bar{i}^{\text{alt}}_n$, $\bm{g}^{i_n,\text{alt}}_n$ and $\bm{t}^{i_n,\text{alt}}_C$ based on Eqs.\eqref{contactgap}\eqref{contactpenalized}; [see Fig. \ref{contactdetect}(c)]\\
  }
}
\uElseIf{$i_n$ is on $j_{m}\rightarrow j_{m+1}$}{
   Let $\bar{i}_n$ = $i_n$, $\bm{g}^{i_n}_n = \bm{0}$,$\bm{t}^{i_n}_C = \bm{0}$;\\
  \If{(``In0" and ``Proj0") or (``In0" and ``WProj0")}{
     Compute $\bar{i}^{\text{alt}}_n$, $\bm{g}^{i_n,\text{alt}}_n$ and $\bm{t}^{i_n,\text{alt}}_C $ using Eqs.\eqref{contactgap}\eqref{contactpenalized}; [see Fig. \ref{contactdetect}(d)]\\
  }
}
\Else{
  \eIf{``Proj0" or ``Proj1"}{
        \uIf{``Proj0" but not ``Proj1"}{
          Compute $\bar{i}_n$, $\bm{g}^{i_n}_n$ using Eq. \eqref{contactgap} and set $\bm{t}^{i_n}_C = \bm{0}$;\\
           \If{``In0"}{
              Compute $\bm{t}^{i_n}_C$ using Eq. \eqref{contactpenalized};\\
              \If{``In1" and ``Wproj1"}{
                  Compute $\bar{i}^{\text{alt}}_n$, $\bm{g}^{i_n,\text{alt}}_n$ and $\bm{t}^{i_n,\text{alt}}_C$ using Eqs.\eqref{contactgap}\eqref{contactpenalized}; [see Fig. \ref{contactdetect}(e)]\\
              }
           }
        }
        \uElseIf{``Proj1" but not ``Proj0"}{
          Compute $\bar{i}_n$, $\bm{g}^{i_n}_n$ using Eq. \eqref{contactgap} and set $\bm{t}^{i_n}_C = \bm{0}$;\\
           \If{``In1"}{
              Compute $\bm{t}^{i_n}_C$ using Eq. \eqref{contactpenalized};\\
              \If{``In0" and ``Wproj0"}{
                  Compute $\bar{i}^{\text{alt}}_n$, $\bm{g}^{i_n,\text{alt}}_n$ and $\bm{t}^{i_n,\text{alt}}_C$ using Eqs.\eqref{contactgap}\eqref{contactpenalized}; [see Fig. \ref{contactdetect}(f)]\\
              }
           }
        }
        \Else{
           Compute $\bar{i}_n$, $\bar{i}^{\text{alt}}_n$, $\bm{g}^{i_n}_n$ and $\bm{g}^{i_n,\text{alt}}_n$ s.t. $||\bm{g}^{i_n}_n||_2 \leq ||\bm{g}^{i_n,\text{alt}}_n||_2$ using Eq. \eqref{contactgap};\\ 
           Set $\bm{t}^{i_n}_C = \bm{t}^{i_n,\text{alt}}_C = \bm{0}$;\\
           \If{``In0" and ``In1"}{
               Compute $\bm{t}^{i_n}_C$ and $\bm{t}^{i_n,\text{alt}}_C$ using Eq. \eqref{contactpenalized}; [see Fig. \ref{contactdetect}(g)]\\
           }
        }
  }{
    Let $\bar{i}_n = j_m$, $\bm{g}^{i_n} = \bm{p}_1$ and set $\bm{t}^{i_n}_C = \bm{0}$;\\ 
    \If{``In0" and ``In1"}{
       Compute $\bm{t}^{i_n}_C$ using Eq. \ref{contactpenalized}, see Fig. \ref{contactdetect}(h);\\
    }
  }
}
Return $\bar{i}_n$, $\bar{i}^{\text{alt}}_n$, $\bm{g}^{i_n}_n$, $\bm{g}^{i_n,\text{alt}}_n$, $\bm{t}^{i_n}_C$, and $\bm{t}^{i_n,\text{alt}}_C$.
\caption{Compute projections and tractions of a give node $i_n$.}
\end{algorithm}
\section{Algorithm 4 and Algorithm 5 checking and correcting projection if needed}
\begin{algorithm}[H]
\SetAlgoLined
\KwData{$\mathcal{I}_{i,C}$, $\mathcal{I}_{\bar{i},C}$, $\mathcal{I}^{'}_{\bar{i},C}$, $\mathcal{F}_{i,C}$, $\mathcal{F}^{'}_{i,C}$, $\mathcal{G}_{i,C}$ and $\mathcal{G}^{'}_{i,C}$.}
\For{every $i_n \in \mathcal{I}_{i,C}$}{
  \If {$i_{n-2}$, $i_{n-1}$, $i_{n+1}$ and $i_{n+2} \in \mathcal{I}_{i,C}$}{
    Locate $\bm{t}^{i_{n-2}}_C$, $\bm{t}^{i_{n-1}}_C$, $\bm{t}^{i_{n}}_C$, $\bm{t}^{i_{n+1}}_C$ and $\bm{t}^{i_{n+2}}_C$ from $\mathcal{F}_{i,C}$, and $\bm{t}^{i_n,\text{alt}}_C$ from $\mathcal{F}^{'}_{i,C}$;\\
    \If {$||\bm{t}^{i_{n}}_C||_2 > 0$ or $||\bm{t}^{i_{n},\text{alt}}_C||_2 > 0$}{
         \If {$||\bm{t}^{i_{n-1}}_C||_2 = 0$ and $||\bm{t}^{i_{n+1}}_C||_2 > 0$ and $||\bm{t}^{i_{n+2}}_C||_2 > 0$}{
             Locate $\bar{i}_n$, $\bar{i}^{\text{alt}}_n$, $\bm{g}^{i_n}_n$, $\bm{g}^{i_{n+1}}_n$ and $\bm{g}^{i_n,\text{alt}}_n$ from $\mathcal{I}_{\bar{i},C}$, $\mathcal{I}^{'}_{\bar{i},C}$, $\mathcal{G}_{i,C}$ and $\mathcal{G}^{'}_{i,C}$;\\
             \If{$\bm{g}^{i_{n+1}}_n \cdot \bm{g}^{i_n,\text{alt}}_n > \bm{g}^{i_{n+1}}_n \cdot \bm{g}^{i_n}_n$}{
                Correct projection $\bar{i}_n \leftarrow \bar{i}^{\text{alt}}_n, \bm{g}^{i_n}_n \leftarrow \bm{g}^{i_n,\text{alt}}_n$ and $\bm{t}^{i_{n}}_C \leftarrow \bm{t}^{i_n,\text{alt}}_C$;\\
             }
         }
         \If {$||\bm{t}^{i_{n-2}}_C||_2 > 0$ and $||\bm{t}^{i_{n-1}}_C||_2 > 0$ and $||\bm{t}^{i_{n+1}}_C||_2 = 0$}{
             Locate $\bar{i}_n$, $\bar{i}^{\text{alt}}_n$, $\bm{g}^{i_n}_n$, $\bm{g}^{i_{n-1}}_n$ and $\bm{g}^{i_n,\text{alt}}_n$ from $\mathcal{I}_{\bar{i},C}$, $\mathcal{I}^{'}_{\bar{i},C}$, $\mathcal{G}_{i,C}$ and $\mathcal{G}^{'}_{i,C}$;\\
             \If{$\bm{g}^{i_{n-1}}_n \cdot \bm{g}^{i_n,\text{alt}}_n > \bm{g}^{i_{n-1}}_n \cdot \bm{g}^{i_n}_n$}{
                Locate $\bm{t}^{i_n,\text{alt}}_C$ from $\mathcal{F}^{'}_{i,C}$;\\
                Correct projection $\bar{i}_n \leftarrow \bar{i}^{\text{alt}}_n, \bm{g}^{i_n}_n \leftarrow \bm{g}^{i_n,\text{alt}}_n$ and $\bm{t}^{i_{n}}_C \leftarrow \bm{t}^{i_n,\text{alt}}_C$;\\
             }   
         }
     }
  }
}
\caption{Contact projection correction for each node $i_n \in  \mathcal{I}_{i,C}$.}
\end{algorithm}
\begin{algorithm}[H]
\SetAlgoLined
\KwData{$\bm{g}^{m}_n$, $\bm{g}^{m,\text{alt}}_n$, $\bm{g}^{i_{n-1}}_n$, $\bm{g}^{i_n}_n$, $\bm{t}^{m}_C$, $\bm{t}^{m,\text{alt}}_C$, $\bm{t}^{i_{n-1}}_C$ and $\bm{t}^{i_n}_C$.}
\If{$||\bm{t}^{m}_C||_2 > 0$ or $||\bm{t}^{m,\text{alt}}_C||_2 > 0$} {
    \If {$||\bm{t}^{i_{n-1}}_C||_2 > 0$ and $||\bm{t}^{i_{n}}_C||_2 = 0$}{
        \If{$\bm{g}^{i_{n-1}}_n\cdot \bm{g}^{m,\text{alt}}_n > \bm{g}^{i_{n-1}}_n\cdot \bm{g}^{m}_n$}{
           Correct projection $\bm{g}^{m}_n \leftarrow \bm{g}^{m,\text{alt}}_n, \bm{t}^{m}_C \leftarrow \bm{t}^{m,\text{alt}}_C$;\\
        }
    }
    \If {$||\bm{t}^{i_{n-1}}_C||_2 = 0$ and $||\bm{t}^{i_{n}}_C||_2 > 0$}{
        \If{$\bm{g}^{i_n}_n\cdot \bm{g}^{m,\text{alt}}_n > \bm{g}^{i_n}_n\cdot \bm{g}^{m}_n$}{
           Correct projection $\bm{g}^{m}_n \leftarrow \bm{g}^{m,\text{alt}}_n, \bm{t}^{m}_C \leftarrow \bm{t}^{m,\text{alt}}_C$;\\
        }
    }
}
\caption{Contact projection correction for a material point $m$ on a connection $i_{n-1} \rightarrow i_n$.}
\end{algorithm}
\bibliographystyle{unsrt}
\bibliography{References}
\end{document}